\documentclass[useAMS,usenatbib]{mn2e}
\usepackage{graphicx,amsmath,txfonts}

\newcommand{\st}{{\it St}}
\newcommand{\alphadiff}{\alpha_{\rm diff}}
\newcommand{\abin}{a_{\rm bin}}
\newcommand{\Tbin}{T_{\rm bin}}
\newcommand{\Torb}{T_{\rm orb}}
\newcommand{\racc}{r_{\rm acc}}
\newcommand{\Hd}{H_{\rm d}}
\newcommand{\msp}{m_{\rm p}}
\newcommand{\tc}{t_{\rm c}}
\newcommand{\tf}{t_{\rm f}}
\newcommand{\ts}{t_{\rm s}}
\newcommand{\tg}{t_{\rm g}}
\newcommand{\cs}{c_{\rm s}}

\newcommand{\omegas}{\omega_{\rm s}}
\newcommand{\Omegap}{\Omega_{\rm p}}
\newcommand{\BV}{Brunt-V\"ais\"al\"a\ }
\newcommand{\ed}{e_{\rm d}}
\newcommand{\rp}{r_{\rm p}}
\newcommand{\ap}{a_{\rm p}}
\newcommand{\qp}{q_{\rm p}}
\newcommand{\taucor}{\tau_{\rm cor}}
\newcommand{\rhod}{\rho_{\rm d}}
\newcommand{\Ep}{E_{\rm p}}
\newcommand{\Ek}{E_{\rm k}}
\newcommand{\vK}{v_{\rm K}}
\newcommand{\vp}{v_{\rm p}}
\newcommand{\vc}{v_{\rm c}}

\title[Hydrodynamical turbulence in circumbinary discs and the formation of circumbinary planets]{Hydrodynamical turbulence in eccentric circumbinary discs and its impact on the in situ formation of circumbinary planets}
\author[A. Pierens, C.~P. McNally, R.~P. Nelson]{Arnaud Pierens $^{1}$, Colin P. McNally$^{2}$, Richard P. Nelson$^{2}$ \\
$^1$Laboratoire d'Astrophysique de Bordeaux, CNRS and Universit{\'e} de Bordeaux, All{\'e}e Geoffroy St. Hilaire, 33165 Pessac, France  \\
 $^2$  Astronomy Unit, Queen Mary University of London, Mile End Road, London, E1 4NS, UK}
\date{Released 2012 Xxxxx XX}

\pagerange{\pageref{firstpage}--\pageref{lastpage}} \pubyear{2014}

\def\LaTeX{L\kern-.36em\raise.3ex\hbox{a}\kern-.15em
    T\kern-.1667em\lower.7ex\hbox{E}\kern-.125emX}
\begin{document}
\label{firstpage}
\maketitle
\begin{abstract}
Eccentric gaseous discs are unstable to a parametric instability involving the resonant interaction between inertial-gravity waves and the eccentric mode in the disc. We present 3D global hydrodynamical simulations of inviscid circumbinary discs that form an inner cavity and become eccentric through interaction with the central binary. The parametric instability grows and generates turbulence that transports angular momentum with stress parameter $\alpha \sim 5 \times 10^{-3}$ at distances $\lesssim 7 \;\abin
$, where $\abin$ is the binary semi-major axis. Vertical turbulent diffusion occurs at a rate corresponding to $\alphadiff\sim 1-2\times 10^{-3}$.  We examine the impact of turbulent diffusion on the vertical settling of pebbles, and on the rate of pebble accretion by embedded planets. In steady state, dust particles with Stokes numbers $\st \lesssim 0.1$ form a layer of finite thickness $\Hd \gtrsim 0.1 H$, where $H$ is the gas scale height. Pebble accretion efficiency is then reduced by a factor $\racc/\Hd$, where $\racc$ is the accretion radius, compared to the rate in a laminar disc. For accreting core masses with $\msp \lesssim 0.1\; M_\oplus$, pebble accretion for particles with $\st \gtrsim 0.5$ is also reduced because of velocity kicks induced by the turbulence. These effects combine to make the time needed by a Ceres-mass object to grow to the pebble isolation mass, when significant gas accretion can occur, longer than typical disc lifetimes. Hence, the origins of circumbinary planets orbiting close to their central binary systems, as discovered by the Kepler mission, are difficult to explain using an in situ model that invokes a combination of the streaming instability and pebble accretion.

\end{abstract}

\section{Introduction}
The Kepler mission discovered 11 circumbinary planets orbiting both components of close binary systems. Many of these planets have sub-Jovian masses and are on orbits close to the region of dynamical instability near the central binary, as defined by Holman \& Wiegert (1999). Examples include Kepler-16b (Doyle et al 2011), Kepler-35b (Welsh et al 2012), Kepler-38b (Orosz et al. 2012a), Kepler-47b (Orosz et al. 2012b) and Kepler-413b (Kostov et al. 2013). The observed orbital locations of these planets relative to the central binaries are in agreement with expectations derived from hydrodynamical simulations of planets embedded in circumbinary discs (Nelson 2003; Pierens \& Nelson 2007, 2008a,b, 2013; Kley \& Haghighipour 2014, 2015), which demonstrate that the inwards migration of a circumbinary planet with sub-Jovian mass is naturally halted at the outer edge of the central cavity formed by the binary.  The fact that only sub-Jovian circumbinary planets have so far been detected in these short period orbits, and the only known Jovian mass circumbinary planet detected so far, Kepler-1647b (Kostov et al. 2016), has a long orbital period ($\sim 1100$ days), is in agreement with the predictions of hydrodynamic simulations (Pierens \& Nelson 2008).

The inner regions of circumbinary protoplanetary discs should be highly disturbed, and hence these systems provide a testbed that should allow us to constrain competing theories of planet formation. An alternative to the migration hypothesis is that circumbinary planets orbiting close to the central binary formed in situ at their observed locations, through either planetesimal or pebble accretion. 

Various studies have highlighted the difficulty of building circumbinary planets in situ through planetesimal accretion (e.g. Paardekooper et al. 2012).  They show that km-size planetesimals are excited onto highly eccentric orbits due to their interaction with the central binary (Meschiari 2012 a,b; Bromley \& Kenyon 2015) and/or surrounding circumbinary disc  (Marzari et al. 2008; Kley \& Nelson 2010; Lines et al. 2014), and combined with size-dependent pericentre alignment  (Scholl et al. 2007) this leads to mutual collisions between planetesimals being destructive.

The efficacy of pebble accretion as a means of forming circumbinary planets in situ, which is the focus of this paper, has not yet been assessed. The inner tidally-truncated cavity can act as a  trap for solid particles that drift inwards due to gas drag, naturally leading to an increase in the local dust-to-gas ratio, potentially triggering the streaming instability (Youdin \& Goodman 2005; Johansen et al. 2009; Simon et al. 2016). The streaming instability (SI) concentrates particles with Stokes numbers (or dimensionless stopping times) $\st\sim0.001-0.1$ into clumps that can subsequently gravitationally collapse to form planetesimals with sizes up to $\sim 10^3$ km. Such objects can grow further through merging or/and by capturing inwards drifting pebbles (Johansen \& Lacerda 2010; Lambrechts \& Johansen 2012), namely solids with Stokes numbers $\st\sim0.01-1$ that are marginally coupled to the gas. Once a mass of $\sim 0.01$ $M_\oplus$ is reached, pebble accretion can become very efficient, depending on the local conditions in the disc, and it has been shown that in a protoplanetary disc similar to the Minimum Mass Solar Nebula (MMSN), 10 Earth mass planets can be formed at $5$ AU in $\sim 10^4$ years (Lambrechts \& Johansen 2012).

The conditions for triggering the SI, and the efficiency of pebble accretion, are sensitive to the level of turbulence operating in the disc. Small particles are lofted away from the midplane by turbulent mixing, such that the local dust-to-gas ratio, and hence the growth rate of the SI, are reduced. Although it has been shown that particles with Stokes numbers $\st\sim0.2-1$ can be subject to the SI in turbulent flows characterised by turbulent stress parameter $\alpha\sim 10^{-3}$ (Johansen et al. 2007), recent work suggests that for smaller particles with $0.01<\st<0.05$ the SI results in only modest overdensities of factors $\sim 4-20$ (Umurhan et al. 2019). 

Turbulent stirring can also lead to a significant reduction in pebble accretion efficiency. Efficient pebble accretion requires the pebble vertical scale height to be smaller than the Hill radius of the accreting core (Lambrechts \& Johansen 2012). For an accreting Ceres mass object, the pebble scale height must to be $\lesssim 1\%$ of the gas pressure scale height, and therefore any turbulence present in the disc must be weak. According to the most sophisticated MHD models of protoplanetary discs, the  condition for weak magnetised turbulence might be fulfilled at distances where the temperature is too low for thermal ionisation to be effective, namely in the region outside $\sim 0.2$ AU where the disc remains essentially laminar and  accretion is mainly driven by a wind launched from high altitudes (Bai \& Stone 2013; Gressel et al. 2015, Bethune et al. 2017). 

MHD instabilties are not the only possible sources of turbulence, however, and hydrodynamical instabilities such as the Vertical Shear Instability might operate in the outer regions of the disc (Nelson et al. 2013). The vertical shear instability (VSI) leads to $\alpha \sim 10^{-4}$ (although its mixing properties are highly anisotropic), and it has  been shown that  particles with Stokes number $\st \sim 10^{-3}$ can undergo clumping through the SI, provided that the dust-to-gas ratio $Z\gtrsim 0.02-0.05$ (Lin 2019). Pebble accretion in a VSI-active disc has been examined by Picogna et al. (2018), who found that the efficiency of pebble accretion onto cores of a few Earth masses is reduced to $\sim 50\%$ of its value in a laminar disc.

Parametric instabilities are also good candidates for generating turbulence in the disc. Among these instabilities, the Spiral Wave Instability (SWI, Bae, Nelson \& Hartmann 2016a) occurs because of the resonant interaction of inertial-gravity waves with a background spiral wave. In a protoplanetary disc where the SWI is triggered by a spiral wave launched by a giant planet, the pebble accretion efficiency can be significantly reduced because of turbulence for pebbles with sizes up to a few centimeters (Bae, Nelson \& Hartmann 2016b). For circumbinary discs, it is not clear whether or not this instability can operate and induce turbulence because the resonant interaction of inertial waves with the binary can only occur in a narrow range of radii where the doppler-shifted frequency of the spiral wave is not too large (Bae et al. 2018b). In a  circumbinary disc that  becomes eccentric due to its interaction with the central binary, however, another possible means to generate turbulence is through the onset of a parametric instability caused by the resonant coupling between inertial-gravity waves and a global $m=1$ eccentric mode in the disc (Papaloizou 2005a; Barker \& Ogilvie 2014). For an isothermal, non-stratified disc around a single star, in which a free $m=1$ global mode is present, non-linear evolution of the instability leads to turbulence with $\alpha\sim 10^{-3}$ (Papaloizou 2005b).

The aim of this paper is to examine if this eccentric parametric instability can operate in circumbinary discs, and if so to determine the nature of the turbulent flow in the  affected regions near the central binary. As we are interested in the non-linear outcome of the instability, we perform three dimensional (3D) global simulations of circumbinary discs, which we find generate turbulence with $\alpha \sim 10^{-3}$.  A particular question of interest is whether or not short-period circumbinary planets can be formed in situ via pebble accretion despite the presence of turbulence. Hence, we also present the results of 3D hydrodynamical simulations that include particles that are aerodynamically coupled to the gas, which we use to examine the effects of turbulent mixing on dust settling, and on the accretion of pebbles by embedded protoplanets. In this initial study, we neglect the effects of the back reaction from the grain particles onto the gas, and the effects of this on our results will be considered in a forthcoming follow up paper.

This paper is organized as follows. In Sect.~2,  we describe the hydrodynamical model and numerical setup. In Sect.~3, we present the results of our 3D simulations. In Sect.~4, we discuss the impact of turbulence on dust settling, and the consequences of turbulence for pebble accretion.  Finally, we discuss the implications of our results for the formation of circumbinary planets in Sect.~5 and summarize in Sect.~6.

\section{The hydrodynamic model}
\subsection{Numerical setup}
The simulations presented in this paper were performed using FARGO3D (Benitez-Lamblay \& Masset 2016) in a modified version that includes 
a particle module (McNally et al. 2019).  In this particle module, a kick-drift-kick integration scheme was employed to compute the particle trajectories, similar to that used in Nelson \& Gressel (2010). 
We solve the hydrodynamic equations for the conservation of mass,  momentum, and internal energy in spherical coordinates $(r,\theta,\phi)$ (radial, polar, azimuthal), with the origin of the frame located at the centre of mass of the binary:
 \begin{equation}
 \frac{\partial \rho}{\partial t}+{\bf \nabla}\cdot(\rho {\bf v})=0,
 \end{equation}
 \begin{equation}
 \rho\left(\frac{\partial {\bf v}}{\partial t}+{\bf v}\cdot{\bf \nabla} {\bf v}\right)=-{\bf \nabla} P-\rho {\bf \nabla} (\Phi+\Phi_{\rm p}),
 \end{equation}
 \begin{equation}
  \frac{\partial e}{\partial t}+{\bf \nabla}\cdot(e {\bf v})=-(\gamma-1) e {\bf \nabla}\cdot {\bf v}+{\cal Q}_{cool} ,
  \label{eq:energy}
 \end{equation}
where $\rho$ is the density, $P$ the pressure, ${\bf v}$ the velocity,  $e$ the internal energy,   $\gamma$ the adiabatic index, which is set to 
$\gamma=1.4$, and ${\cal Q}_{\rm cool}$ is the cooling rate.  $\Phi$ is the binary gravitational potential, which can be written as:
 \begin{equation}
 \Phi=-\frac{GM_1}{|\mathbf{r}-\mathbf{r_1}|}-\frac{GM_2}{|\mathbf{r}-\mathbf{r_2}|},
 \end{equation}
where $M_1$, $M_2$ are the masses and ${\bf r_1}$, ${\bf r_2}$ are the radius vectors of the primary and secondary stars, respectively.  When a planetary companion is included, it contributes to the gravitational potential through the expression 
\begin{equation}
\Phi_{\rm p}=-\frac{G\msp}{(|\mathbf{r}-\mathbf{r}_{\rm p}|^2+b^2)^{1/2}},
\end{equation}
where $\msp$ is the planet mass, ${\bf r}_{\rm p}$ is the planet radius vector and $b$ is the smoothing length which is set to $b=0.5 R_{\rm H}$, where $R_{\rm H}$ is the planet Hill radius. The indirect terms arising from the planet and disc are taken into account but we do not include disc self-gravity. Previous work (Mutter et al. 2017) for flat 2D discs showed that for disc masses $\lesssim 5$ MMSN (where MMSN refers to the minimum mass solar nebula model of Hayashi (1981)), the disc structure is only weakly affected by self-gravity. 

In the energy equation (Eq. \ref{eq:energy}), we include a simple cooling function ${\cal Q}_{\rm cool}$ of the form:
\begin{equation}
{\cal Q}_{\rm cool}=-\frac{1}{\tc}\left(e-\frac{\rho e_i}{\rho_i}\right),
\end{equation}
where $e_i$ is the initial internal energy, $\rho_i$ the initial density and $\tc$  the cooling time-scale $\tc=\beta \Omega^{-1}$, with $\Omega$ the Keplerian frequency. We adopt the following values for $\beta$: 0.1, 1, 10, 100.

Computational units are chosen such that the total mass of the binary is $M_\star=M_1 + M_2 =1$, the gravitational constant $G=1$, and the radius $R=1$ in the computational domain corresponds to the binary semi-major axis for the  Kepler-16 system ($\abin= 0.22$ au, see Table 1). When presenting the simulation results, unless otherwise stated we use the binary orbital period $\Tbin=2\pi\sqrt{\abin^3/GM_\star}$ as the unit of time. The computational domain in the radial direction extends from $R_{\rm in}=1.5\;\abin$ to $R_{\rm out}=18\;\abin$ and we employ $894$ logarithmically spaced grid cells. In the azimuthal direction the simulation domain extends from $0$ to $2\pi$ with $700$ uniformly spaced grid cells. In the meridional direction, the simulation domain covers $4$ disc pressure scale heights above and below the disc midplane, and we adopt $144$ uniformly spaced grid cells. 

\subsection{Initial conditions}
\label{sec:init}
The initial radial profile of the sound speed, $\cs$, is given by: 
\begin{equation}
\cs(R)=h_0\left(\frac{R}{R_0}\right)^{q/2},
\end{equation}
where $R=r\sin \theta$ is the cylindrical radius and $h_0$ the disc aspect ratio at $R=R_0=1$. We adopt $h_0=0.05$ and $q=-1$ such that the aspect ratio, $h$, is constant with $h=h_0=0.05$.

The initial density and azimuthal velocity profiles are given by:
\begin{equation}
\rho(R,Z)=f_{\rm gap}\, \rho_0\left(\frac{R}{R_0}\right)^p \exp\left(\frac{GM_\star}{\cs^2}\left[\frac{1}{\sqrt{R^2+Z^2}}-\frac{1}{R}\right]\right)
\label{eq:rho0}
\end{equation}
and
\begin{equation}
v_\varphi(R,Z)=\left[(1+q)\frac{GM_\star}{R}+(p+q)\cs^2-q\frac{GM_\star}{\sqrt{R^2+Z^2}}\right]^{1/2},
\end{equation}
\label{sec:initial}
where $Z=r\cos \theta$ is the altitude and $\rho_0$ the density at $R=R_0$. The power-law index for the density is set to $p=-5/2$ so that the slope of the surface density profile $\Sigma$ corresponds to that of the MMSN, namely $\Sigma \propto R^{-3/2}$ (Hayashi 1981). In Eq.~\ref{eq:rho0}, $\rho_0$ is defined such that $\sim 0.005\;M_\star$ is contained within 40 AU, resulting in a disc mass that is slightly smaller than half of the MMSN (Lines et al. 2015; Thun et al. 2018). $f_{\rm gap}$ is a gap-function used to initiate the disc with an inner cavity (assumed to be created by the binary), and is given by:
\begin{equation}
f_{\rm gap}=\left(1+\exp\left[-\frac{R-R_{\rm gap}}{0.1R_{\rm gap}}\right]\right)^{-1},
\end{equation}
 where $R_{\rm gap}=2.5\abin$ is the analytically estimated gap size (Artymowicz \& Lubow 1994). The initial radial and meridional velocities are set to zero. 
 
We do not include the gravitational back-reaction from the disc onto the binary, so the binary orbit remains fixed. The binary semi-major axis and eccentricity are chosen to match those of Kepler-16 (see Table 1).  

\begin{table}
\caption{Binary parameters for the Kepler-16 system (from Doyle et al. 2011).}              
\label{table1}      
\centering                                      
\begin{tabular}{c c}          
\hline\hline                        
Parameter label  & Kepler-16    \\ 
\hline 
$M_1(M_\odot)$ & $0.69$\\
$M_2(M_\odot)$ & $0.2$\\
$q_{\rm bin}=M_2/M_1$ & $0.29$\\                        
$\abin$ (AU)  & $0.22$  \\
$e_{\rm bin}$ & $0.16$  \\

\hline                                             
\end{tabular}
\end{table}

\subsection{Boundary conditions}
We employ closed radial boundary conditions at both the inner and outer edges of the disc.  Previous 2D studies (Mutter et al. 2017; Thun et al. 2017) have emphasized the strong dependence of the circumbinary disc structure on the choice of the inner boundary condition. Boundary conditions that have been employed in numerical simulations of circumbinary discs include: i)  closed boundaries for which gas is not allowed to flow across the disc edge, ii)  open boundaries for  which only gas outflow is allowed, iii) viscous boundaries where the radial velocity of the gas is set to the viscous drift velocity. We remark that since we deal with inviscid discs, our choice for a  closed inner boundary is obviously equivalent to a viscous boundary.  Compared to an open boundary, using a closed boundary leads to a larger density maximum and appears to be more sensitive to numerical issues (Thun et al. 2017). Employing an open boundary leads to a quasi-stationary disc structure after ${\cal O}(10
^4)$ binary orbits, but only provided that the location of the inner disc edge is small enough with, $R_{min}\lesssim a_{bin}$ (Mutter et al. 2017). In a 3D simulation, however, an evolution time of ${\cal O}(10
^3)$ binary orbits is the maximum that is computationally feasible. Choosing $R_{min}> a_{bin}$ is also required to allow a timestep that is large enough to makes 3D simulations tractable. Therefore, we do not expect the disc structure at the end of our simulations to have reached a quasi-steady state, even in the case where an open boundary is employed.  Although this needs to be checked by dedicated simulations in 3D, we also note that in previous 2D simulations we have found that employing an open boundary, together with an inner radius $R_{min}> a_{bin}$, can lead to the formation of an artificially large inner hole when the central binary has a moderate to large eccentricity. 

At the outer boundary, we also make use of a wave-killing zone for $R> 16$ to avoid wave reflection.  Ordinarily we would adopt outflow conditions at the inner edge to allow mass to accrete onto the binary, and hence for a steady state disc structure to develop. In these 3D calculations we consider inviscid conditions, such that a steady structure in which viscous stresses and gravitational torques come into balance does not exist. Furthermore, the computational expense of running 3D simulations would not allow us to achieve such a steady state even if we relaxed the inviscid assumption. We therefore do not expect the inner boundary condition to play an important role in determing the outcome of our simulations.

At the meridional boundaries, an outflow boundary condition is used for the velocities and internal energy, and for which all  quantities in the ghost zones have the same values as in the first active zone, except the meridional velocity whose value is set to $0$ if it is directed towards the disc midplane to prevent inflow of material. For the density, we follow Bae et al. (2016a) and maintain vertical stratification by solving the following condition for hydrostatic equilibrium in the meridional direction:
\begin{equation}
\frac{1}{\rho}\frac{\partial}{\partial \theta} (\cs^2 \rho)=\frac{v_\phi^2}{\tan \theta}.
\end{equation}

 \begin{figure}
\centering
\includegraphics[width=\columnwidth]{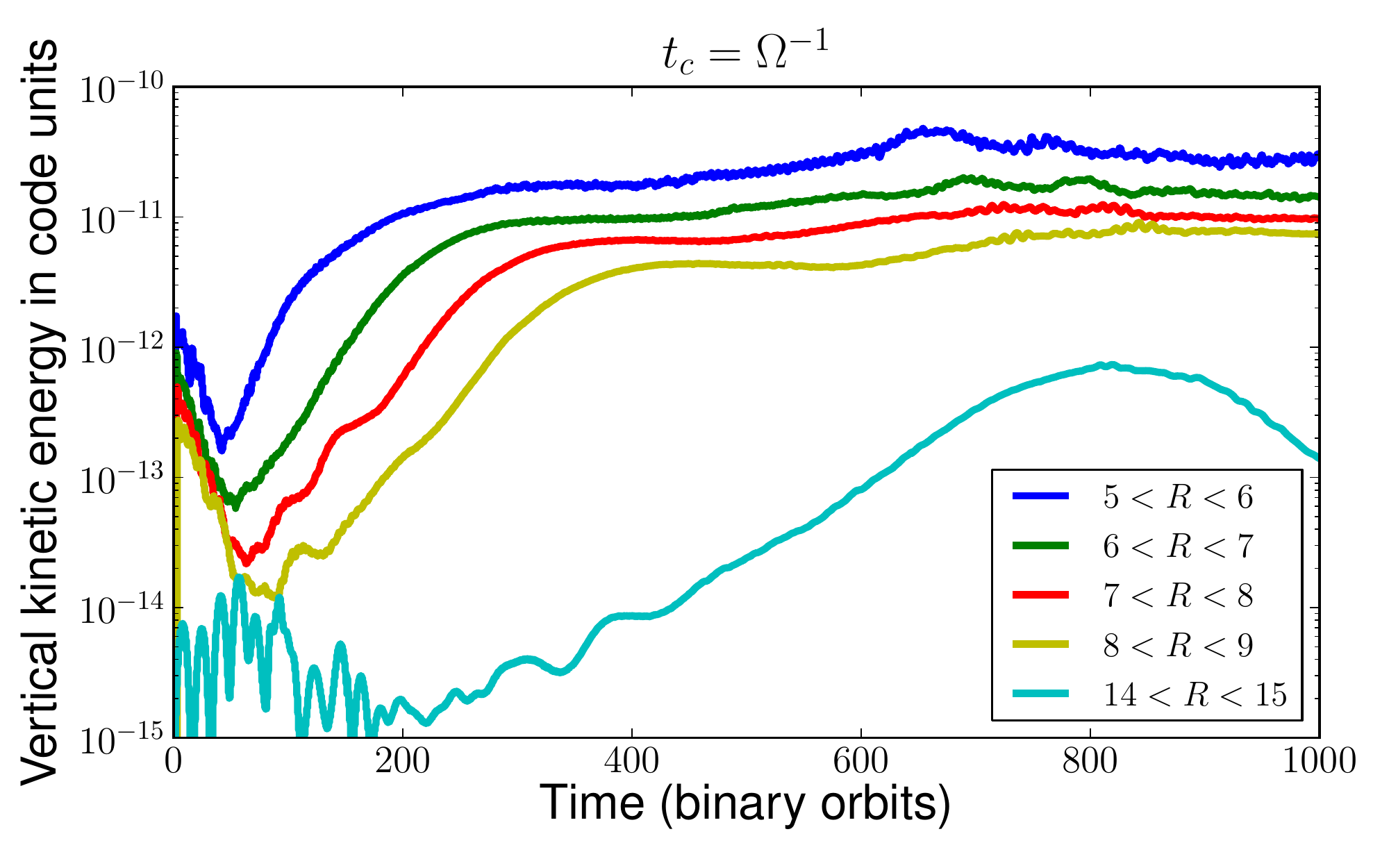}
\includegraphics[width=\columnwidth]{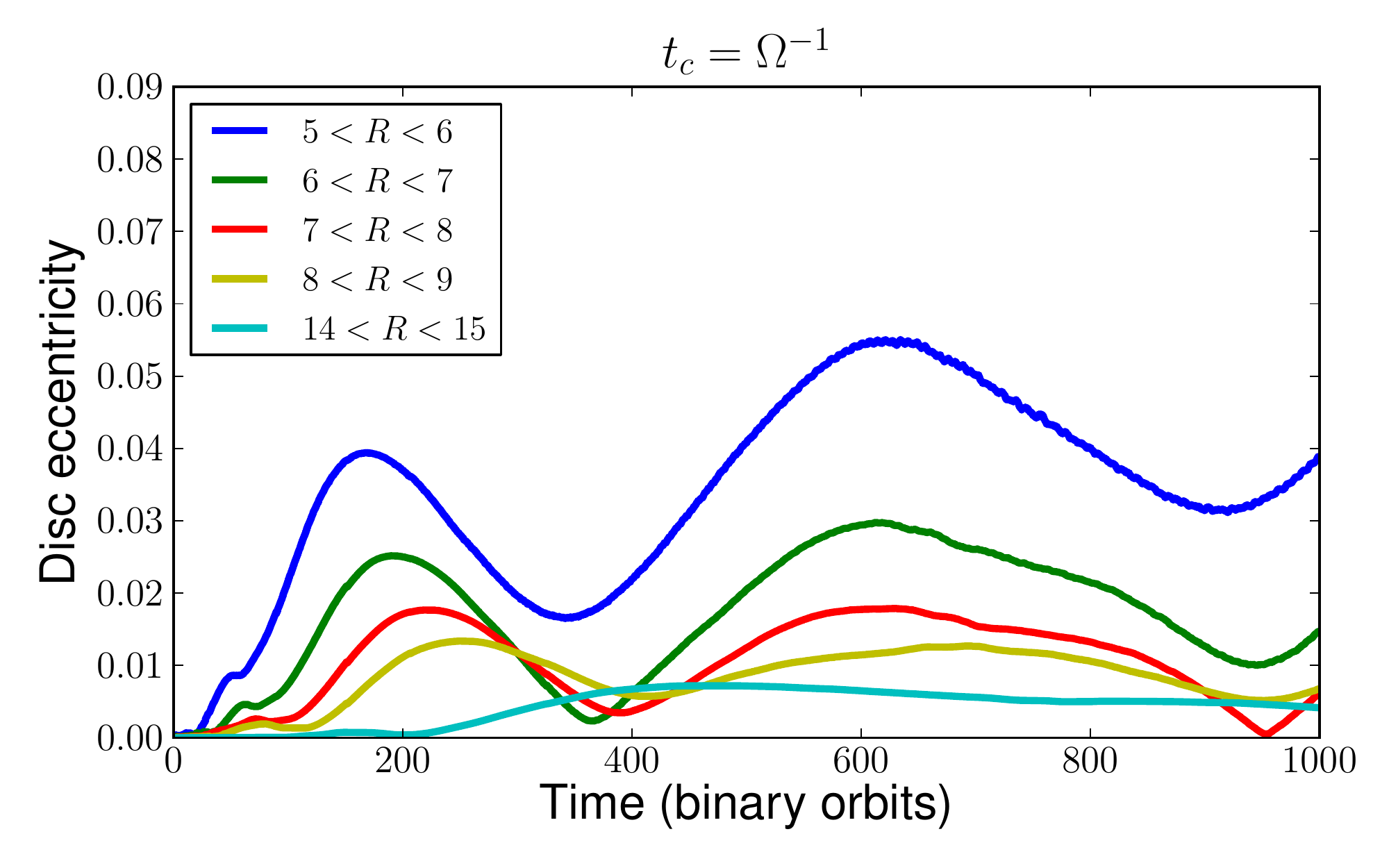}
\caption{{\it Top:} Time evolution of the meridional kinetic energy for the fiducial run with $\tc=\Omega^{-1}$ and for various radial bins. {\it Bottom: } for the same model, time evolution of the disc eccentricity at these bins.}
\label{fig:ekfid}
\end{figure}

 \begin{figure}
\centering
\includegraphics[width=\columnwidth]{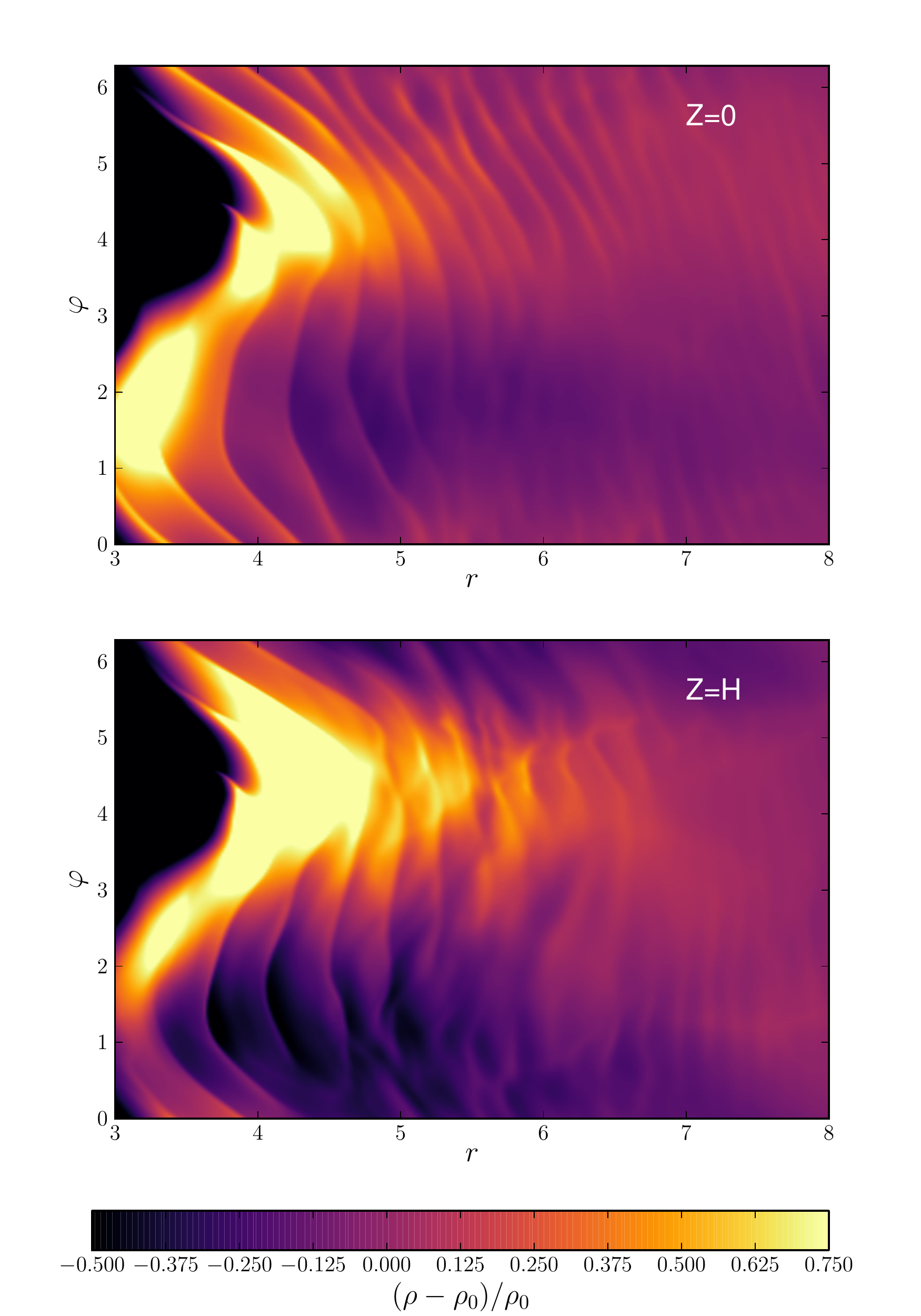}
\caption{Normalized  density perturbation for the fiducial run with $\tc=\Omega^{-1}$ at $Z=H$ and Time=715. }
\label{fig:rphi2d}
\end{figure}

 \begin{figure*}
\centering
\includegraphics[width=\textwidth]{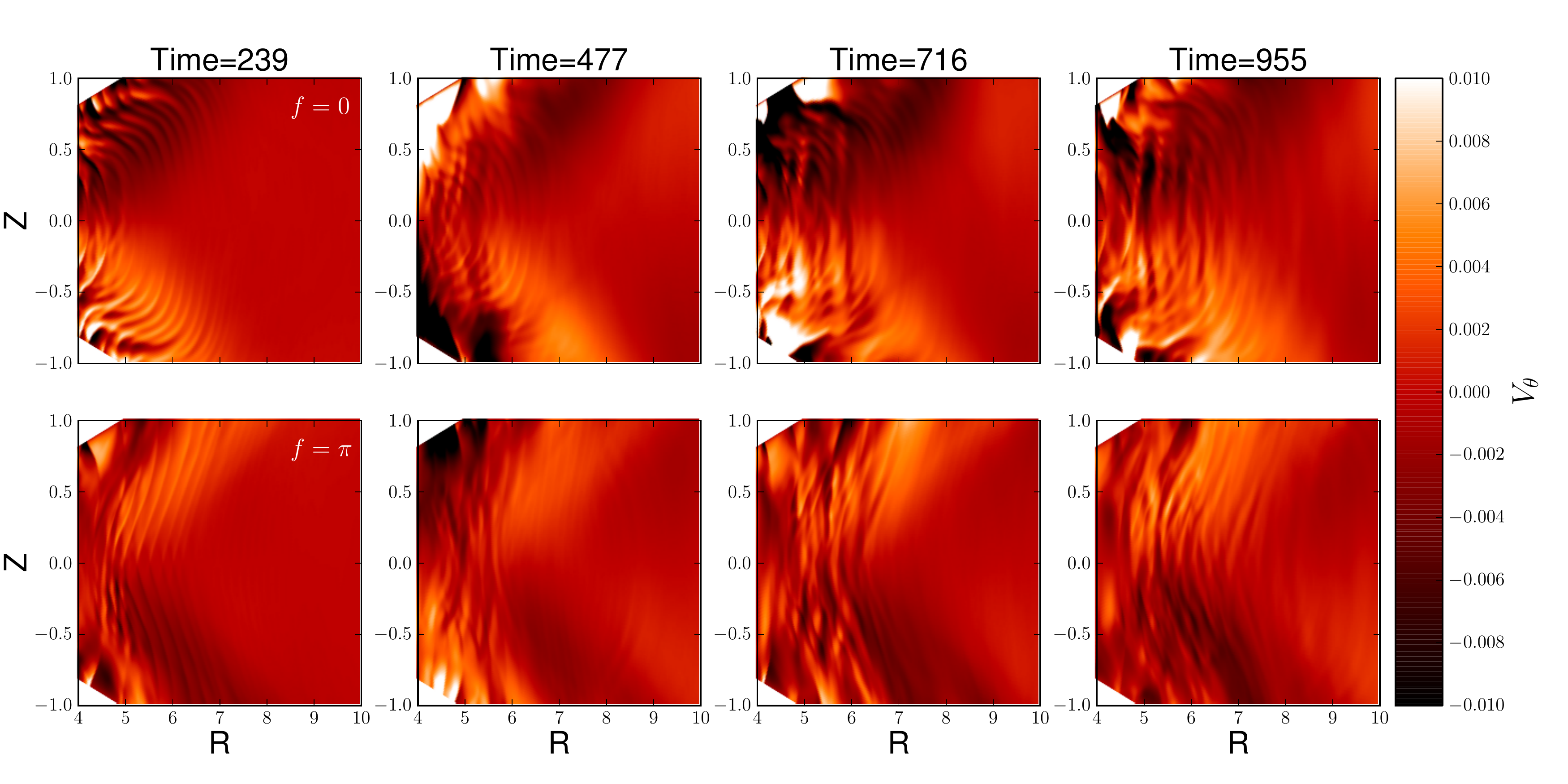}
\caption{For the reference run with $\tc=\Omega^{-1}$, distribution of the meridional velocity at different times in a vertical plane located at an azimuth corresponding to disc pericenter (top) and apocenter (bottom). }
\label{fig:vtheta}
\end{figure*}

\begin{figure}
\centering
\includegraphics[width=\columnwidth]{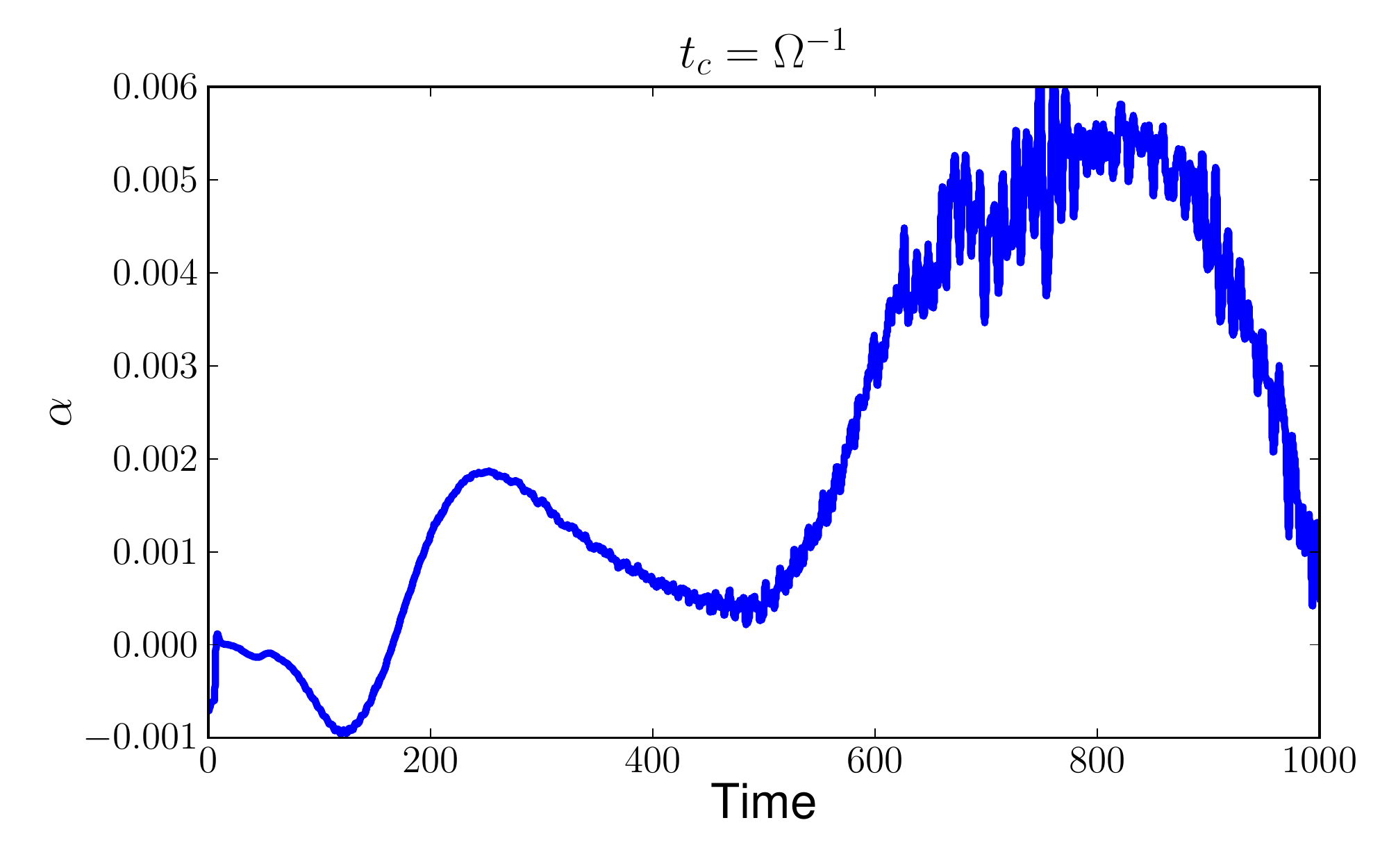}
\caption{Time evolution of the Shakura-Sunyaev stress parameter (see Eq. \ref{eq:alpha} ) for the fiducial run $\tc=\Omega^{-1}$ and averaged in between 
$r=5$ and $r=7$.}
\label{fig:alpha}
\end{figure}

\section{Theoretical expectations}
\label{sec:theory}
In a differentially rotating disc, linear perturbation analysis of the fluid equations in the shearing-sheet approximation results in the following dispersion relation for local perturbations of the form $\exp i(k_R R+k_Z Z-\omega t)$, where $k_R$ and $k_Z$ are the radial and vertical wavenumbers respectively and 
$\omega$ the mode frequency (Goodman 1993):
\begin{equation}
\frac{\omega^2/\cs^2}{\omega^2-N^2}- \frac{k_Z^2}{\omega^2-N^2}- \frac{k_R^2}{\omega^2-\kappa^2}=0
\label{eq:dispersion}
\end{equation}
where $\kappa$ is the epicyclic frequency given by
\begin{equation}
\kappa^2=\frac{1}{R^3}\frac{d}{dR}(R^2\Omega)^2,
\label{eq:epicyclic}
\end{equation}
and $N$ is the vertical \BV frequency defined by
\begin{equation}
N^2=\left(\frac{g}{\gamma}\frac{d}{dz}\ln\left(\frac{P}{\rho^\gamma}\right)\right)^{1/2},
\end{equation}
where $g=\Omega^2z$ is the vertical gravity, $P$ the pressure, and $\gamma$ the adiabatic index.  In the limit where $N^2\sim 0$, which applies to an isothermal gas or near the disc midplane,  Eq.~\ref{eq:dispersion} becomes
\begin{equation}
(\omega^2-k_z^2\cs^2)(\omega^2-\kappa^2)=\omega^2 k_R^2 \cs^2.
\label{eq:dispersions}
\end{equation}
The high frequency branch ($\omega \gg \kappa$) corresponds to sound waves with $\omega \sim \sqrt{k_r^2+k_z^2} \cs$, whereas the low frequency branch corresponds to inertial waves which are supported by the Coriolis force. Parametric instability occurs when the frequency of a forcing disturbance in the disc, $\omegas$, viewed in a frame corotating with a fluid element, is equal to twice the mode frequency of an inertial wave. Here, we are interested in a $m$=1 perturbation, where $m$ is the azimuthal wavenumber, arising from an eccentric disc with small pattern frequency $\Omegap \ll \Omega$ and for which $\omegas=\Omega-\Omegap\sim\Omega$. In that case, the condition for parametric instability becomes
\begin{equation}
\omega =\frac{\Omega}{2}.
\end{equation}
Substituting $\omega=\Omega/2$ into Eq.~\ref{eq:dispersions} and assuming $\kappa \sim \Omega$ yields
\begin{equation}
3(4k_z^2H^2-1)=4 k_R^2 H^2,
\end{equation}
where $H$ is the disc scale height.  Given that the minimum value for $k_z$ is of the order of $k_{z,{\rm min}}\sim 2 \pi /H$, the previous expression can be  approximated as
\begin{equation}
\frac{k_R}{k_z}\sim \sqrt{3},
\end{equation}
consistent with Papaloizou (2005a). Unstable inertial modes with vertical wavelength $\lambda_z\sim H$ therefore have a radial wavelength $\lambda_R\sim 0.6 H$.  For the numerical resolution adopted in the simulations presented here, such modes are resolved by $\sim 18$ grid cells in the vertical direction and $\sim 11$ grid cells in the radial one. 

\section{Results}
\subsection{A fiducial run}
\label{sec:fiducial}
We treat the model with  cooling timescale $\tc=\Omega^{-1}$ as the fiducial run. The volume-integrated meridional kinetic energy, $e_\theta$, is given by
\begin{equation}
e_\theta=\frac{1}{2}\int_V \rho v_\theta^2 {\rm d}V,
\end{equation}
where the volume integration is generally performed over a narrow range of radii. For the fiducial model, the time evolution of the meridional kinetic energy is shown for various 
radial bins in the upper panel of Fig.~\ref{fig:ekfid}. The meridional kinetic energy is observed to damp at first, and to then undergo exponential growth until non-linear saturation occurs. During the linear growth phase, the growth rate of the kinetic energy is $\sim 0.02 \Tbin^{-1}$, which is equivalent to $\sim 0.3 \Torb^{-1}$ at $R=6$, where $\Torb$ is the local orbital period. Similar growth rates have been reported in the context of turbulence generated by the VSI, where inertial waves are destabilised by the background rotation profile (Nelson et al. 2013; Stoll \& Kley 2018). 

The bottom panel of Fig.~\ref{fig:ekfid} shows the time evolution of the disc eccentricity, $\ed$, at the same radial locations, which is computed through the expression:
\begin{equation}
\ed=\frac{\int_V \rho \;e_{\rm c} {\rm d}V}{\int_V \rho \, {\rm d} V}
\end{equation}
where $e_{\rm c}$ is the eccentricity computed at the center of each grid cell. Comparing with the upper panel of Fig.~\ref{fig:ekfid},  we see the growth of the meridional kinetic energy and the disc eccentricity are correlated at early times. On longer time scales the eccentricity exhibits oscillatory behaviour with a period equal to the disc precession period (Thun et al. 2018), measured to be $\sim 400$ $\Tbin$. The correlated growth of disc eccentricity and the meridional kinetic energy suggests the instability we capture here is the parametric instability of inertial waves that resonantly couple with the eccentric disc (Papaloizou 2005; Barker \& Ogilvie 2014). As mentioned above, onset of the instability occurs whenever the  inertial wave frequency $\omega$ matches the resonance condition (Wienkers \& Ogilvie 2018):
\begin{equation}
\omega\sim \frac{\Omega}{2}.
\end{equation}
In discs with no vertical stratification,  the growth rate of this instability is $\sigma=\frac{3}{16} \ed\Omega^{-1}$ (Papaloizou 2005a), whereas $\sigma=\frac{3}{4}\ed\Omega^{-1}$ in stratified disc models (Barker \& Ogilvie 2014). As mentioned earlier, we find a growth rate of $\sim 0.3 \Torb^{-1}\sim 0.045 \Omega^{-1}$ at $R=6$ which would lead 
to $\ed\sim 0.06$ using the expression of Barker \& Ogilvie (2014) for the growth rate. The bottom panel of Fig.~\ref{fig:ekfid} shows this is close to the maximum value reached by the disc eccentricity at $R=6$. \\

Figure~\ref{fig:rphi2d} presents contours showing the normalized density perturbation $\delta \rho/\rho_0$, with $\delta \rho=\rho-\rho_0$, at time $t=715 \, \Tbin$. Each panel shows the density perturbation at different heights in the disc. Just outside the inner cavity, where the disc is significantly eccentric, the spiral waves excited by the binary look perturbed and fragmented because of turbulent motions generate by the instability. Figure~\ref{fig:vtheta} shows contours of the 
meridional velocity in vertical planes with true anomaly $f=0$ (corresponding to the disc pericentre) and $f=\pi$ (corresponding to the disc apocentre).  For $f=0$, the $t=477$ panel reveals a checkerboard pattern one scale height above the midplane, characteristic of the excitation of inertial modes (Fromang \& Papaloizou 2007). The radial and vertical wavelengths associated with these inertial modes are 
$\lambda_R\sim 1.4 H$ and $\lambda_Z \sim 0.5H$, and are resolved by $\sim 25$ and $\sim 9$ grid cells in the radial and vertical directions, respectively. As time proceeds, breaking of these inertial waves, and perhaps mode-mode interactions, causes energy to cascade to smaller scales, and the $t=955$ panels show this results in a disordered, turbulent flow. We note that sustained turbulence resulting from the parametric instability in eccentric discs has also been highlighted in previous numerical simulations of cylindrical disc models with no vertical stratification (Papaloizou 2005b), as well as in local numerical 
models (Wienkers \& Ogilvie 2018). Here, the distribution of the meridional velocity in a vertical plane slicing through the disc apocentre suggests that the dominant modes have $k_R/k_Z \gg 1$, similar to the most unstable modes that grow as a result of the VSI (Nelson et al. 2013).

To estimate the angular momentum transport induced by the turbulence, we calculate the local Shakura-Sunyaev stress parameter, $\alpha(r)$, which is 
defined as the azimuthal- and meridional-averaged stress-tensor normalized by the mean pressure $\left<P\right>$:
\begin{equation}
\alpha(r)=\frac{\left<\rho v_r \delta v_\varphi\right>}{\left<P\right>},
\label{eq:alpha}
\end{equation}
where $\left<\right>$ refers to an arithmetic average performed over $\theta$ and $\varphi$, and $\delta v_\varphi=v_\varphi - \left<v_\varphi\right>$ is the deviation of the azimuthal velocity from its mean. Figure~\ref{fig:alpha} shows the time evolution of $\alpha$ averaged in the domain $r\in [5,7]$. Here, $\alpha$ was 
further averaged over $\sim 10$ binary orbits using $100$ snapshots. At times $\lesssim 500$, there is a contribution to the Reynolds stress arising from the spiral waves excited by the binary as they propagate outwards, and which is evaluated to be $\alpha\sim 0.001-0.002$, whereas at later times the maximum value for $\alpha$ is observed to $\alpha\sim 5-6\times 10^{-3}$. This implies that the parametric instability induces a Reynolds stress 
corresponding to $\alpha \sim 4-5\times 10^{-3}$, consistent with Papaloizou (2005b).

\subsection{Evidence for an eccentricity induced parametric instability}
In this section we investigate whether the instability described in Sect.~\ref{sec:fiducial} really arises because of the parametric instability described in Sect.~\ref{sec:theory}. In particular, the aim is to check that the observed growth of vertical kinetic energy is not related to other instabilities that might drive turbulence and transport angular momentum with a similar $\alpha$ parameter. Since we adopt a thermally relaxing model with constant aspect ratio, 
the disc might, for example, be unstable to the VSI (Nelson et al. 2013). For a thermally relaxing equation of state and disc aspect ratio $h\sim 0.05$, the VSI is expected to be triggered in  the limit of small cooling timescales $\tc \ll \Omega^{-1}$, much shorter that the cooling timescale considered in our fiducial model. The Spiral Wave Instability (SWI; Bae et al. 2016) is another possible candidate for inducing hydrodynamic turbulence in the disc. The SWI is a parametric instability that arises because of the resonant interaction of inertial waves and a background spiral wave. It has been suggested the SWI might operate in circumbinary discs in a narrow range of radii where the doppler-shifted frequency of the spiral wave matches half the frequency of the inertial waves (Bae et al. 2016a).

We have performed simulations to examine whether or not the SWI can grow in our circumbinary discs, but none of these resulted in triggering of the SWI. We considered a setup where only one component of the binary potential $\Phi_{ml}$ was included, with $m$ being the azimuthal wavenumber and $l$ the time-harmonic wavenumber (Artymowicz \& Lubow 1994). For a circumbinary disc, only outer Lindblad resonances (OLR) corresponding to $(m,l)=(m,1)$ components of the binary potential are likely to reside outside the truncated cavity. This is because the OLR associated with the  $(m,l)$ component of the binary potential is located at $R\sim\left(\frac{m+1}{l}\right)^{2/3} \abin$ while the inner edge of the cavity typically resides at $R\sim 2.5 \abin$, corresponding to the location of the $(4, 1)$ resonance. Given that the $(m,l)$ Fourier component of the gravitational potential scales as $e^{-|m-l|}$ (Goldreich \& Tremaine 1980),  the associated Fourier amplitude is very small, such that the SWI does not operate. For example, the amplitude of the $(5,1)$ term is $A_{ml}\sim 10^{-6}$ for binary parameters corresponding to Kepler-16, while $A_{ml}\sim 10^{-11}$ for the $(8,1)$ term.  Because of such small Fourier amplitudes, it is not surprising that our simulations failed to generate turbulence through the SWI. In fact, this instability appears to be a poor candidate for inducing hydrodynamical turbulence in circumbinary discs.  
 
\begin{figure}
\centering
\includegraphics[width=\columnwidth]{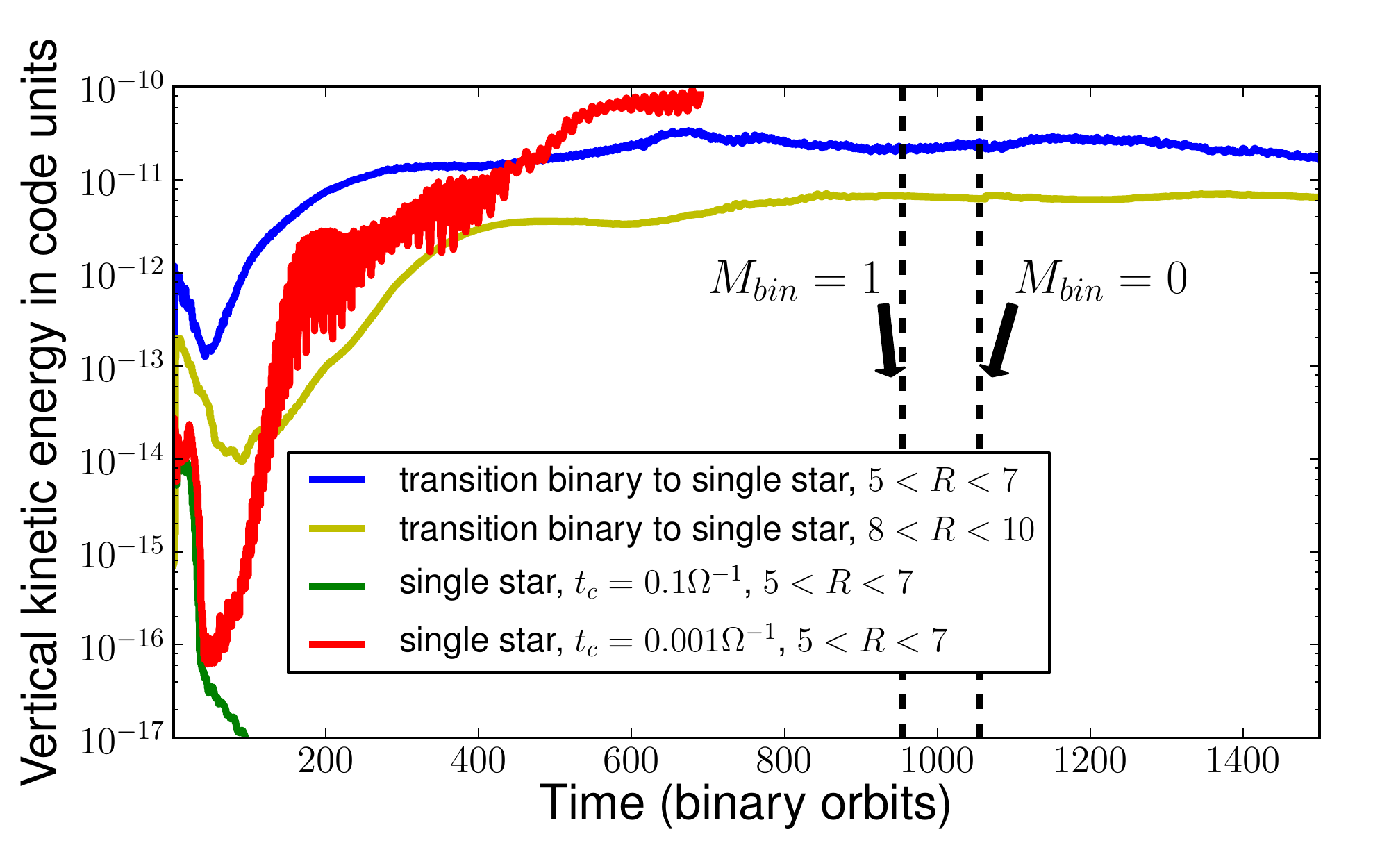}
\caption{Time evolution of vertical kinetic energy for a restart simulation in which the gravitational potential is given by Eq.~\ref{eq:pottrans} (blue, yellow). Vertical dashed lines mark the restart time and the time from which the gravitational potential is that of a single star.  The green and red lines correspond to discs orbiting a single star initially.}
\label{fig:ektransit}
\end{figure}

\begin{figure}
\centering
\includegraphics[width=0.49\columnwidth]{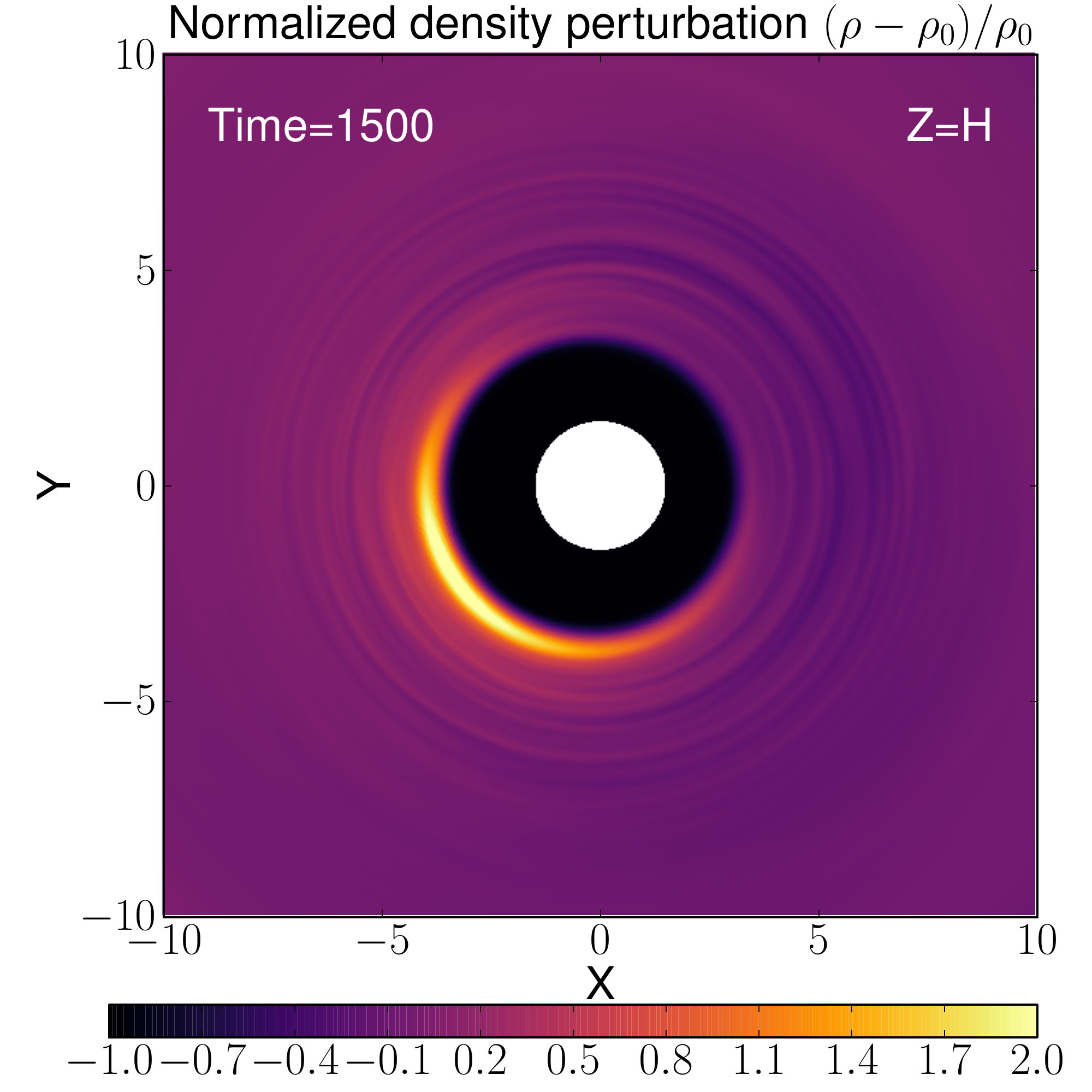}
\includegraphics[width=0.49\columnwidth]{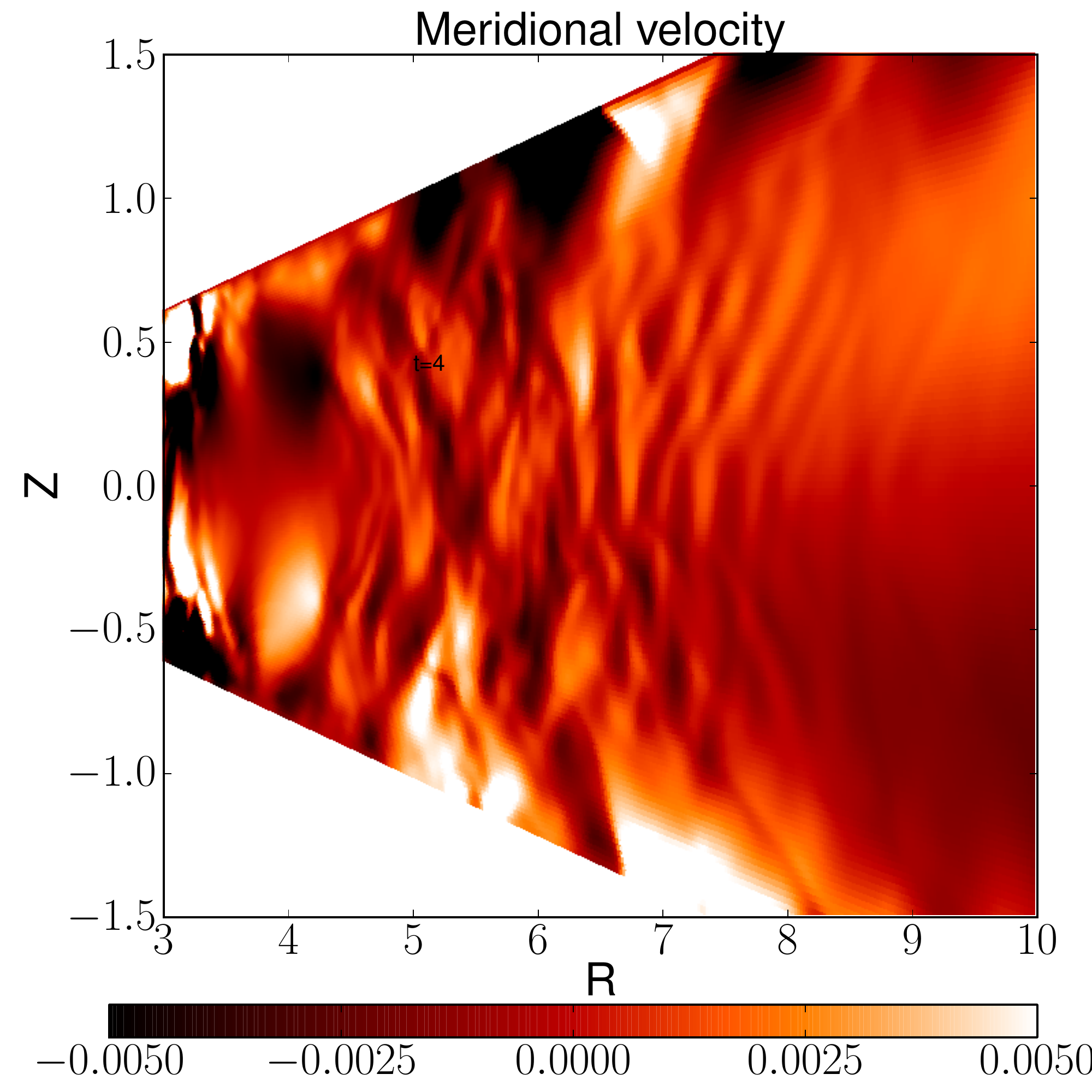}
\caption{{\it Left:}  Normalized density perturbation one scale height above the midplane at time=1500 for the restart simulation in which the gravitational potential is given by Eq.~\ref{eq:pottrans}. {\it Right:} Corresponding meridional velocity in a $[R,Z]$ plane located at azimuth $\varphi=\pi$.}
\label{fig:denstransit}
\end{figure}

 To definitively assess whether or not the growth of the vertical kinetic energy observed in our simulations is associated with the disc eccentricity, we performed 3 additional simulations using the parameters of the fiducial model, except that: i) In the first simulation, the disc orbits a single star and the cooling timescale is set to $\tc=0.1 \Omega^{-1}$; ii) In a second run, the disc orbits a single star and the cooling timescale is set to $\tc=0.001 \Omega^{-1}$; iii) In the third run, we restarted the fiducial model at time $t_0=955$ but with the system slowly transitioning from a central binary system into a single star located at the centre of mass. This is done by treating the gravitational potential as the sum of two terms, 
 one corresponding to the binary plus one corresponding to a single star, with weighting factors that change with time. More precisely, the gravitational potential for this simulation is given by (Mutter et al. 2017):
 \begin{equation}
 \Phi_{\rm trans}=(1-W (t))\Phi_{\rm bin}-\frac{GM_\star}{r} W(t)
 \label{eq:pottrans}
 \end{equation}
 with
 \begin{align}
 W(t) &=
\begin{cases}
 1        & \text{if } \;  t-t_0 \geq 100\Tbin \\
 \frac{1}{2}\left[1-\cos\left(\frac{\pi (t-t_0)}{100\Tbin}\right) \right]        & \text{if} \; t-t_0 < 100\Tbin
 \end{cases}
 \end{align}
 The aim of simulations i) and ii) is to check that our fiducial disc is stable to the VSI whereas run iii) is useful to examine whether the turbulent flow 
 is maintained under unforced conditions. For these three calculations, the time evolution of the vertical kinetic energy is shown in Fig.~\ref{fig:ektransit}. Considering simulations i) and ii), exponential growth of the vertical kinetic energy due to growth of the VSI only occurs for the case with $\tc=0.001\Omega^{-1}$. This is consistent with the results of Richard et al. (2016), who found that for $h=0.05$ the disc remains stable to the VSI provided $t_c > 0.05\Omega^{-1}$.  It also demonstrates that our fiducial model, which has $\tc=\Omega^{-1}$, is stable to the VSI. 
 For simulation iii), the saturation level of the vertical kinetic energy remains essentially unchanged after the switch ($t=1055$). Contours of the normalized  density perturbation one scale height above the midplane, and of the meridional velocity in a $[R,Z]$ plane located at azimuth $\varphi=\pi$, are presented in Fig.~\ref{fig:denstransit} at $t=1500$. They clearly reveal that both the eccentric mode and the turbulence can be maintained under unforced conditions. The continued existence of the eccentric mode at that time is not surprising since $m=1$ free  eccentric modes can be long lived (Papaloizou 2005). The persistence of the turbulent flow is a strong indication that the instability originates from the presence of an eccentric mode in the disc.

\subsection{Dependence on cooling timescale}
\begin{figure}
\centering
\includegraphics[width=\columnwidth]{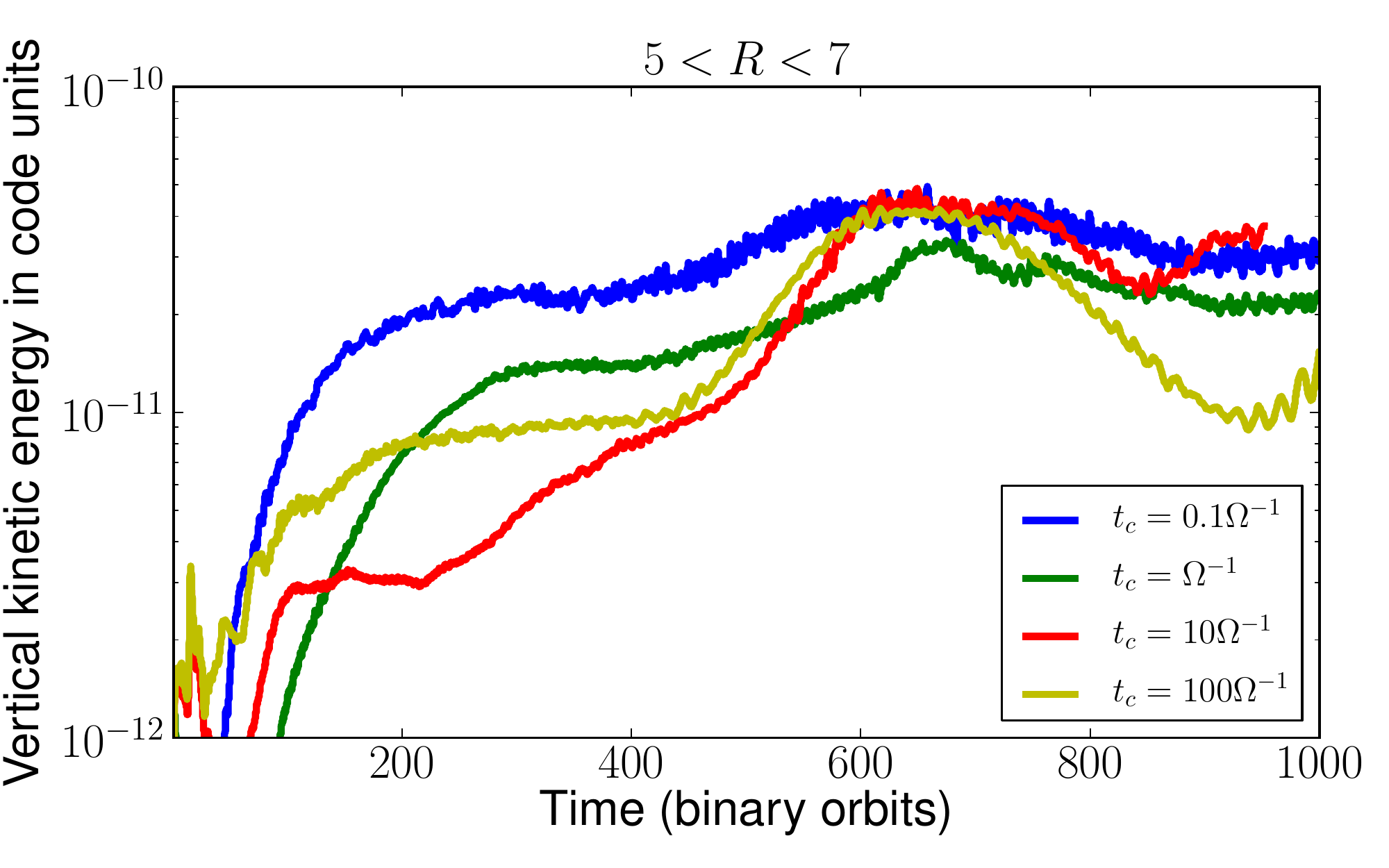}
\includegraphics[width=\columnwidth]{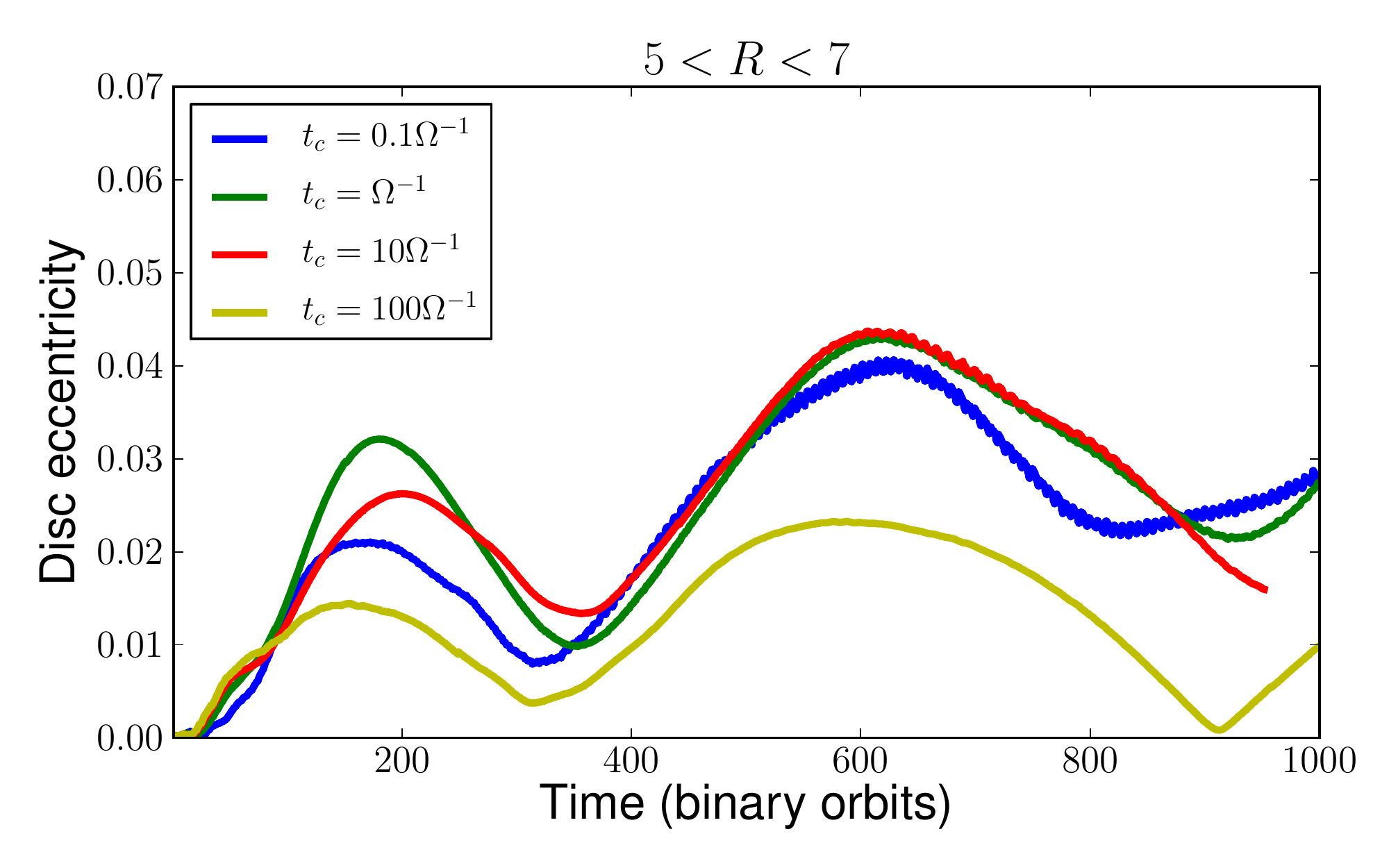}
\caption{{\it Top:} Time evolution of the meridional kinetic energy for models  with $\tc=0.1, 1, 10, 100 \Omega^{-1}$ and for various radial bins. {\it Bottom: } For the same models, time evolution of the disc eccentricity in these bins.}
\label{fig:ektc}
\end{figure}

 To test the robustness of the instability as a function of cooling timescale, we have considered the evolution of models with cooling timescales $\tc=0.1, 10, 100 \,\Omega^{-1}$. The evolution of the vertical kinetic energies and disc eccentricities are plotted in the top and bottom panels of Fig.~\ref{fig:ektc}, respectively. Exponential growth of the vertical kinetic energy occurs in each case, with a tendency for shorter growth timescales and higher saturation amplitudes to occur for smaller values of the cooling timescale. The onset of instability is robust regarding the disc thermodynamics, and may occur in both optically thin and optically thick circumbinary discs. 
 
\begin{figure*}
\centering
\includegraphics[width=\textwidth]{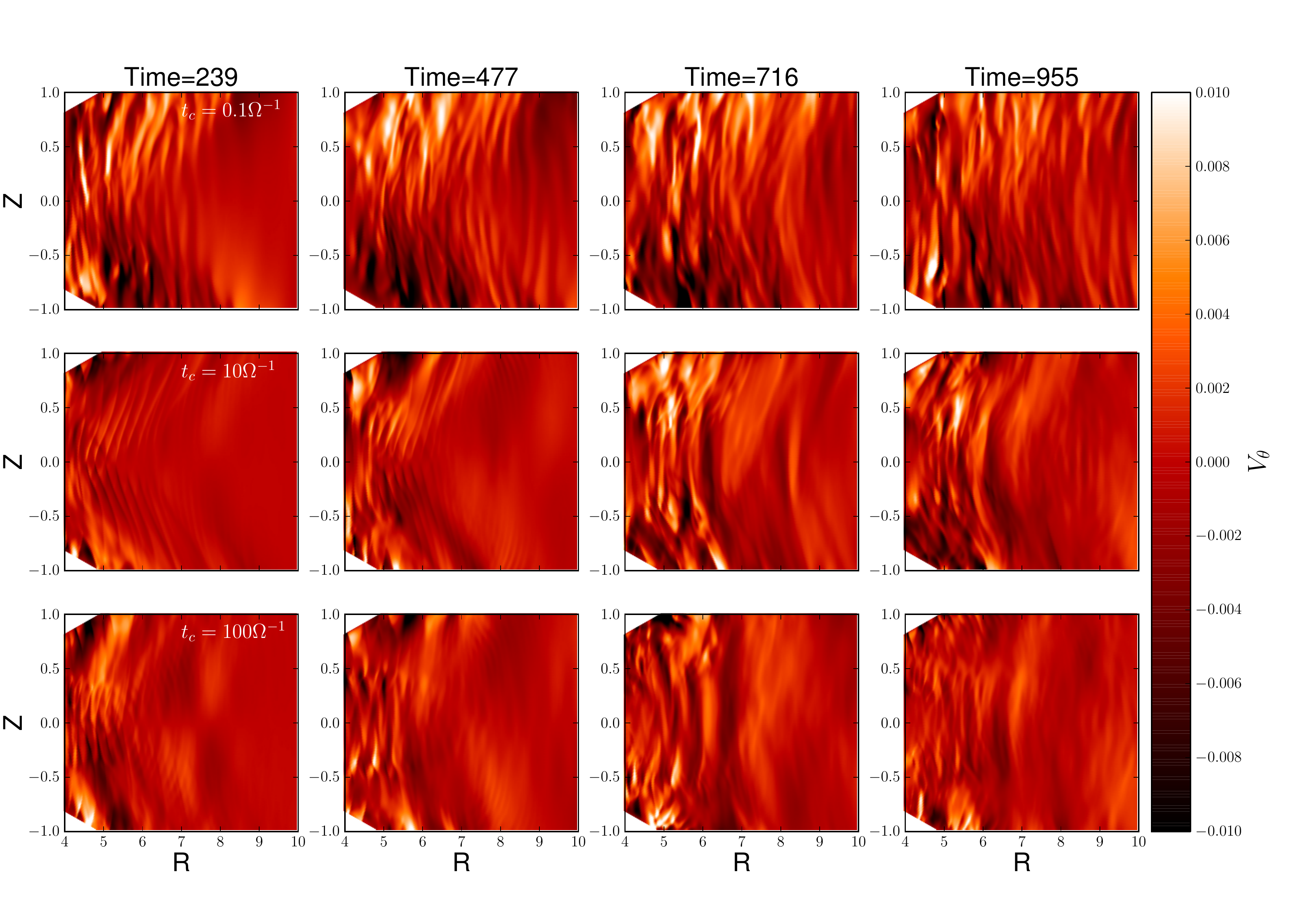}
\caption{For cooling timescales $\tc=0.1, 10, 100\ \Omega^{-1}$, contour plots of the meridional velocity at different times in a vertical plane with azimuth corresponding to the disc apocenter.}
\label{fig:vtheta_tc}
\end{figure*}

\begin{figure*}
\centering
\includegraphics[width=\textwidth]{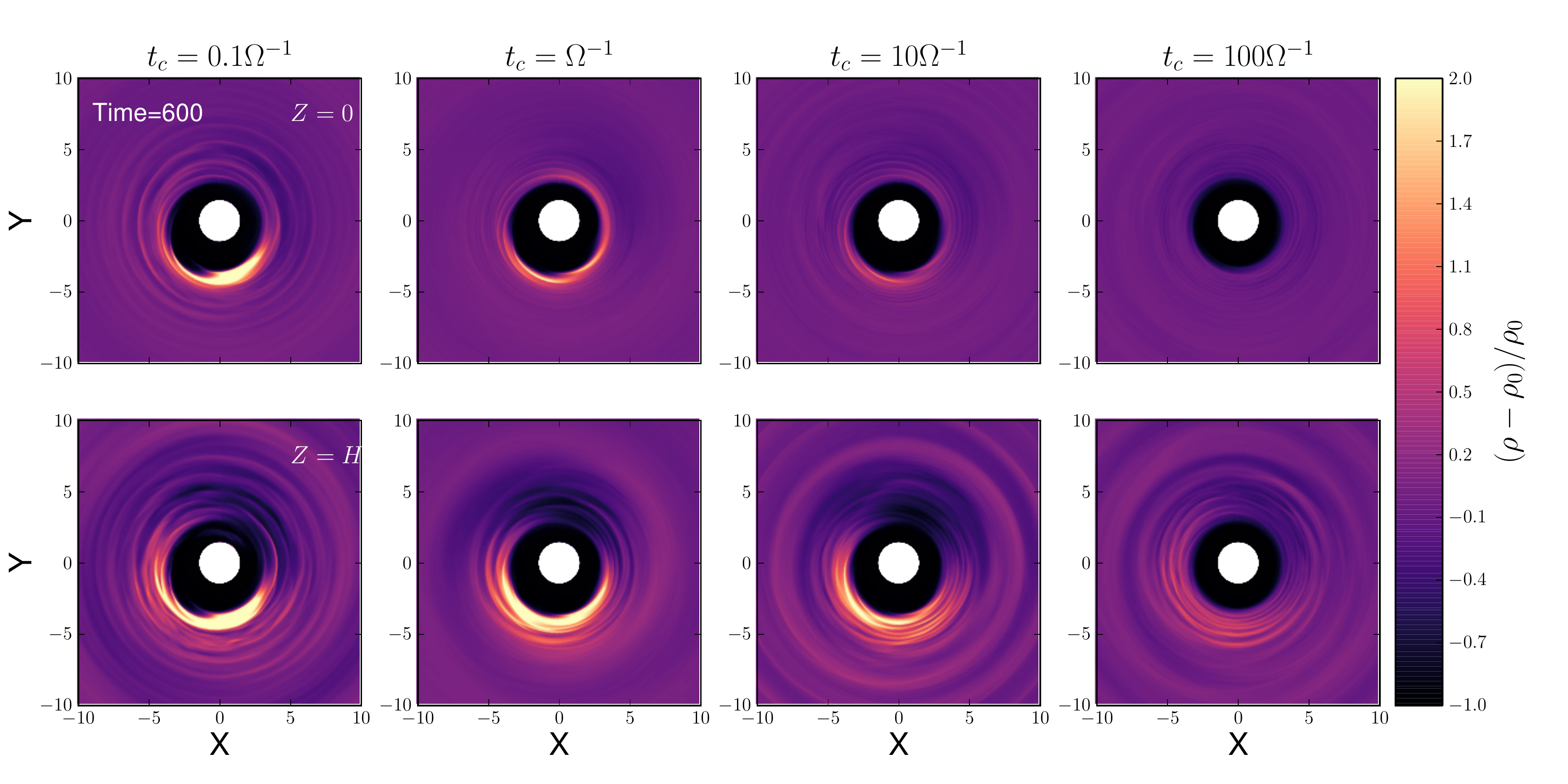}
\caption{Normalized density perturbation at $Z=0$ (top) and two heights above the disc midplane (bottom) at $t=600$, for different values of the cooling timescale $t_c$.}
\label{fig:ring}
\end{figure*}

 For $\tc=0.1$, 10 and 100 $\Omega^{-1}$, contour plots of the meridional velocity at different stages of the disc evolution are presented in Fig.~\ref{fig:vtheta_tc}. There is a clear trend for the enhanced development of vertically elongated flow structures as the cooling timescale is increased. Previous numerical simulations of the VSI (Nelson et al. 2013; Stoll \& Kley 2014) have reported similar features, which in this context correspond to fundamental corrugation
 modes where the perturbed vertical velocities are symmetric about the midplane. A major  difference here is that these disturbances emerge for long cooling timescales ($\tc=10$, $100\ \Omega^{-1}$), whereas the VSI does not develop in such nearly adiabatic discs. As these corrugation-type modes correspond to axisymmetric $m=0$ modes, these can naturally lead to the temporary formation of ring structures. This is illustrated in Fig.~\ref{fig:ring} where the perturbed density distributions in horizontal planes located at $Z=0$ and $Z=H$ are presented at $t=600$, for the different values of the cooling timescale. For $\tc=10$ and $100\,\Omega^{-1}$, axisymmetric ring-like structures can be clearly distinguished at 
  $Z=H$ in the region  $R\in [5,7]$ where the coherent vertical flows that can be seen in the bottom right panel of Fig.~\ref{fig:vtheta_tc} operate. A feature worth emphasizing is the trend for the amplitude of the eccentric mode in the disc to decrease when moving from the midplane to higher altitudes. This is likely a consequence of a higher temperature in the midplane due to shock heating, resulting in a lower perturbed density required to maintain pressure equilibrium. As the cooling timescale is decreased and the disc behaves more and more isothermally, however, the dependence of disc eccentricity with height is reduced while the global disc eccentricity increases, consistent with the time evolution of the disc eccentricity plotted in the lower panel of Fig.~\ref{fig:ektc}. The bottom left panel of Fig.~\ref{fig:ring} suggests ring-like features can also be  present in the $\tc=0.1\Omega^{-1}$ case, although the corresponding vertical velocity distribution in the upper row of Fig. \ref{fig:vtheta_tc} does not exhibit any obvious axisymmetric modes for such small values of the cooling timescale. One possibility is that zonal flows, namely long-lived concentric axisymmetric structures, are developing in the disc. It has been suggested by Wienkers \& Ogilvie (2018) that in eccentric discs, zonal flows may indeed emerge as a result of radial variations in the Reynolds stress. As also noticed by these authors, growth of zonal flows may have important consequences on the long-term evolution of the turbulent flow as they can extract energy stored in the disc eccentricity.
  
\begin{figure}
\centering
\includegraphics[width=0.49\columnwidth]{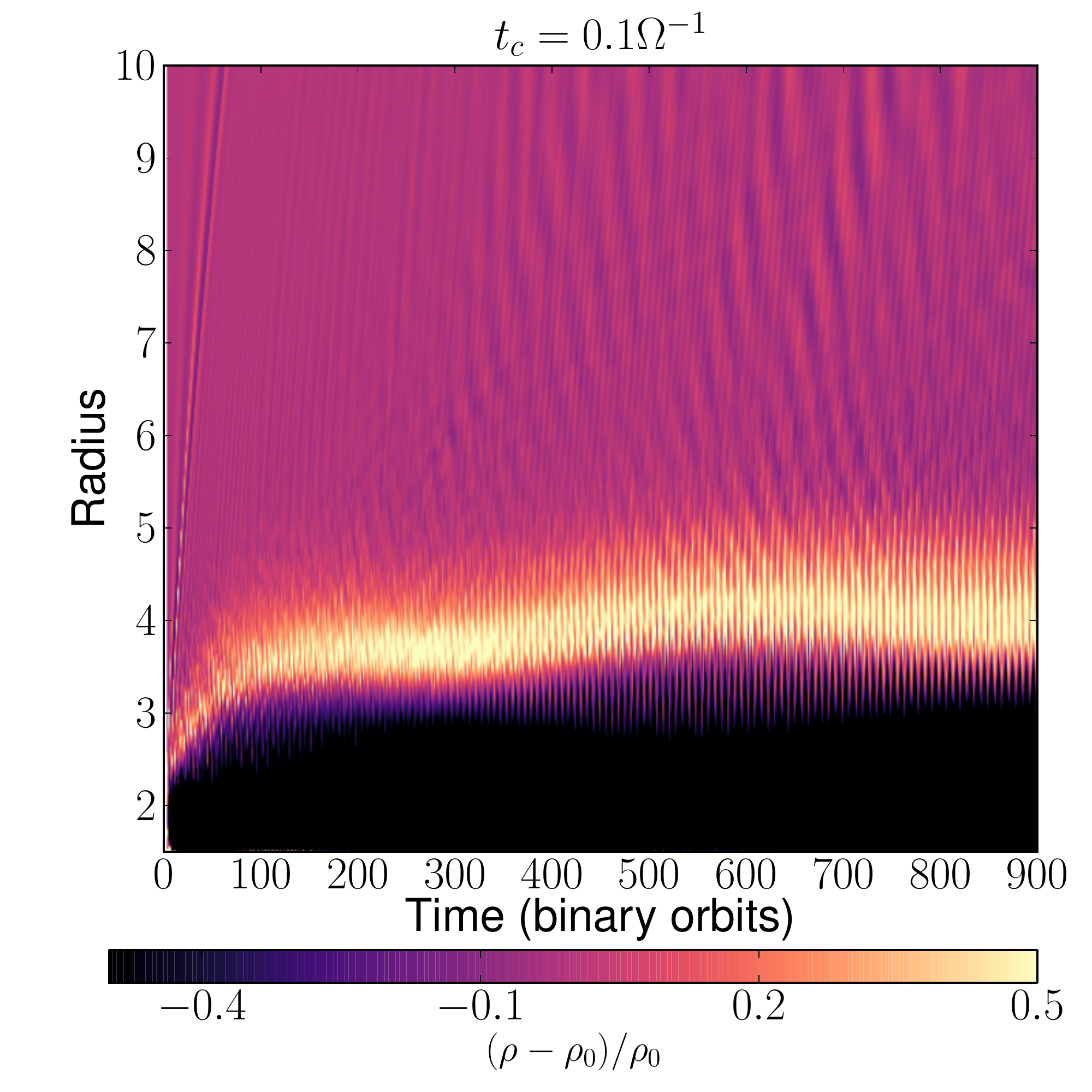}
\includegraphics[width=0.49\columnwidth]{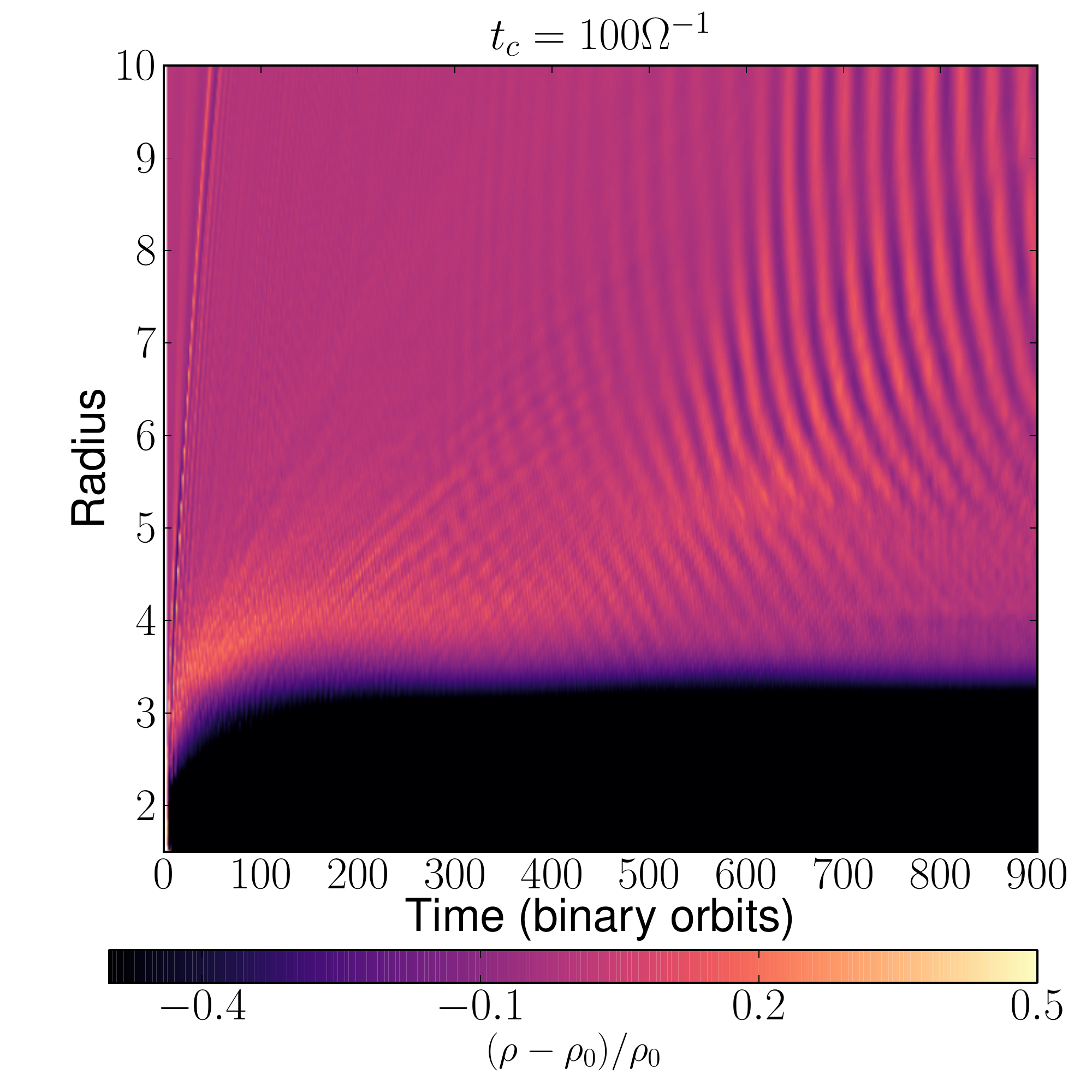}
\caption{Space-time plots of the normalized density perturbation $\delta \rho/\rho=(\rho-\rho_0)/\rho_0$ for the runs with $\tc=0.1\Omega^{-1}$ (left) and $\tc=100\Omega^{-1}$ (right). Here, $\delta \rho/\rho$ has been averaged over the azimuthal and meridional directions.}
\label{fig:zonal}
\end{figure}

 To examine whether or not zonal flows  are present in our simulations, we show space-time plots of the normalized density perturbation $\delta \rho/\rho_0$ for the runs with $\tc=0.1\Omega^{-1}$ and $100\Omega^{-1}$ in Fig.~\ref{fig:zonal}. Here, $\delta\rho/\rho_0$ has been averaged over the azimuthal and meridional directions and is presented in Fig.~\ref{fig:zonal} as a function of time and radius. The persistent axisymmetric structure visible at $R=3-4$ in the case $\tc=0.1\Omega^{-1}$ corresponds simply to the density peak located at the edge of the inner truncated cavity. 
 
 We note in passing that the size of the cavity inferred from 3D simulations seems to be smaller compared to 2D discs, which would lead to final positions of migrating planets that are in better agreement with observations. For instance, in the work of Pierens \& Nelson (2013), the size of the cavity for a Kepler-16 simulation with $\alpha\sim 10^{-4}$ was $\sim 6 \abin$, much larger than for our 3D inviscid calculations. In the 3D simulations of Kepler-413 by Pierens  \& Nelson (2018), it has been reported that the location of the gap edge was also consistent with the observed orbital location of Kepler-413b.
  
  In Fig.~\ref{fig:zonal}, the other axisymmetric features that emerge in the outer disc appear to be only intermittent structures with lifetimes of a few tens of orbits.  Therefore, these are more likely products of the modes excited by the instability rather than long-lived zonal flows. Nevertheless, these finite lifetime structures might be able to  temporarily capture particles as they correspond to pressure bumps (Stoll \& Kley 2016; Flock et al. 2017).

\section{Consequences for dust settling}
\label{sec:settling}

\subsection{Vertical particle distribution as a function of Stokes number}
We now examine how the settling of solid particles is impacted by the turbulence generated in the eccentric circumbinary disc. A steady state vertical distribution of dust grains is expected to arise, because gravitational settling balances turbulent diffusion, which can be characterized by a dust scale height $\Hd$. This is expected to be a function of the Stokes number, or equivalently the particle size. For small particles with $\st \ll 1$, $\Hd$ is given by (Zhu et al. 2015):
\begin{equation}
\Hd=\frac{H}{\sqrt{H^2\,\Omega\, \st/D_z+1}},
\end{equation}
where $D_z$ is the diffusion coefficient associated with the gas in the vertical direction. A more general expression, valid for any Stokes number, is given by (Youdin \& Lithwick 2007):
\begin{equation}
\Hd^2=\frac{D_z}{\Omega \st}\frac{1+\st}{1+\st+\st (\taucor\Omega)^2},
\label{eq:YL07}
\end{equation}
where $\taucor$ is the correlation timescale of the vertical velocity fluctuations. For a given value of the Stokes number, an important aim here 
is to compare the dust scale height inferred from our simulations with these two previous expressions. To achieve this, we restarted the fiducial 
run with $\tc=\Omega^{-1}$ at $t=955$, but including $7$ particle species with Stokes numbers $\st \in [10^{-3},1]$. For the disc model we considered, the corresponding particle sizes $a$ are such that $a\le 9/4 \lambda$, where $\lambda$ is the gas mean free path, so that all particles undergo Epstein drag, with a friction timescale, $\tf$, given by:
\begin{equation}
\tf=\frac{\rhod a}{\rho \cs},
\end{equation}
where $\rhod$ is the material density of particles. Given that $\Sigma \sim \sqrt{2 \pi} \rho H$, for particles close to the discs midplane and $\tf=\st \Omega^{-1}$ we can express the particle size as a function of Stokes number:
\begin{equation}
a\sim \frac{1}{\sqrt{2 \pi}} \frac{\Sigma}{\rhod} \st.
\end{equation}
Assuming silicate particles with $\rhod \sim 3$ g cm$^{-3}$, we list in Table~2 the particle sizes corresponding to particular values of the Stokes number, for the adopted disc model at orbital distance $R=6\;\abin$ from the central binary.  Because the surface density can exhibit strong density gradients, especially close to the disc edge, we notice that these estimations may vary significantly with disc location. For this reason and for illustrative purpose, we have also listed in Table
~2 particles sizes corresponding to a given Stokes number at $R=4\;\abin$, close to the disc edge. 

Moreover, since the particle scale height tends to be much smaller than the gas pressure scale height, we can reasonably assume that the Stokes number does not depend on the gas properties at the particle position. In this study, we therefore treat the Stokes number of a given particle as a constant, resulting in a drag force, $\mathbf{f_d}$, which can be expressed as:
\begin{equation}
\mathbf{f_d}=\frac{1}{ \Omega^{-1}\st}(\mathbf{v} -\mathbf{v_{\rm d}}),
\end{equation}
where $\mathbf{v_{\rm d}}$ is the particle velocity.

\begin{table}
\caption{Stokes number and corresponding particle size in the midplane of the  disc.}              
\label{table2}      
\centering                                      
\begin{tabular}{c c c}          
\hline\hline                        
 $\st$  & $a$($R=6\;\abin$) & $a$($R=4\;\abin$) \\ 
          & (cm)        &      (cm)  \\
\hline 
$0.001$  & $0.04$ & $0.14$ \\
$0.005$ &  $0.2$ & $0.73$ \\
$0.01$ &  $0.38$ & $1.39$ \\                       
$0.05$   &  $1.91$& $7.01$ \\
$0.1$ &  $3.8$ & $13.9$ \\
$0.5$ &  $19.1$ & $70.1$ \\
$1$ &  $38.2$ & $140.3$\\

\hline                                             
\end{tabular}
\end{table}

\begin{figure}
\centering
\includegraphics[width=0.49\columnwidth]{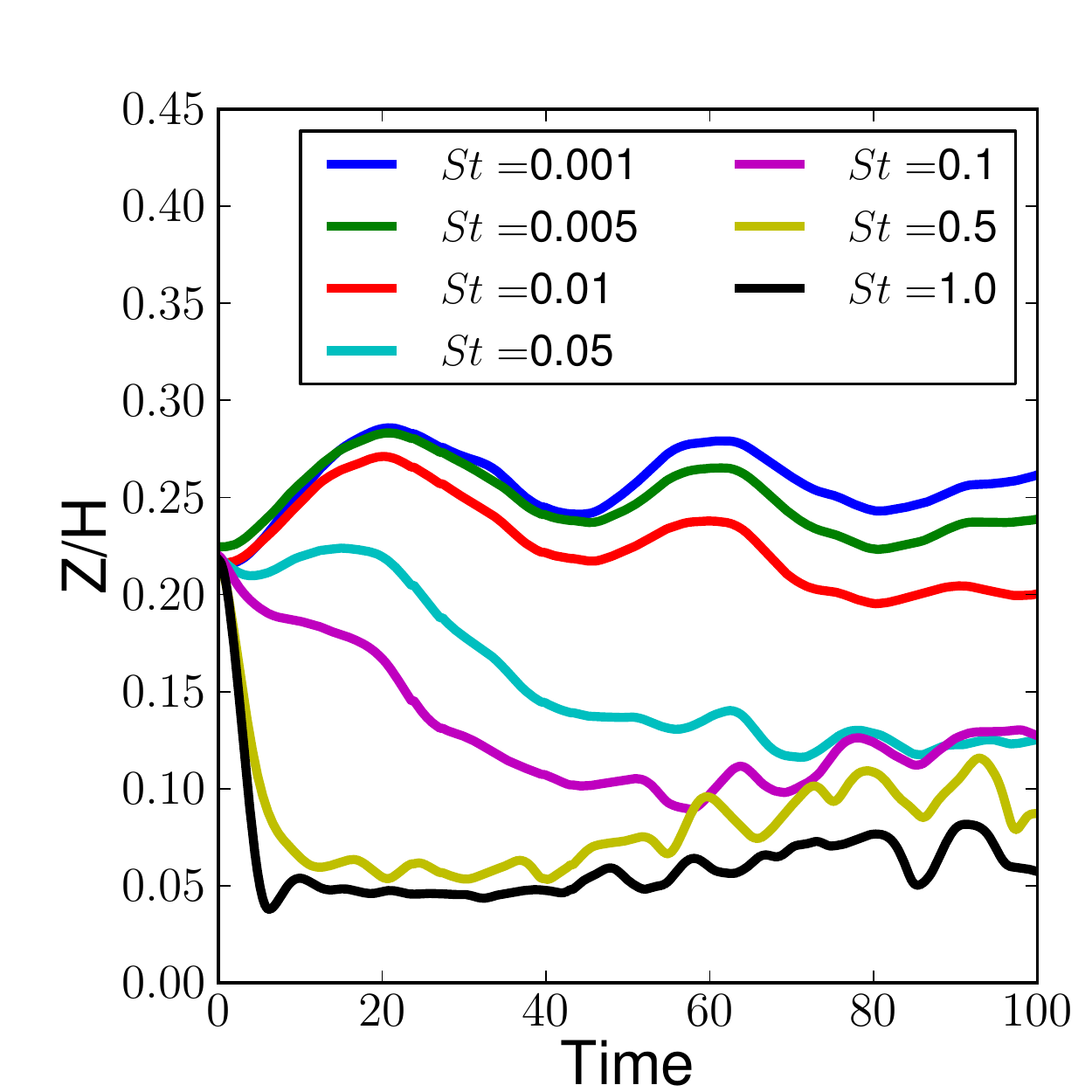}
\includegraphics[width=0.49\columnwidth]{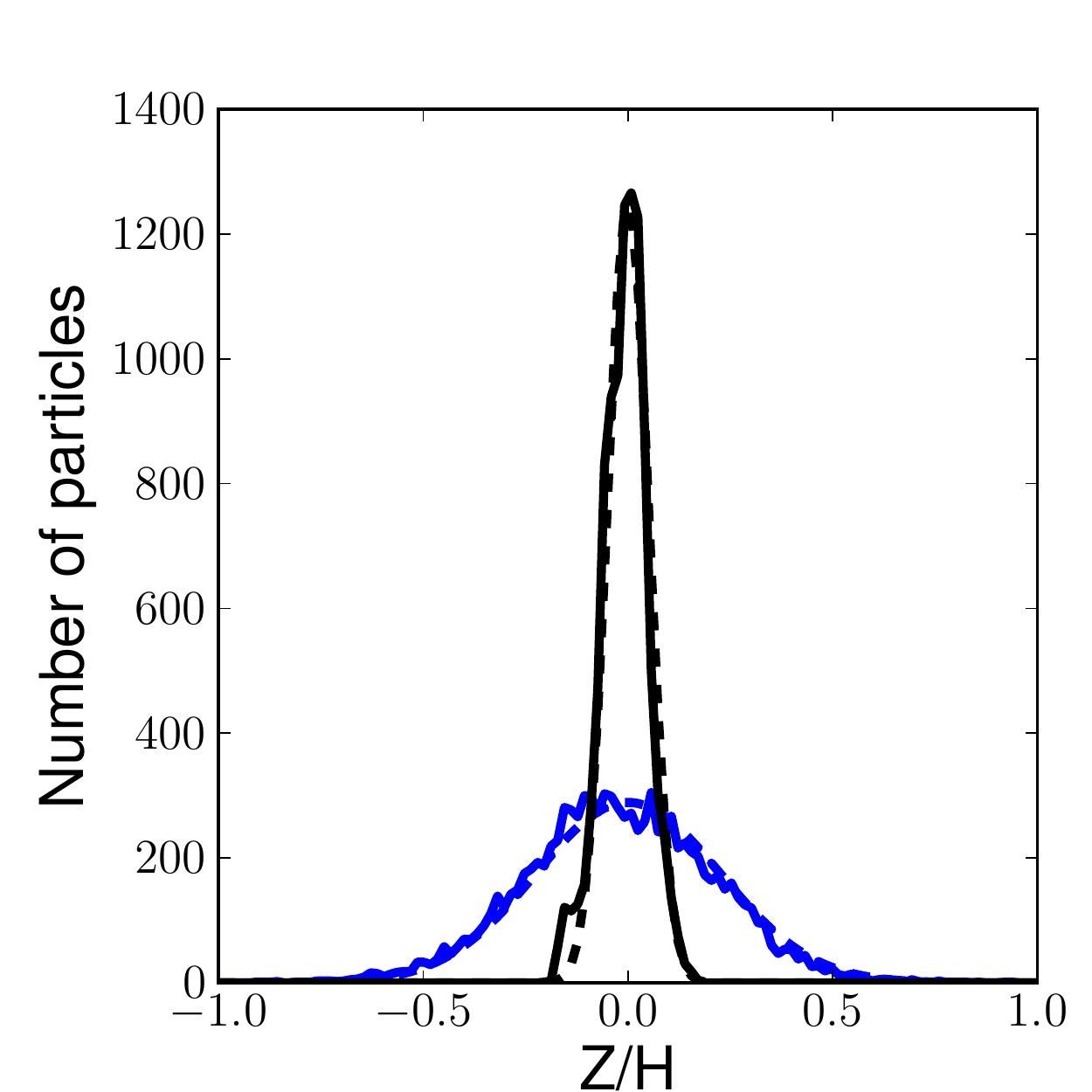}
\caption{{\it Left:} Evolution of the dust scale height as a function of time, for each value of the Stokes number that we considered. {\it Right:} For $\st=0.001$ and $\st=1$, vertical distribution of particles at $t=85$.  }
\label{fig:htime}
\end{figure}

 We consider $10^4$ particles per size that are initially distributed using a Gaussian profile with thickness $\Hd=0.2H$ in the vertical direction, and distributed uniformly in the $0<\varphi<2 \pi$ and $5<R<7$ region. In the left panel of Fig. \ref{fig:htime} we show the time evolution of the dust scale height, assuming that the vertical distribution can be fitted by a Gaussian. This assumption is supported by the right panel of  Fig. \ref{fig:htime}, which shows the vertical distribution of particles with $\st=0.001$ and $\st=1$ together with a Gaussian fit to the simulation data at $t=85$, at which indicated the particle distribution has 
reached a quasi-stationary state.\\

\begin{figure}
\centering
\includegraphics[width=\columnwidth]{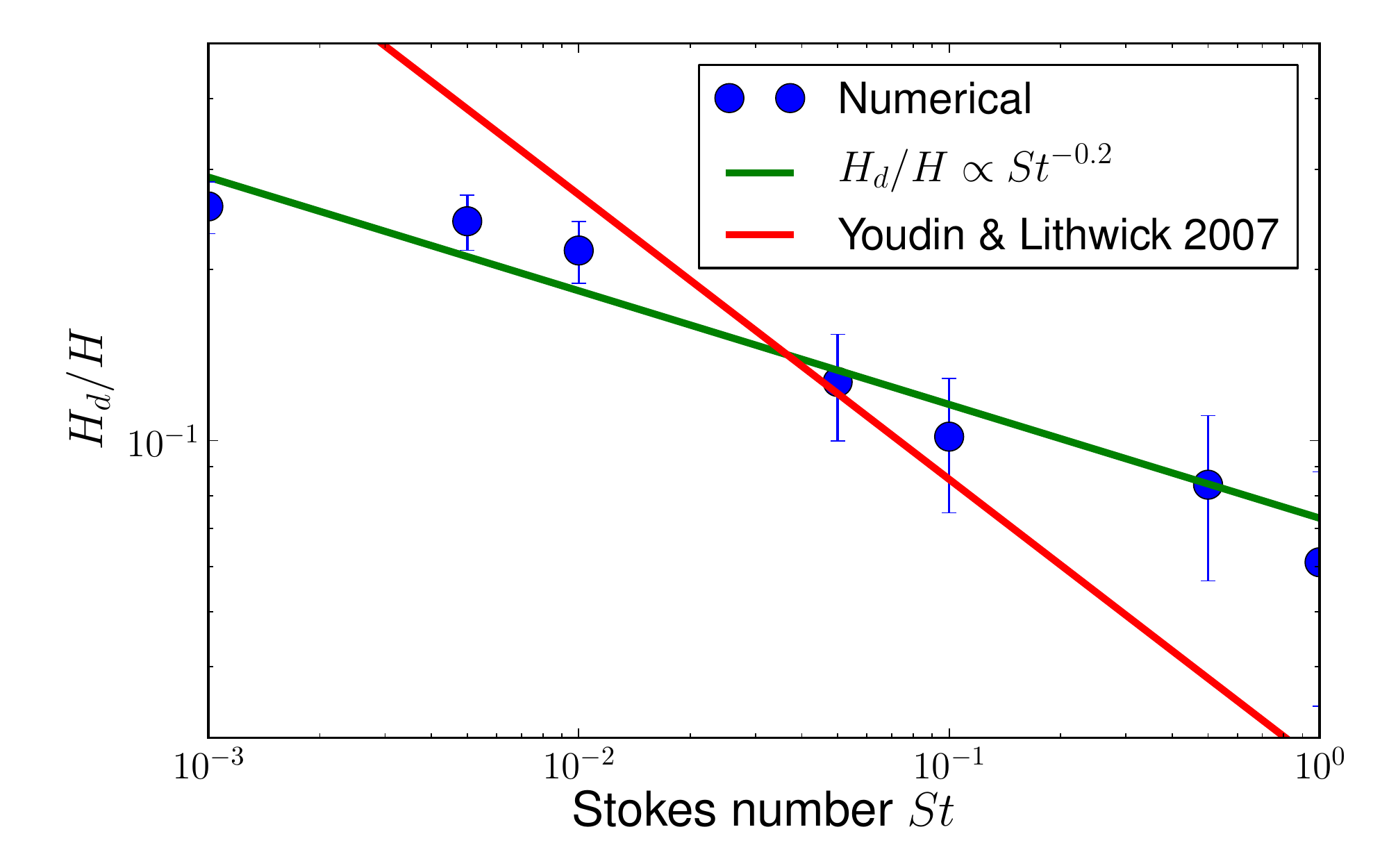}
\caption{Dust scale height, relative to the gas scale height,  as a function of Stokes number deduced from numerical simulations, and averaged  in the domain $R\in [5,7]$ and between $50$ and $80$ orbits. These are represented as blue dots, whereas the solid green line corresponds to a power-law fit such that $\Hd/H\propto \st ^{-0.2}$. The red  line corresponds to the analytical estimate of Youdin \& Lithwick (2007).  Error bars have been estimated from the vertical diffusion coefficient (see text for details).}
\label{fig:hd}
\end{figure}

\begin{figure}
\centering
\includegraphics[width=0.49\columnwidth]{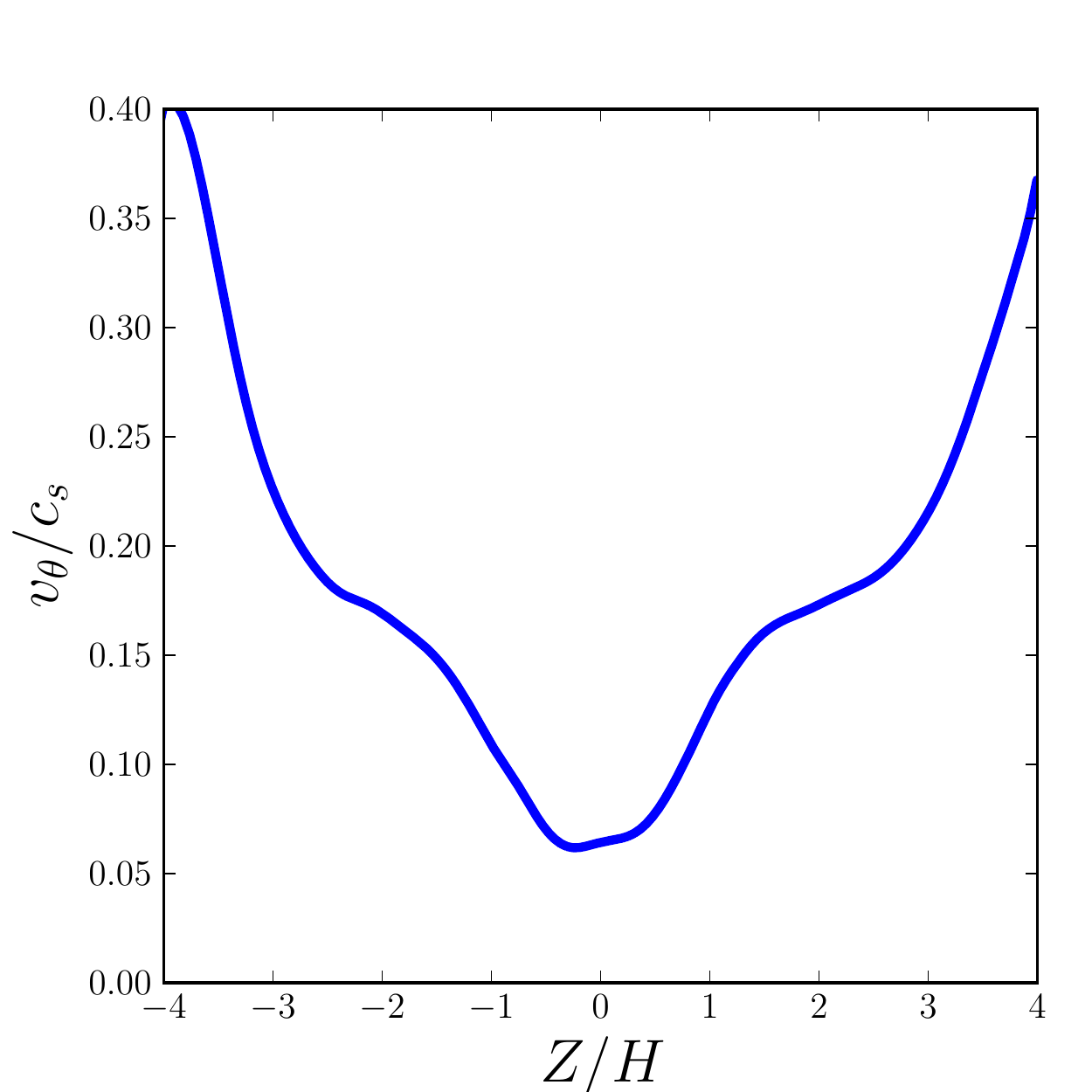}
\includegraphics[width=0.49\columnwidth]{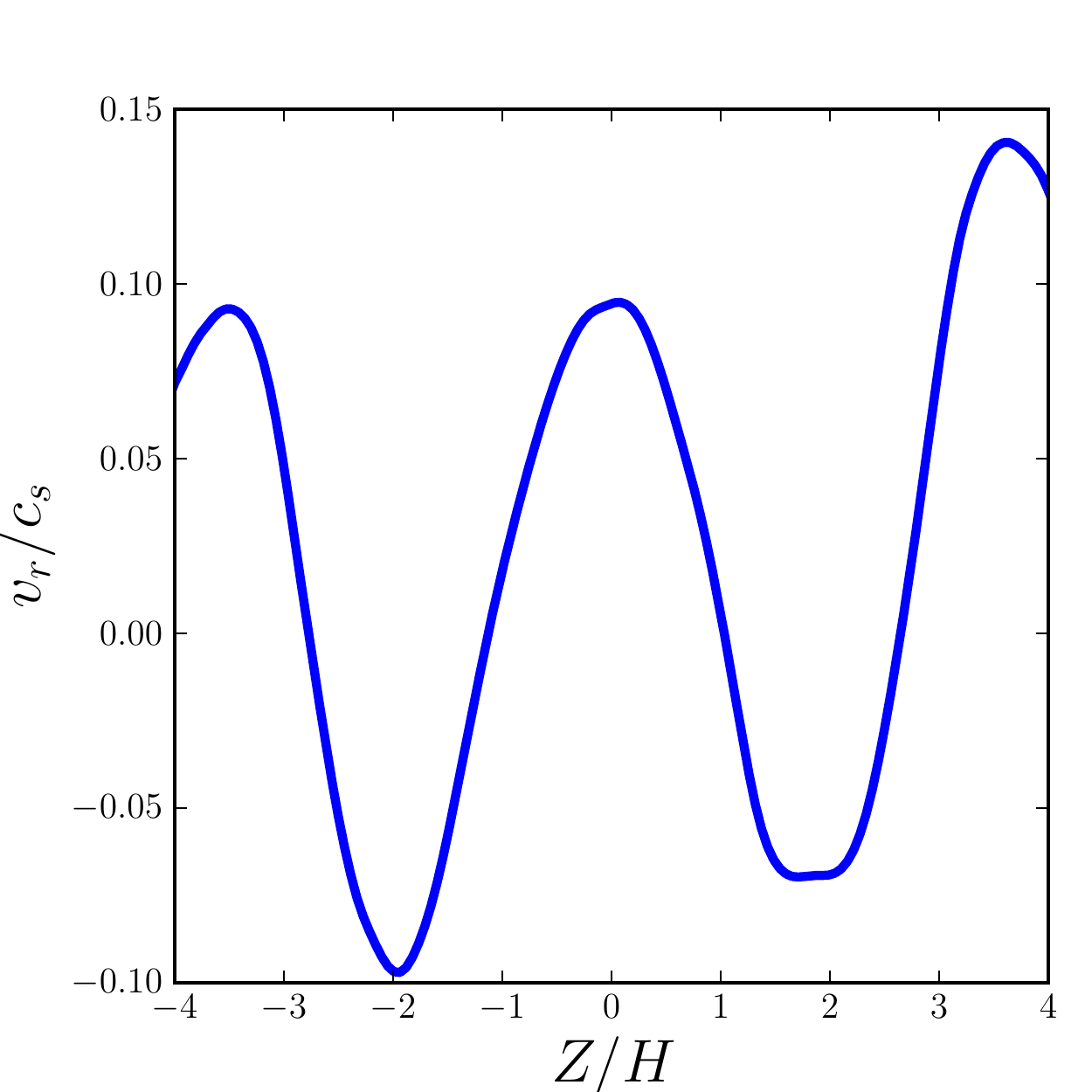}
\caption{{\it Left:}  Vertical velocity fluctuations, relative to the sound speed, as a function of height at $t=955$ and at $R=6$. Data have been averaged over 3 binary orbits. {\it Right:}  Radial velocity fluctuations versus height.}
\label{fig:vzvr}
\end{figure}

For the different values of the Stokes numbers, the steady-state dust scale heights, $\Hd$, inferred from the simulations are represented as blue dots  in Fig. \ref{fig:hd},  with overplotted error bars that have been obtained assuming that the error on the numerical measure is of the order of $\sqrt{D_z \taucor}$, where $D_z$ is the vertical gas diffusion coefficient and $\taucor$ the correlation time. We see that the grains are significantly settled. Strongly coupled particles with $\st=10^{-3}$ have scale heights $\Hd\sim 0.2-0.3H$, and more weakly coupled grains have $\Hd < 0.1 H$. We note in passing that this is consistent with the results of Lin (2018), who reports similar values from simulations of dust settling in VSI-active discs. One anomaly, however, is that the variation of $\Hd$ with $\st$ significantly deviates from the classical result $\Hd \propto \st^{-0.5}$ that has been reported in many studies (Dubrulle et al. 1995; Carballido et al. 2006). As can be seen in Fig.~\ref{fig:hd}, the variation of $\Hd$ with $\st$ can be fitted rather well by a power-law $\Hd\propto \st^{-0.2}$. Interestingly,  Fromang \& Nelson (2009) found a power-law exponent equal to $-0.2$ provides a reasonable fit to simulations of dust settling in MHD turbulence. In their simulations, however, a range of particle sizes were considered and some sizes remained distributed over a number of vertical scale heights, $H$. The deviation from $\Hd \propto \st^{-0.5}$ arises in that case because MHD turbulence is not vertically homogeneous, and is characterized by vertical velocity fluctuations that vary significantly with altitude, the consequence being that the vertical distribution of dust cannot be fitted by a Gaussian profile. As discussed earlier, this situation does not seem to apply to our runs (see right panel of Fig.~\ref{fig:htime}), which suggests that the vertical velocity fluctuations should not vary significantly over one dust scale height.  For $t=955$, velocity fluctuations are displayed as a function of height in the left panel of Fig.~\ref{fig:vzvr}. We see that similarly to MHD turbulence, the turbulence arising in our simulations is not vertically homogeneous, with vertical velocity fluctuations $v_\theta$ of the order of $5\%$ of the sound speed in the disc midplane, whereas they are $\sim 25 \%$ of the sound speed at $Z=3H$. Nevertheless, it is also evident that for each value of $\st$, the variation of $v_\theta$ over one particle scale height is only modest, which explains why the vertical particle distribution can be reasonably fitted by a Gaussian profile. Hence, it is unclear why our runs do not give rise to the expected $\Hd \propto \st^{-0.5}$, and we leave it to future work to explore this in more detail.  Nevertheless, we can speculate that this may be plausibly  related to the coherent structures in vertical velocity that are observed (see Sect. \ref{sec:comparison} below). In VSI-active discs which are  characterized by similar coherent vertical flows,   Picogna et al. (2018) have indeed found a similar trend $\Hd\propto \st^{-0.2}$ for Stokes number $\st \lesssim 0.01$, while the expected scaling $\Hd\propto \st^{-0.5}$ is recovered for larger particles. 
 
 The right panel of Fig.~\ref{fig:vzvr} shows the variation of the gas radial velocity with height at the same time as the left panel. The radial velocity is observed to be positive in the disc midplane, directed inwards at intermediate altitudes, and then positive again near the disc surface. Although this is similar to magnetised discs with net magnetic flux (Bethune et al. 2017), the direction of the flow is opposite to the radial flow generated by the VSI as described by Stoll \& Kley (2016). As the radial velocities are determined by the vertical stress profile, this indicates that the combined effects of the eccentricity-generated turbulence and the spiral waves in our our simulations produce a different stress profile compared to that generated by VSI turbulence.

\begin{figure}
\centering
\includegraphics[width=\columnwidth]{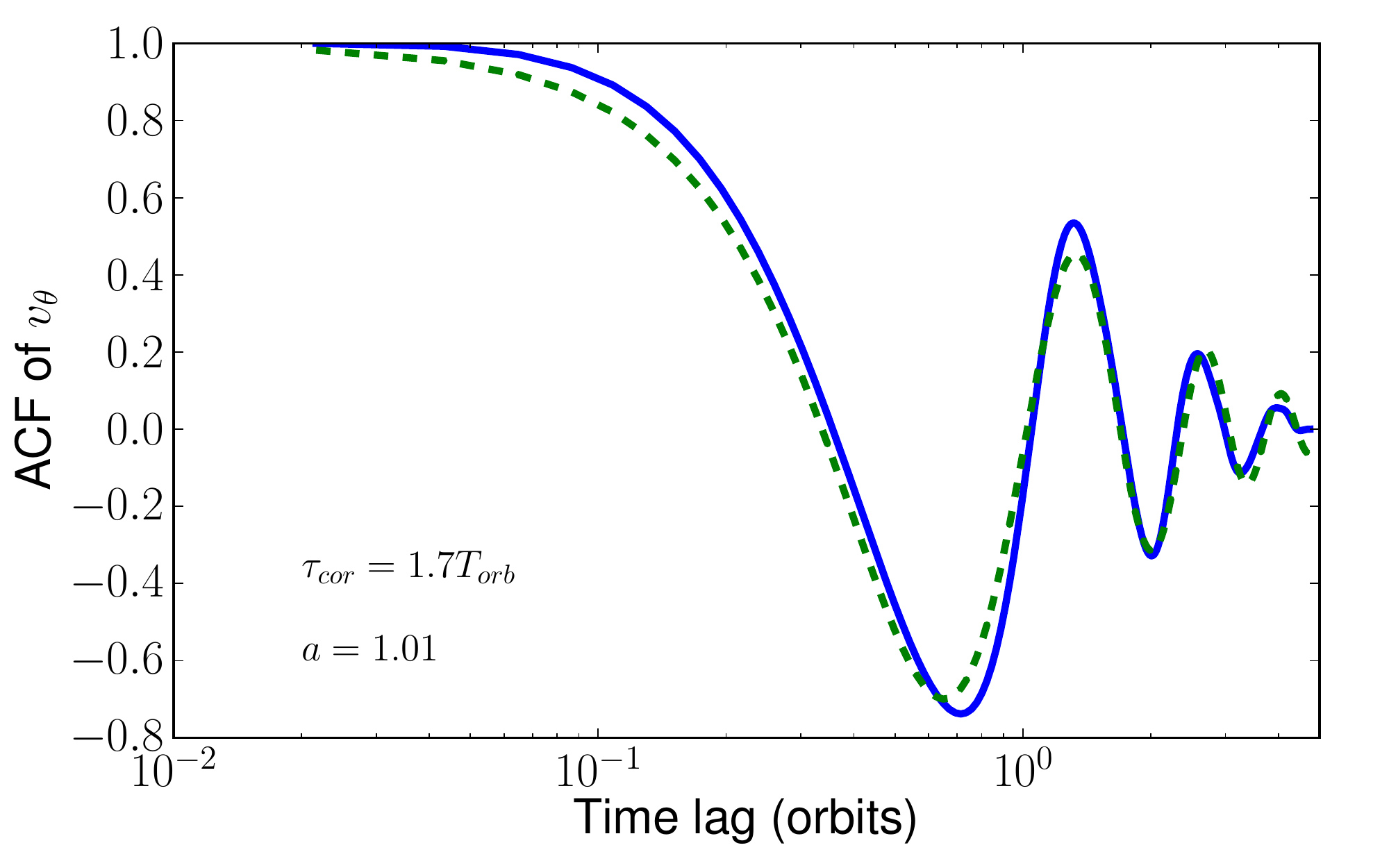}
\caption{Auto-correlation function (ACF) of the vertical velocity fluctuations at $R=6$. The green dashed line corresponds to the fit obtained using Eq.~\ref{eq:fit}. }
\label{fig:acf}
\end{figure}

\begin{figure}
\centering
\includegraphics[width=\columnwidth]{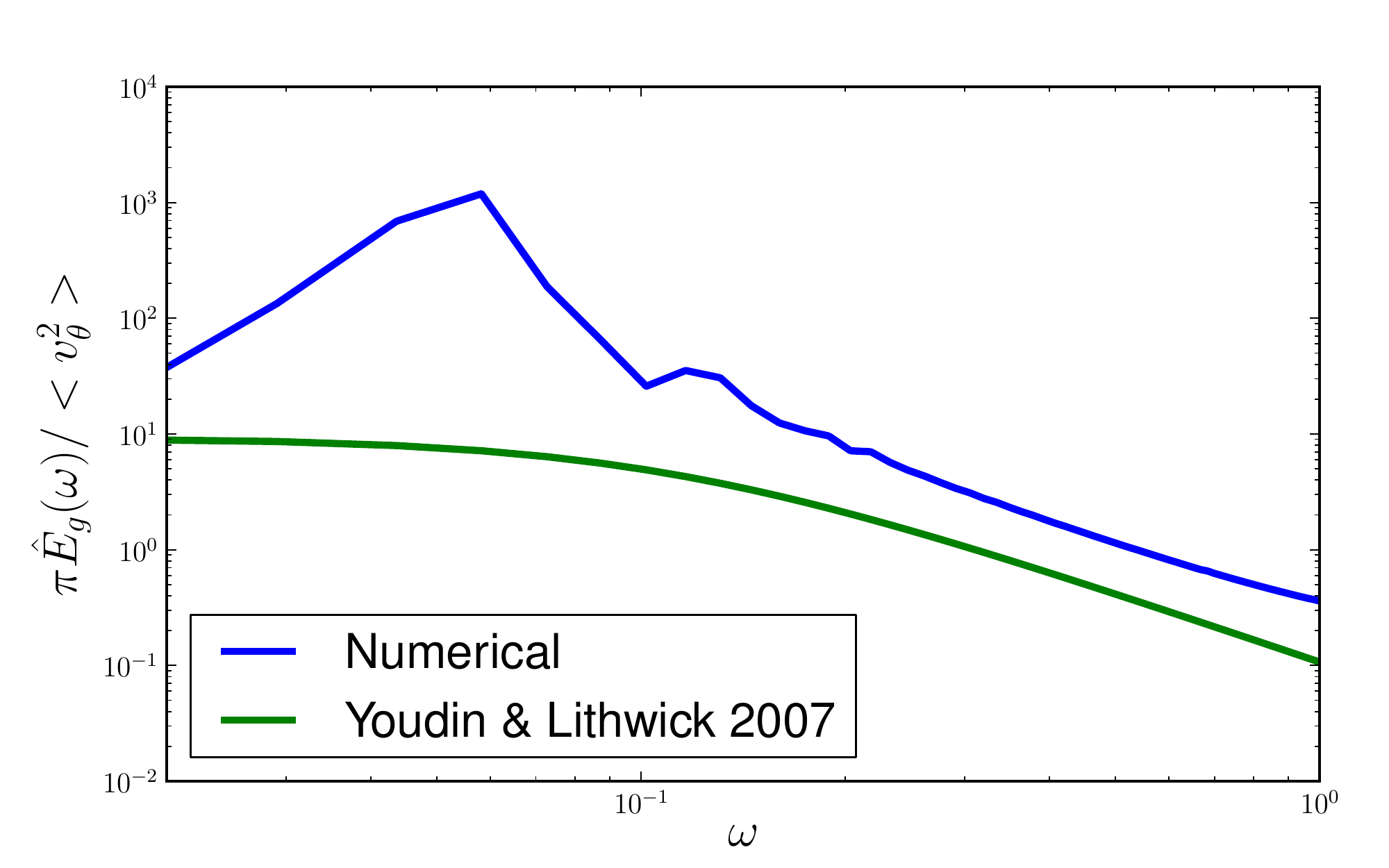}
\caption{Power spectrum of the vertical velocity fluctuations at $R=6$. }
\label{fig:fourier}
\end{figure}

\subsection{Vertical diffusion coefficient}
In order to directly compare the dust scale height deduced from our simulations with Eq.~\ref{eq:YL07}, we first determine the gas vertical diffusion coefficient $D_z=\delta v_\theta^2 \taucor$. To estimate the correlation time $\taucor$, we restarted the run with $\tc=\Omega^{-1}$ at $t=955$, outputting the meridional velocity every $0.02 \Torb(R=6)$, and evaluating the autocorrelation function (ACF) of $v_\theta$ according to:
\begin{equation}
ACF(\tau)=\left< v_\theta(0)v_\theta(\tau) \right>,
\end{equation}
where the ensemble average is produced by averaging $v_\theta(0)v_\theta(\tau)$ over time and over the disc regions corresponding to $|Z|<H$ and $5\le R\le 7$. Three different methods for estimating the correlation time, $\tau_{cor}$, have been presented in the literature: i) Fitting the ACF using the following function (Nelson \& Gressel 2010):
\begin{equation}
S(t)=[(1-A)+A\cos(2\pi \omega t)]\exp(-t/\taucor),
\label{eq:fit}
\end{equation}
where $A$ represents the relative strength of the sinusoidal feature in the ACF and $\omega$ is the frequency associated with the sinusoidal component. As can be seen in Fig. \ref{fig:acf}, a model with $A=1.01$ and $\taucor=1.7 \Torb$ provides a good fit of the ACF. ii)  The second method consists in 
determining the smallest lag value for which the ACF crosses zero with positive slope (Baruteau \& Lin 2010), and this results in $\taucor \sim \Torb$. iii) Finally, $\taucor$ can be estimated by computing the time integral of the ACF (Yang et al. 2009):
\begin{equation}
\taucor\sim \frac{\int_0^\infty ACF(t) \, {\rm d}t}{2\ ACF(0)}.
\label{eq:yang}
\end{equation}
Using this method, one gets $\taucor\sim 0.015 \Torb$ or equivalently $\taucor\sim 0.1 \Omega^{-1}$, which is close to the correlation time typical of MHD turbulence in protoplanetary discs (Fromang \& Nelson 2006; Nelson \& Gressel 2010). 

Assuming $v_z\sim v_\theta$ and given that we have $\left<v_z^2\right>\sim 3.3\times 10^{-6}$ at $R=6$ in code units, this results in a  vertical diffusion coefficient  $D_z=\left<v_z^2\right>\tau_{cor}$ of
$D_z\sim 5\times 10^{-4}, 3\times 10^{-4}, 4.5\times 10^{-6}$ for methods i), ii), iii) respectively. Equivalently, in terms of the vertical Schmidt number ${\it Sc}_z=\frac{\alpha H^2 \Omega}{D_z}$, we obtain ${\it Sc}_z=0.02, 0.04, 3$. We note that ${\it Sc}_z\sim 3$ has  been reported in MHD simulations of dust vertical settling in turbulent discs (Zhu et al. 2015). A value of $D_z\sim 4.5\times 10^{-6}$, equivalent to a vertical diffusion parameter $\alphadiff\sim 1.6 \times 10^{-3}$ based on the $\alpha$ prescription, is also more consistent with the results of our  simulations, which show a high level of vertical settling occurring. As a consequence, this suggests that using Eq.~\ref{eq:yang} provides the best estimate of the correlation time.

\subsection{Comparison with analytical estimate for $H_d$}
\label{sec:comparison}
Returning to Fig.~\ref{fig:hd}, the red line shows the estimate for $\Hd$ using Eq.~\ref{eq:YL07}, which we plugged in our estimated value for $D_z$. 
There is substantial discrepancy between the Youdin \& Lithwick (2007) formula and the particle vertical distribution obtained numerically, which is not surprising
since Eq.~\ref{eq:YL07} predicts $\Hd\sim\sqrt{D_z/\Omega \st}$ in the limit where $\st \rightarrow 0$ and $\taucor\sim \Omega^{-1}$ (Zhu et al. 2015). However, we remark that Eq.~\ref{eq:YL07} has been derived assuming isotropic turbulence and a power spectrum for the turbulence that reads:
\begin{equation}
\hat E_{\rm g}(\omega)=\frac{\left<v_\theta^2\right>}{\pi}\frac{\taucor}{1+\omega^2 \taucor^2},
\label{eq:ps}
\end{equation}
where $\omega$ is the frequency. The power spectrum of the turbulent velocity fluctuations at $R=6$ is compared to the previous expression in  Fig.~\ref{fig:fourier}. We see that there is a fairly good agreement for frequencies above $1/\taucor$, which also demonstrates
that the turbulent spectrum is Kolmogorov in the integral scale. In the inertial scale, however, namely for frequencies below $1/\taucor$, the Youdin \& Lithwick (2007) power spectrum is almost constant while the power spectrum derived from the simulations actually increases with frequency. The main implication is that Eq.~\ref{eq:YL07}, which can be obtained assuming Eq.~\ref{eq:ps} for the power spectrum (see Youdin \& Lithwick 2007), cannot be used in the present work to derive the dust scale height as a function of Stokes number.  A similar result has been reported in the non-ideal MHD simulations of Zhu et al. (2015) that include ambipolar diffusion. In that case, the authors state that this arises because of coherent structures in the vertical velocity 
that persist over hundreds of orbits. Here, it is clear that the coherent vertical flows that can be observed in Fig.~\ref{fig:vtheta_tc} may also contribute to making the power spectrum for turbulence deviate from the one given in Eq.~\ref{eq:ps}.

\section{Pebble dynamics in turbulent circumbinary discs}
\label{sec:pebbles}
We now examine the efficiency of pebble accretion onto protoplanets that are embedded in the fiducial circumbinary disc model, using a suite of simulations containing pebbles and accreting protoplanets. The aim is to determine the effect of binary-induced turbulence on the ability of planets to form via pebble accretion near the inner edge of a circumbinary disc, where numerous circumbinary planets have been discovered by the Kepler mission. Turbulent stirring may reduce the pebble accretion rate because it causes the particle vertical scale height to be larger than the Hill radius of the accreting body (Lambrechts \& Johansen 2012). Furthermore, the turbulent velocity kicks received by particles that enter the Hill sphere of an accreting body may also reduce the accretion efficiency.

 We restart the disc model at $t=955$ with an embedded protoplanet that accretes pebbles. The protoplanet remains on a fixed circular orbit with semimajor axis $\ap=6$, and has a mass in the interval $0.1 \; M_{\oplus} \le \msp \le 10\;M_\oplus$.  The planet-to-binary mass ratio, $\qp$, always satisfies $\qp<h^3$, so the thermal criterion for gap opening (Ward 1997) is never satisfied. The core masses we consider are therefore below the pebble isolation mass, above which the inwards drift of pebbles can be halted outside the planet's orbit (Bitsch et al. 2018; Ataiee et al. 2018). 

We inject $\sim 10^5$ solid particles per size bin in the radial range $R\in[5,7]$, with a Gaussian vertical distribution $\propto \exp(-Z^2/\Hd^2)$, where $\Hd$ takes the value derived in Sect.~\ref{sec:settling}. Pebbles that pass within the estimated radius for pebble accretion (see below) are considered to have been accreted by the protoplanet, provided that both of the following conditions are satisfied (Picogna et al. 2018): 
\begin{enumerate}
\item The total energy of the particle relative to the planet is smaller than the gravitational potential energy at a distance of one Hill radius from the core:
\begin{equation}
\Ek+\Ep<\Ep(R_{\rm H}),
\end{equation}
where $\Ek$ is the kinetic energy, $\Ep$ is the gravitational potential energy, and $R_{\rm H}=\ap(\msp/3M_\star)^{1/3}$ is the protoplanet Hill radius.
\item The gravitational deflection time, $\tg=\Delta v/(G\msp/\rp^2)$, where $\Delta v$ and $\rp$ are the relative velocity and distance between the particle and the protoplanet, is such that $\tg<4\;\st\;\Omega^{-1}$ (Ormel \& Klahr 2010; Picogna et al. 2018) .
\end{enumerate}

The pebble accretion efficiency is defined as the ratio of the pebble accretion rate onto the planet, $\dot M$, and the pebble accretion rate through the disc, $\dot M_{\rm drift}=2\pi R\Sigma_{\rm p} v_{\rm drift}$. To obtain the pebble accretion efficiency, we measure $\dot M$ by monitoring the number of accreted particles over a time interval of $\sim 10$ planetary orbits, and we estimate $\dot M_{\rm drift}$ using the following relation for the radial drift velocity
\begin{equation}
v_{\rm drift}=\frac{-2\st}{1+\st^2}\eta \vK
\end{equation}
with
\begin{equation}
\eta=-\frac{1}{2}h^2(p+q)\, .
\end{equation}
where $\vK$ is the Keplerian velocity around the binary.

 An alternative method of calculating the pebble accretion efficiency would be to consider a pebble ring located outside the planet's orbit, and to record the fraction of pebbles that cross the planet's orbit in a given time interval. In laminar discs, this approach has been employed by Morbidelli \& Nesvorny (2012), and was used by Picogna et al. (2018) in disc models where turbulence was driven by the VSI. Here, however, we found that using this method does not lead to a reliable value for the pebble accretion efficiency, especially for particles that are strongly coupled to the gas. This is  due to the complex random trajectories followed by these particles as a result of disc turbulence, resulting in a solid particle moving back and forth across the planet orbit. To overcome this, we would need to be able to evolve our simulations over time scales where this random walk would average out, such that even tightly coupled particles would experience significant radial drift. Unfortunately this requirement goes beyond our available computational resources. 
 
\begin{figure}
\centering
\includegraphics[width=\columnwidth]{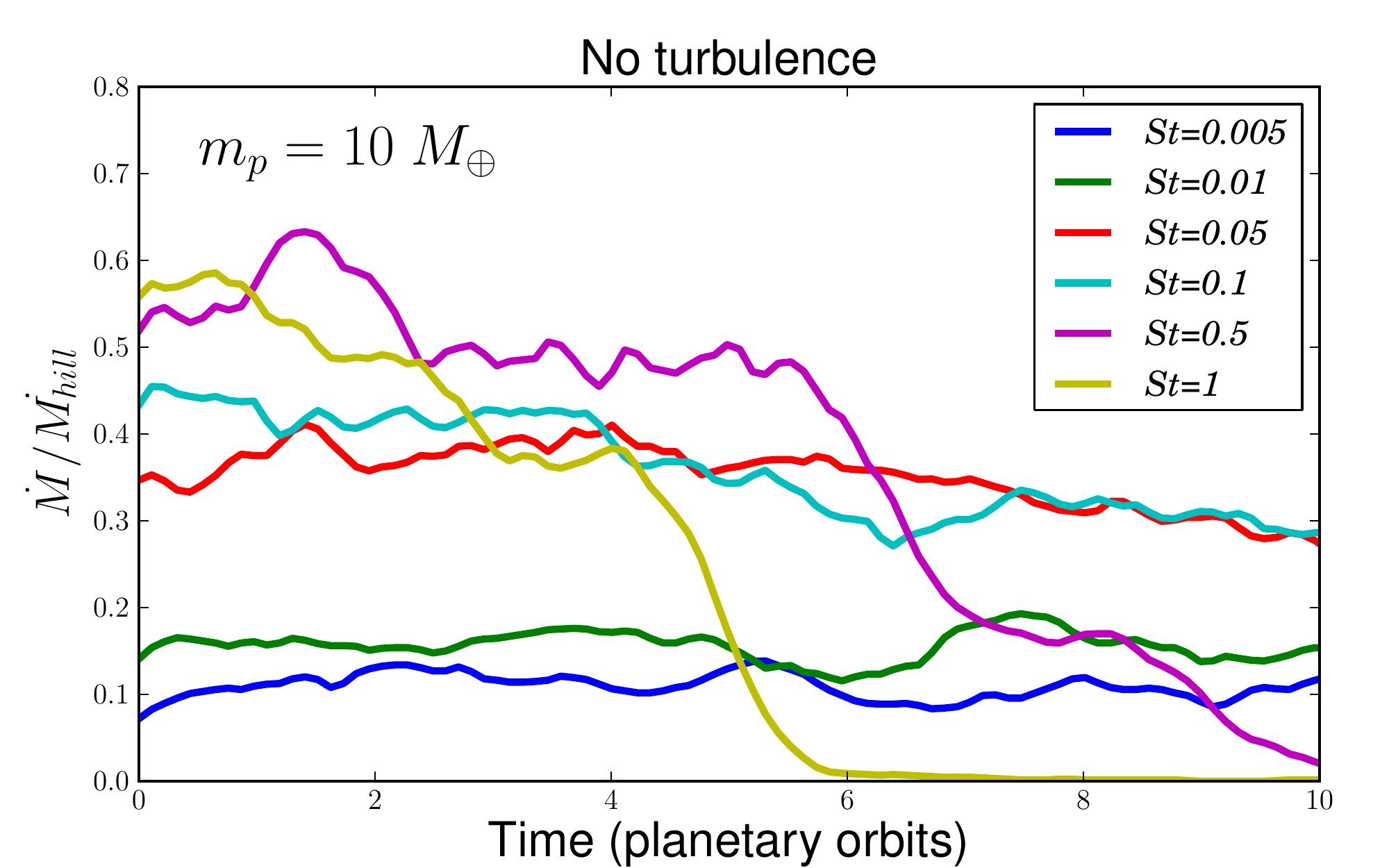}
\includegraphics[width=\columnwidth]{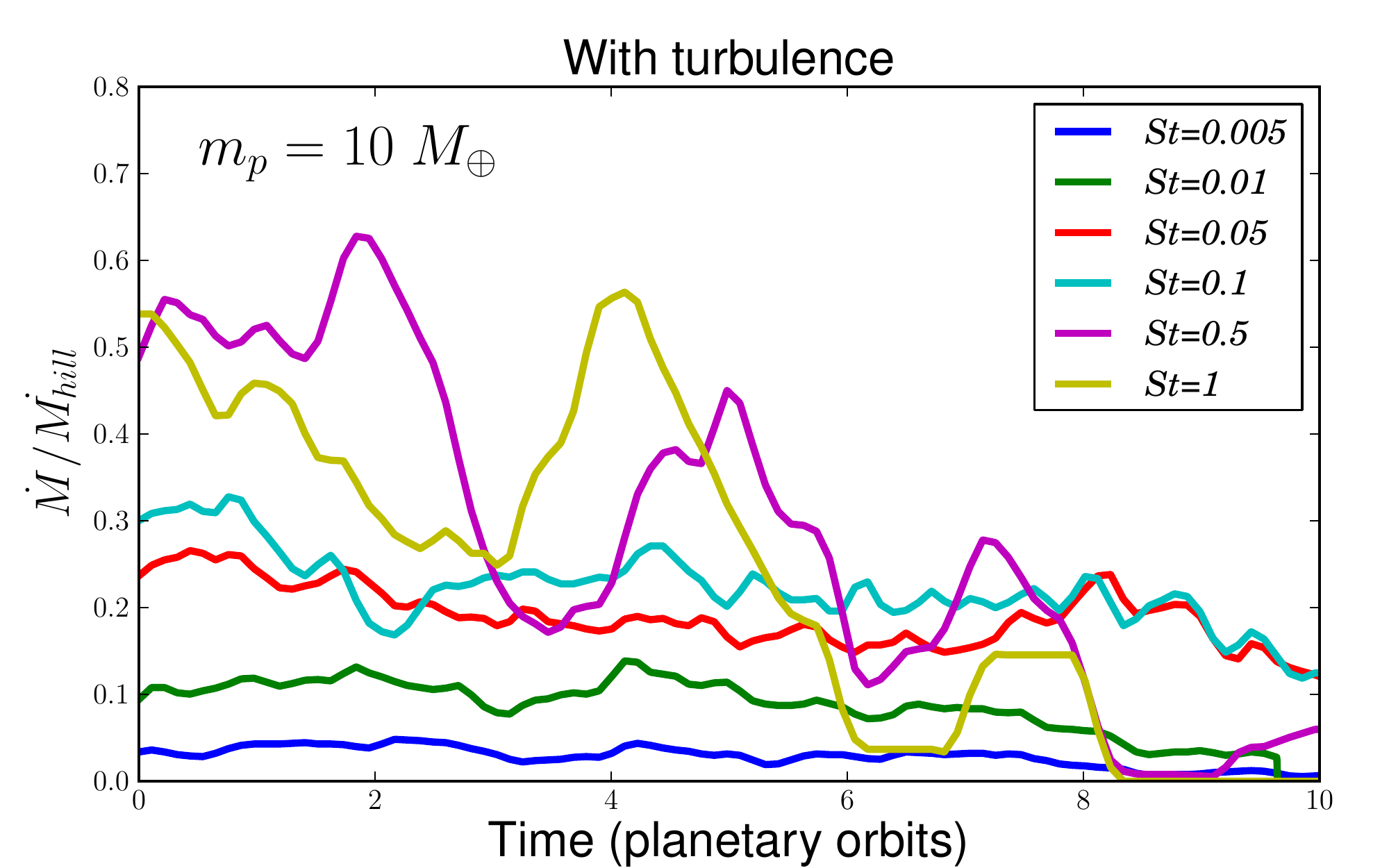}
\caption{{\it Top:}  Time evolution of the pebble accretion rate onto a $10\;M_\oplus$ protoplanet relative to the expected accretion rate in the Hill regime, in simulations where pebbles experience gas drag from the initial and non-evolving disc model. {\it Bottom:} Same but in the case of an evolving disc where turbulence fully develops.  }
\label{fig:mdothill}
\end{figure}

\begin{figure*}
\centering
\includegraphics[width=\textwidth]{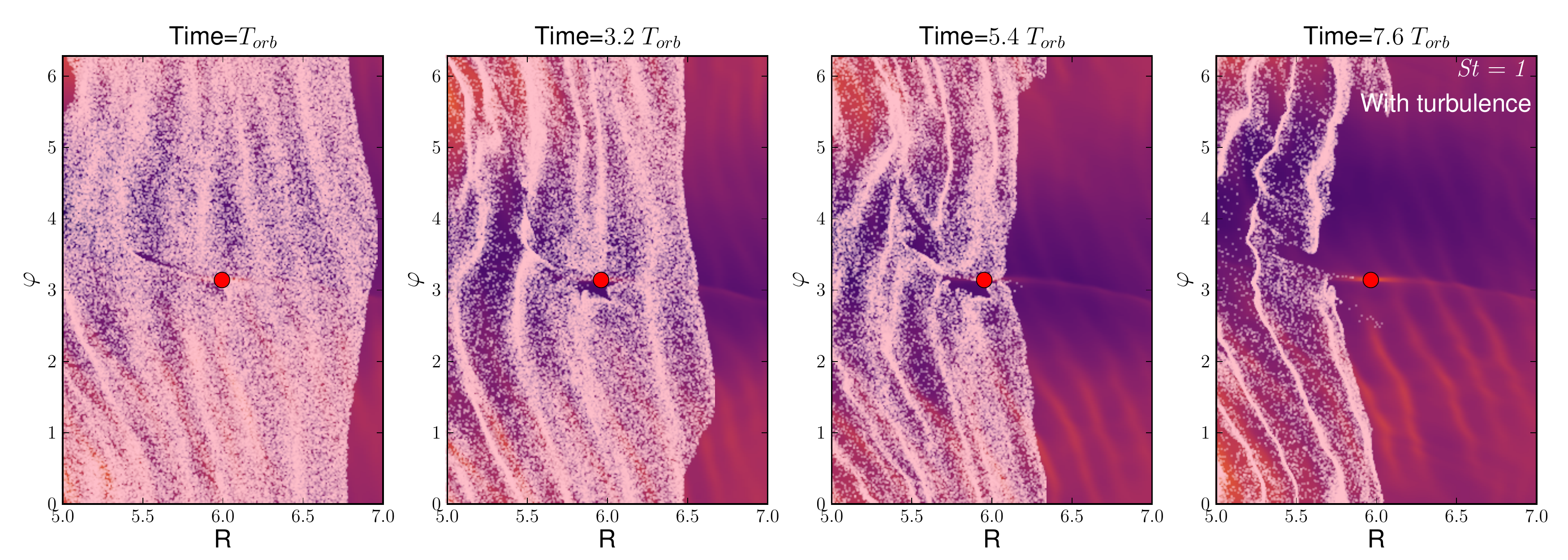}
\caption{Snapshots of the distribution of particles with $\st=1$ at different times for a turbulent run in which pebbles are injected at $t=955$. Note the unit of time adopted in each panel is the orbital period at $r=6 a_{\rm bin}$.}
\label{fig:st1}
\end{figure*}
 
In the Hill regime for pebble accretion, the expected accretion rate is given by (Lambrechts \& Johansen 2012): 
\begin{equation}
\dot M_{\rm Hill}=2\Sigma_{\rm p} R_{\rm H} v_{\rm H},
\end{equation}
where $v_{\rm H}=\Omega R_{\rm H}$ is the Hill velocity.  Figure~\ref{fig:mdothill} shows, relative to $\dot M_{\rm Hill}$, the pebble  accretion rate onto a $10\;M_\oplus$ core as a function of time, for simulations in which:
\begin{enumerate}
\item  Pebbles are injected at $t=0$ in a laminar disc orbiting a single star, whose structure corresponds to the initial conditions described in  Sect.~\ref{sec:init}, and which does not evolve over time. Using simulations in which the non-evolving disc orbits a central binary, we checked  that the direct gravitational interactions between the central binary and the pebbles have little impact on pebble accretion for planets located at $\ap=6 \abin$.
\item Pebbles are injected at $t=955$ and feel the effect of gas drag from the circumbinary disc whose velocity and density fields are not fixed, and in which turbulence is fully developed. 
\end{enumerate}
For pebbles with $\st \le 0.1$,  Fig.~\ref{fig:mdothill} shows that a quasi-stationary value for the accretion rate is quickly reached in both cases. However, for particles with $\st \ge 0.5$ that are moderately coupled to the gas, and therefore experience fast radial drift, the accretion rate rapidly decreases once most of the pebbles available in the region $R\in[6,7]$ have crossed the planet's orbit. This is illustrated in Fig.~\ref{fig:st1}, where snapshots of the distribution of particles with $\st=1$, projected onto the disc midplane, are shown for the turbulent disc. For these particles, the pebble accretion efficiency can be simply obtained by dividing the number of accreted pebbles by the number of pebbles located initially in the region $R\in[6,7]$. 

 For core masses $\msp=0.1$, 1, 5, 10~$M_\oplus$, and the two disc setups, the pebble accretion efficiency as a function of $\st$ is presented in the left column of Fig.~\ref{fig:efficiency}. In this figure, we also compare our estimates with the analytical estimate of Liu \& Ormel (2018) for the pebble accretion efficiency, $\epsilon_{\rm set}$, in the settling limit :
\begin{equation}
\epsilon_{\rm set}=0.32\sqrt{\frac{\qp}{\st \;\eta^2}\frac{v_{\rm rel}}{\vK}},
\label{eq:epsilonset}
\end{equation}
where $v_{\rm rel}$ is the relative velocity between the planet and the pebble, for which Liu \& Ormel (2018) make use of an analytical estimate that combines the Hill and drift regimes:
\begin{equation}
v_{\rm rel}=\left[1+5.7\left(\frac{\qp\; \st}{\eta^3}\right)\right]^{-1} \eta \vK+0.52(\qp \st)^{1/3}\vK.
\end{equation}
We see that for setup i), there is relatively good agreement between the pebble accretion efficiency deduced from our simulations and the relation for $\epsilon_{\rm set}$ given by Eq.~\ref{eq:epsilonset}, which validates our procedure for calculating the pebble accretion rate. In a turbulent disc,  however, pebble accretion tends to be less efficient, especially for masses $\msp\lesssim 1\; M_\oplus$ and small pebble sizes. We show below that the reduced efficiency of pebble accretion in the presence of turbulence occurs because of two effects. First, turbulence can reduce pebble accretion because of a 3D effect. This occurs for pebbles that are lofted up from the midplane by the turbulent flow, such  that the pebble scale height is larger than the planet accretion radius. Compared to a non-turbulent disc, in which the accretion process is two dimensional, the  accretion rate onto the core should be reduced by a factor of $\sim r_{\rm acc}/\Hd$. Second, for small core masses, it is possible that the settling velocity becomes smaller than the typical velocity fluctuations induced by the turbulence, leading to a reduction of the pebble accretion efficiency.

\subsection{3D effects of turbulence on pebble accretion}
 We can provide a quantitative evaluation of this effect by first estimating the accretion radius, $r_{\rm acc}$. A crude value for $r_{\rm acc}$ can be simply obtained by setting $\Delta v\sim v_{\rm set}$ where $v_{\rm set}=G\msp \ts/r_{\rm acc}^2$ is the settling velocity, obtained by balancing the 
gravitational force exerted by the planet with the gas drag force (Liu \& Ormel 2018). This gives :
\begin{equation}
r_{\rm acc}\sim \sqrt{\frac{G\msp\ts}{\Delta v}}.
\label{eq:racc}
\end{equation}
In the drift regime for pebble accretion, the relative velocity between a particle and the accreting core, $\Delta v$, is of the order of the headwind velocity 
$v_{\rm hw}=\eta \vK$. Setting $\Delta v=\eta \vK$ in Eq.~\ref{eq:racc}, it is straightforward to show the accretion radius in that case is given by
\begin{equation}
r_{\rm acc, d}\sim 1.4 \;\qp^{1/6}\st^{1/2}\eta^{-1/2}R_{\rm H}.
\end{equation}
In the Hill regime for pebble accretion, however, $\Delta v$ equals the shear velocity $v_{\rm sh}=r_{\rm acc} \Omega_{\rm K}$. Again, substituting $\Delta v=r_{\rm acc} \Omega_{\rm K}$ in Eq.~\ref{eq:racc} results in: 
\begin{equation}
r_{\rm acc, h}\sim 1.4\;  \st^{1/3}R_{\rm H}.
\end{equation}
Overall, the accretion radius can therefore be defined as (Baruteau et al. 2016):
\begin{equation}
r_{\rm acc}=\min(r_{\rm acc,d}, r_{\rm acc, h}).
\end{equation}
We note in passing that the transition between the drift and Hill regimes is expected to occur at a mass $M_{\rm t}$ for which $v_{\rm hw}\sim v_{\rm sh}$, and which is given by 
  $M_{\rm t}\sim \eta^3 \st^{-1}M_\star$. Alternatively, for a given planet-to-binary mass ratio $\qp$, we expect pebble accretion to proceed in the Hill (resp. drift) regime for Stokes numbers higher (resp. lower) than: 
\begin{equation}
\st \sim \eta^3 \qp^{-1}.
\end{equation}
For $\qp=3\times 10^{-5}$ ($\msp=10\;M_\oplus$), this gives $\st \sim 0.003$ whereas we have $\st \sim 0.3$ for  $\qp=3\times 10^{-7}$ ($\msp=0.1\;M_\oplus$). 

\begin{figure*}
\centering
\includegraphics[width=0.45\textwidth]{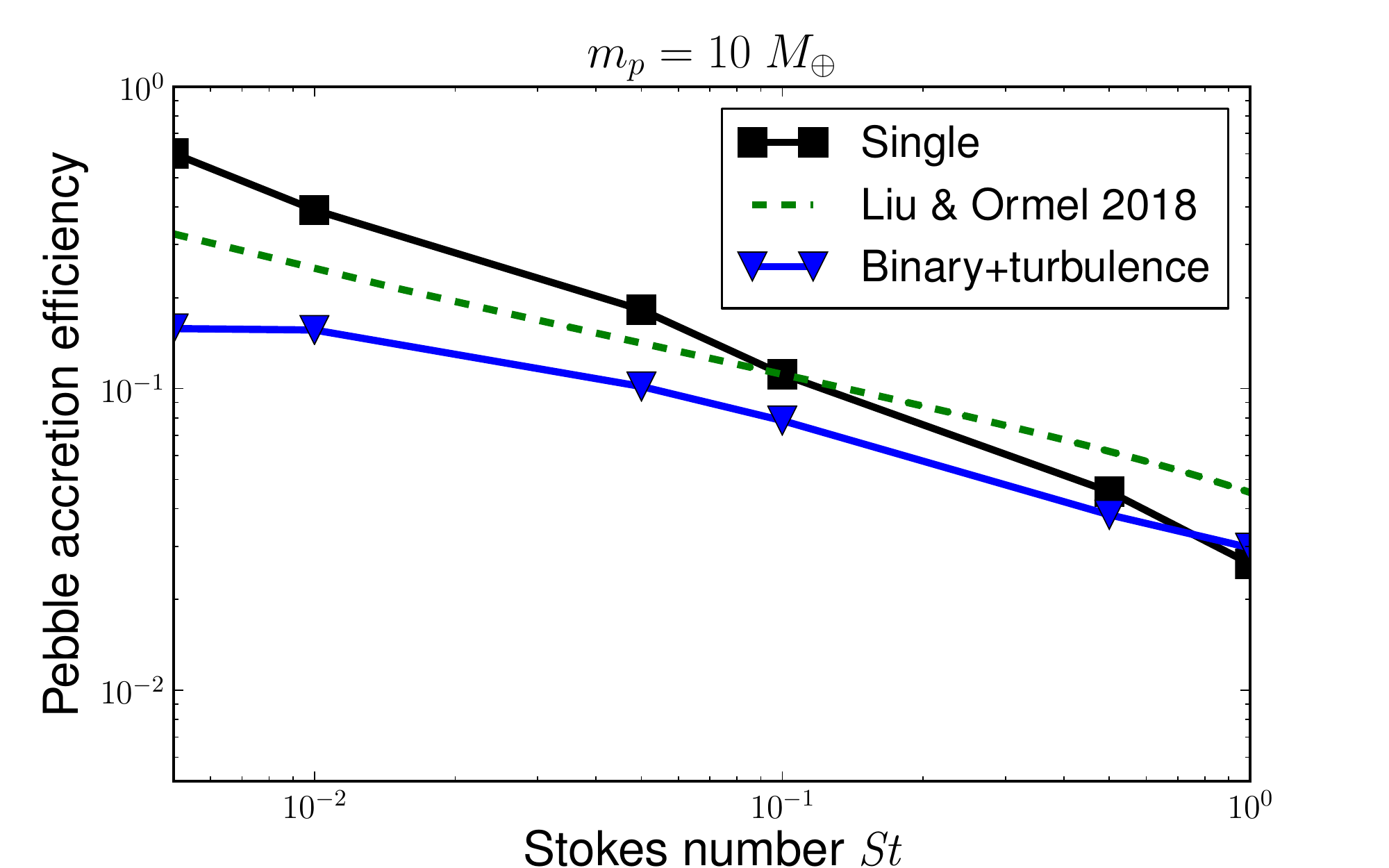}
\includegraphics[width=0.45\textwidth]{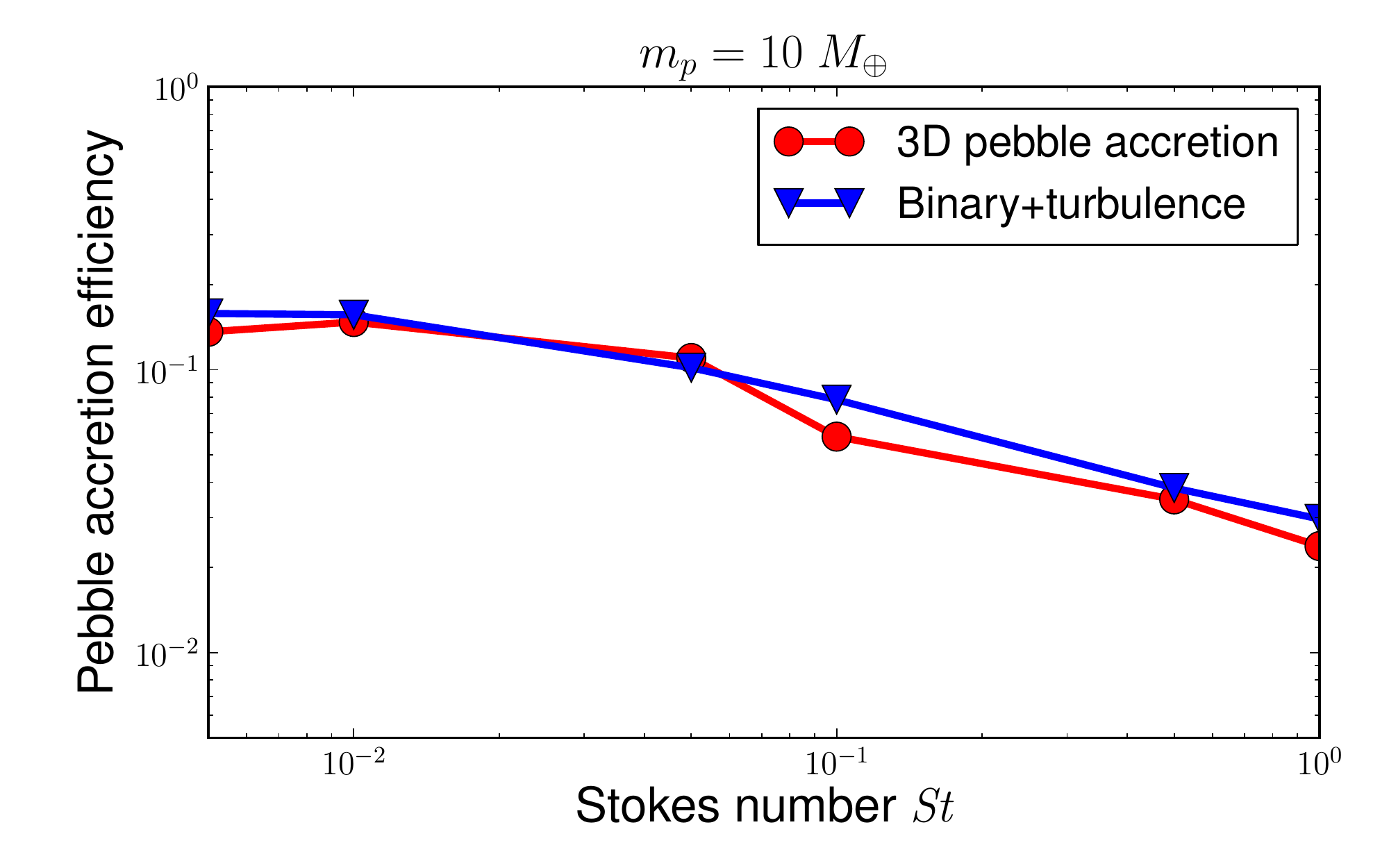}
\includegraphics[width=0.45\textwidth]{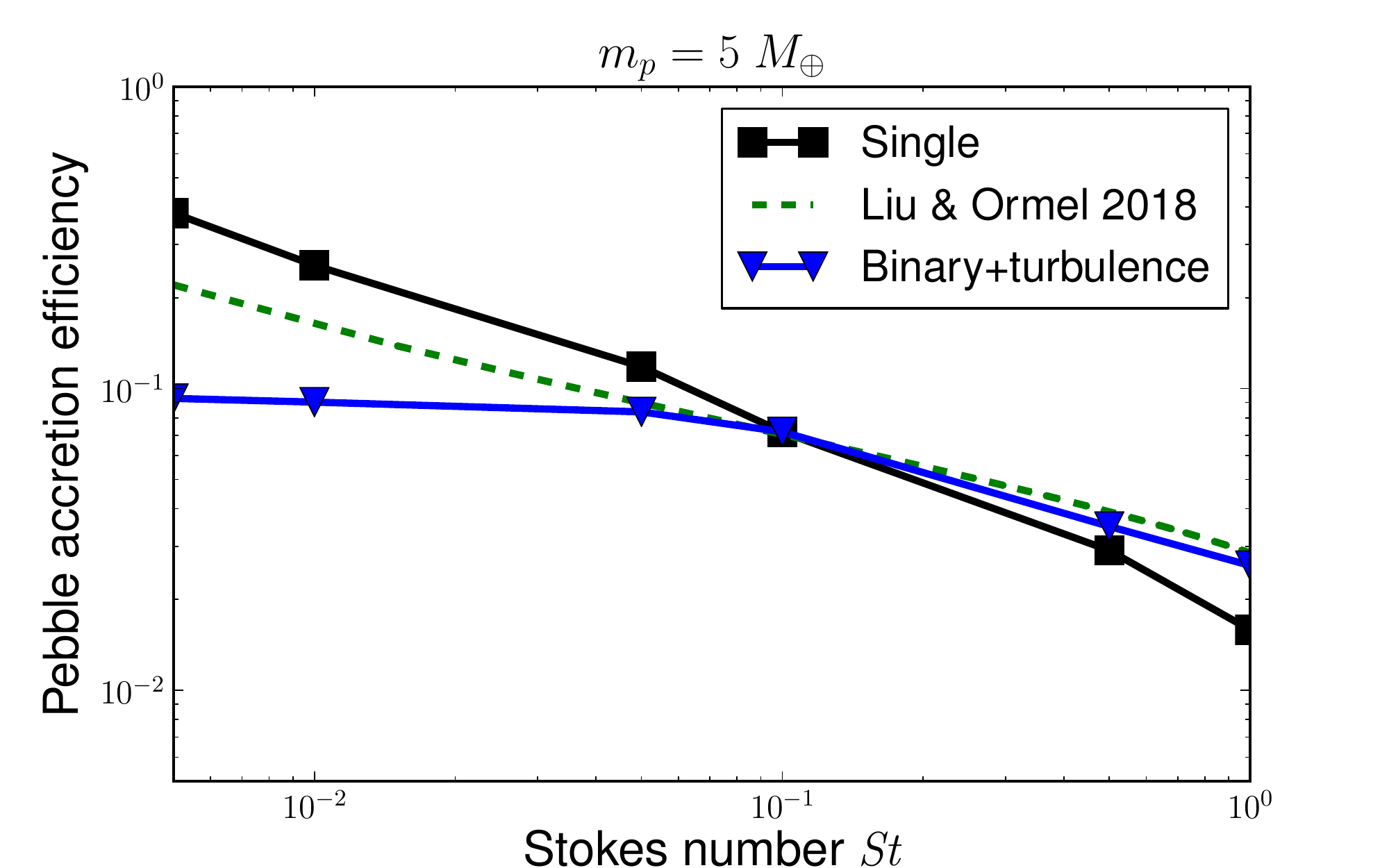}
\includegraphics[width=0.45\textwidth]{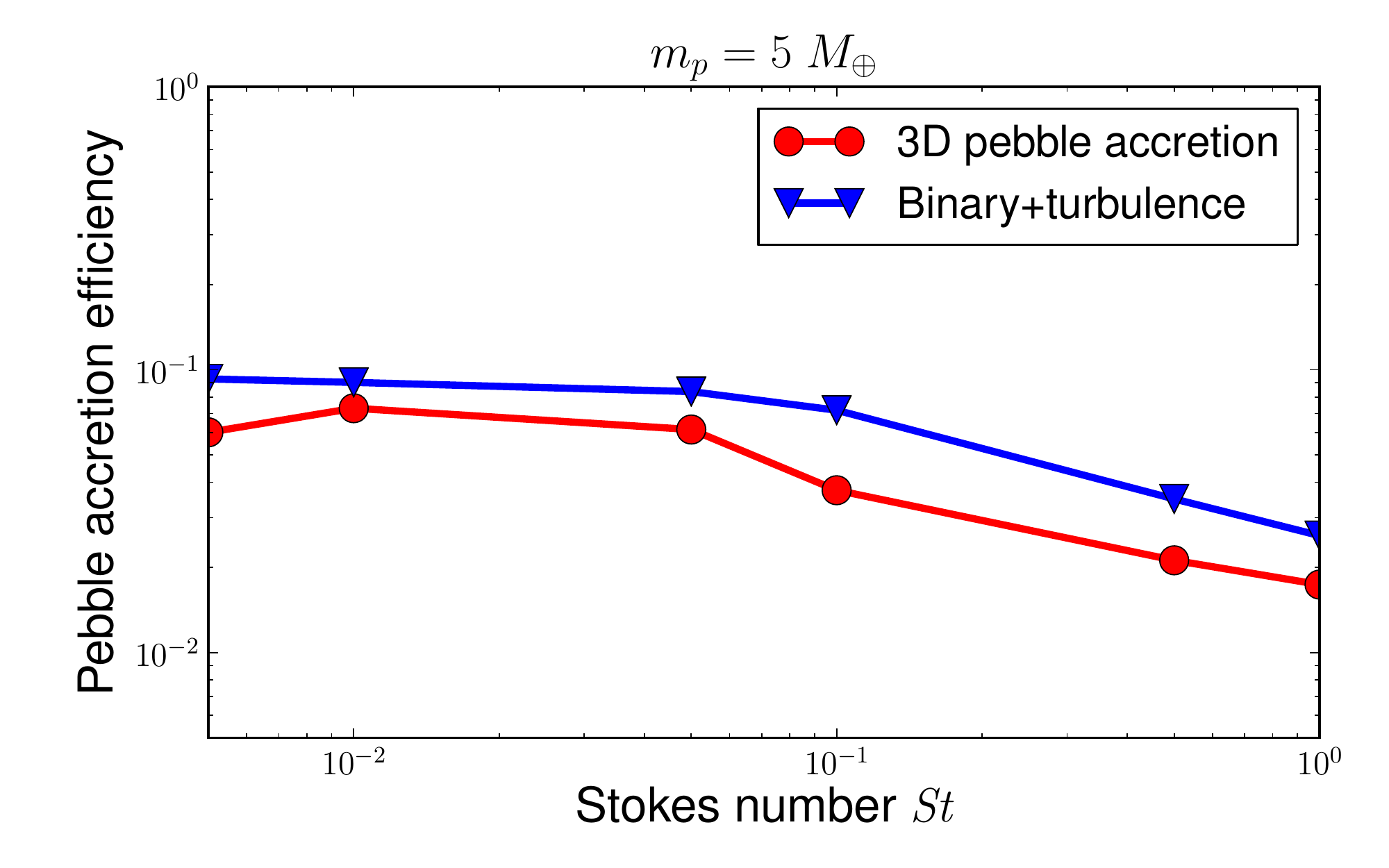}
\includegraphics[width=0.45\textwidth]{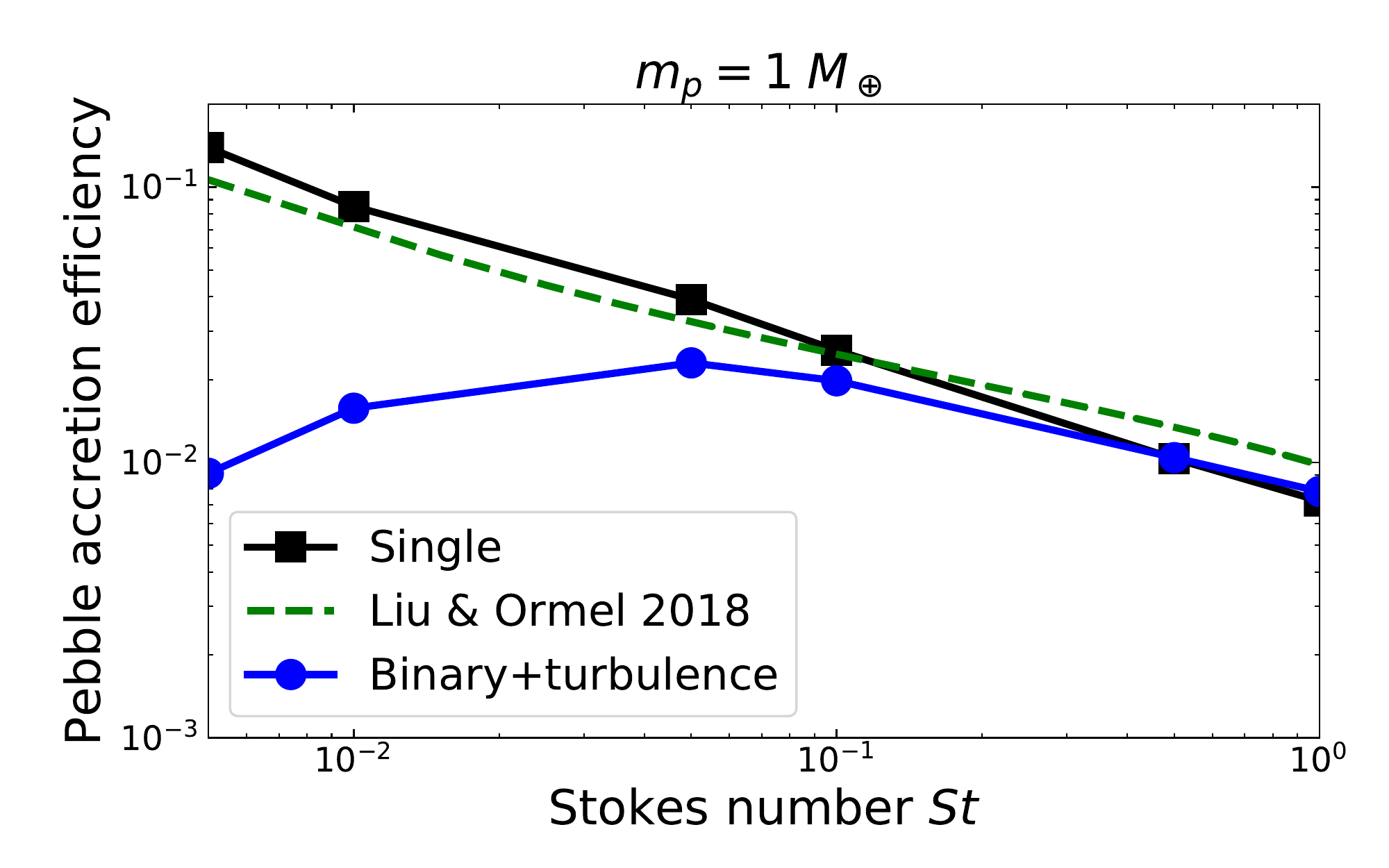}
\includegraphics[width=0.45\textwidth]{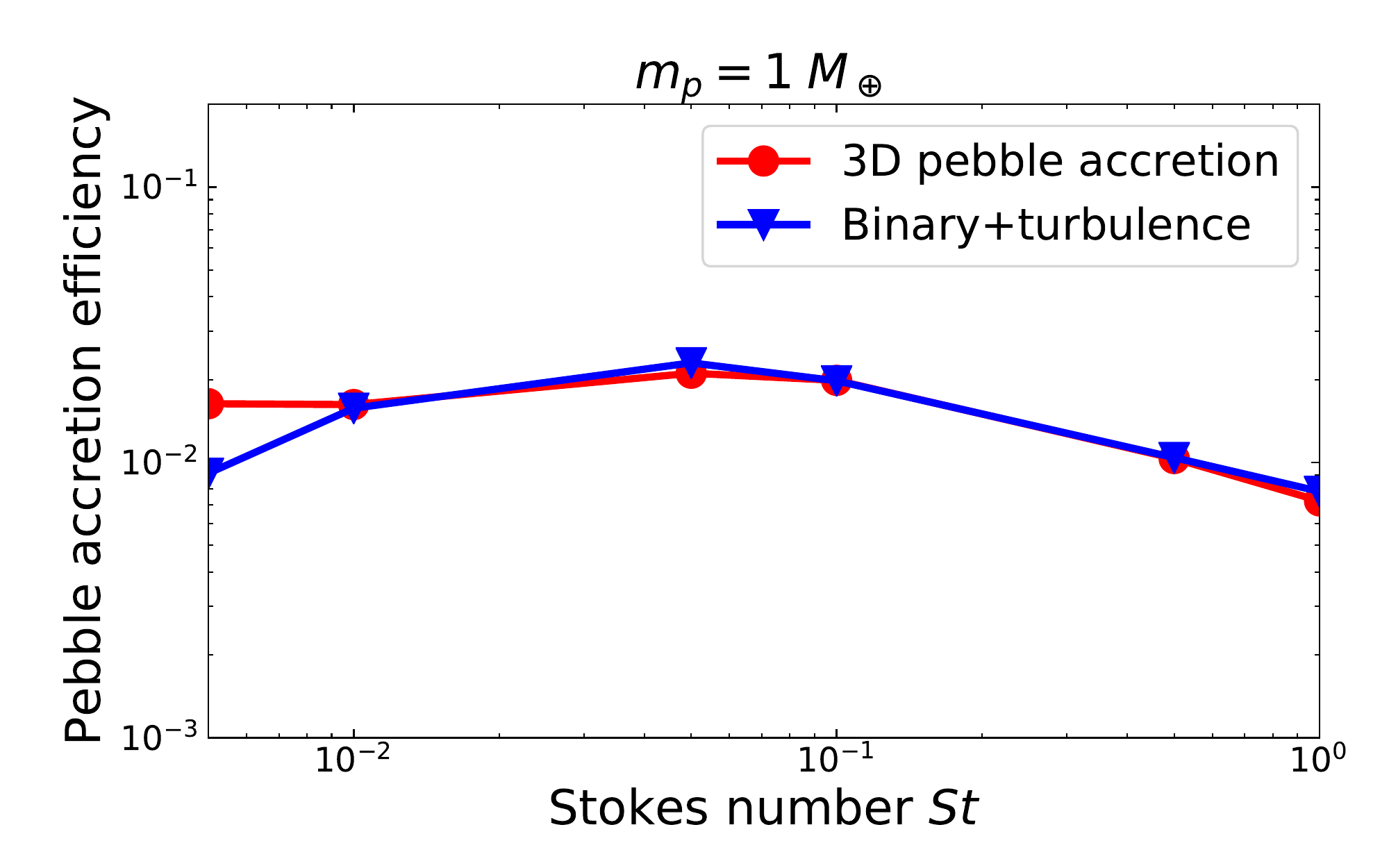}
\includegraphics[width=0.45\textwidth]{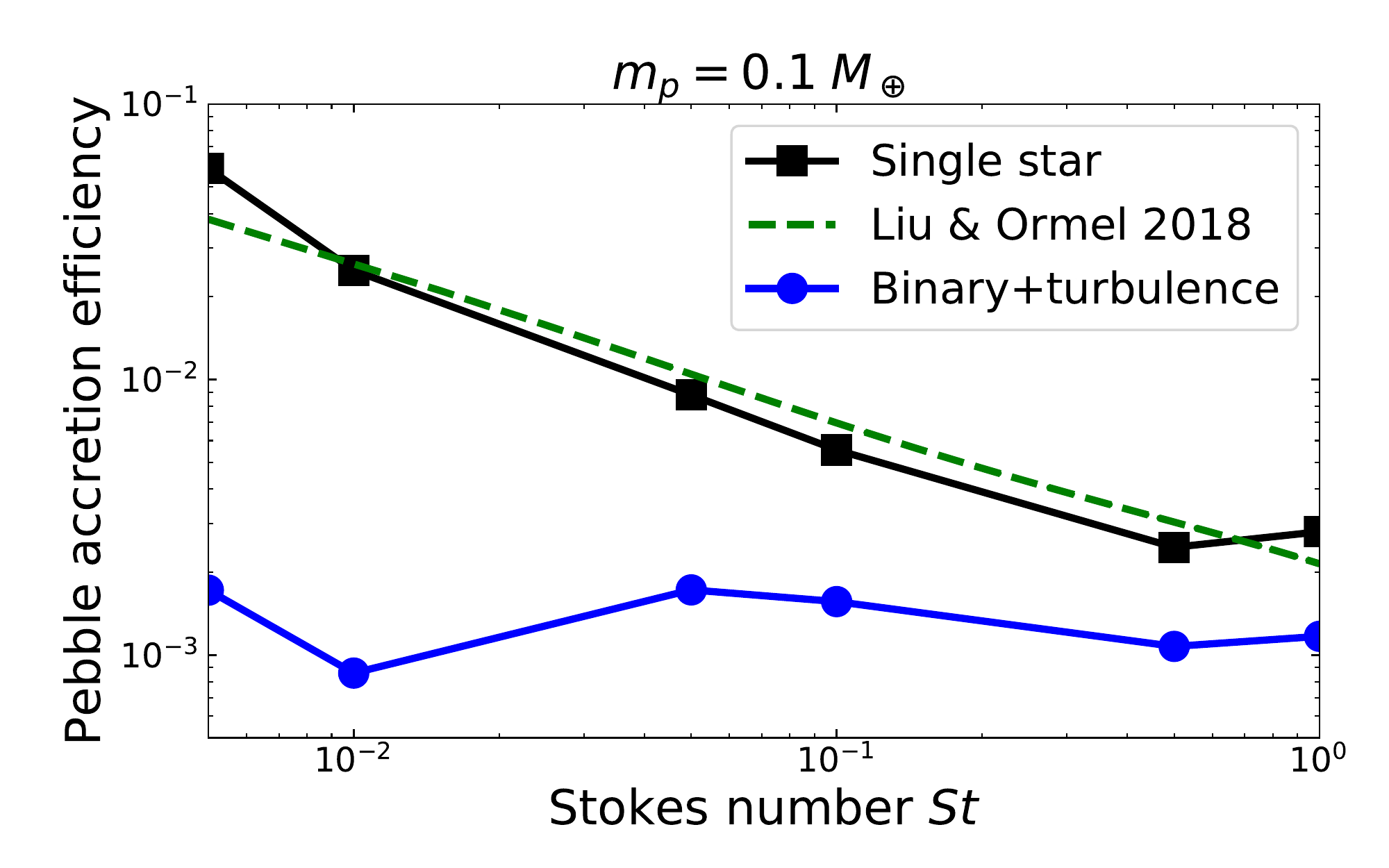}
\includegraphics[width=0.45\textwidth]{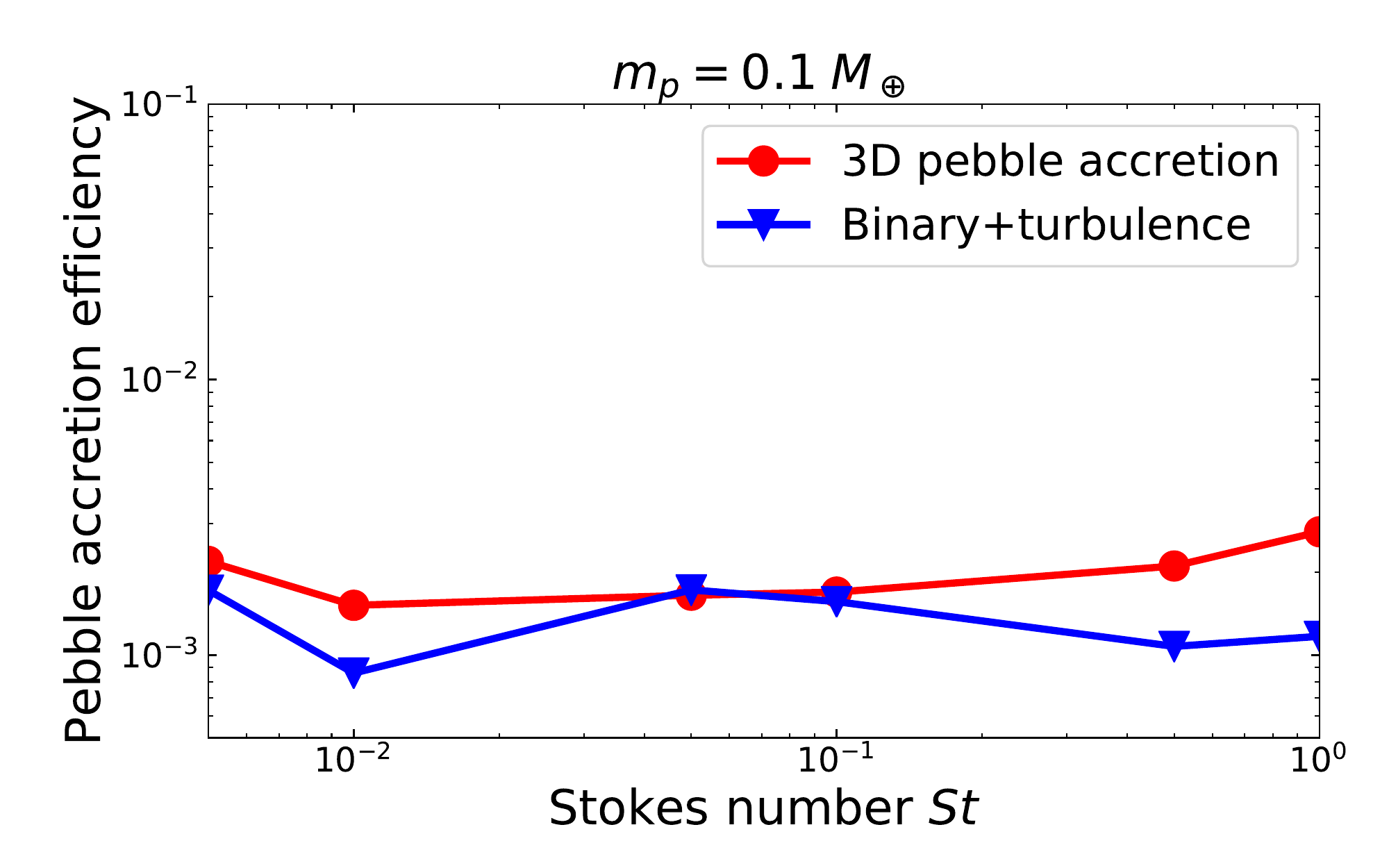}
\caption{{\it Left column:} Pebble accretion efficiency as a function of Stokes number for core masses $\msp=10,5,1,0.1\;M_\oplus$ in turbulent and non-turbulent runs. The dashed green line corresponds to the analytical expression for the pebble accretion efficiency of Liu \& Ormel (2018).  {\it Right column:} Pebble accretion efficiency for the turbulent  and non-turbulent simulations with 3D effects accounted for.}
\label{fig:efficiency}
\end{figure*}

\begin{figure}
\centering
\includegraphics[width=0.45\textwidth]{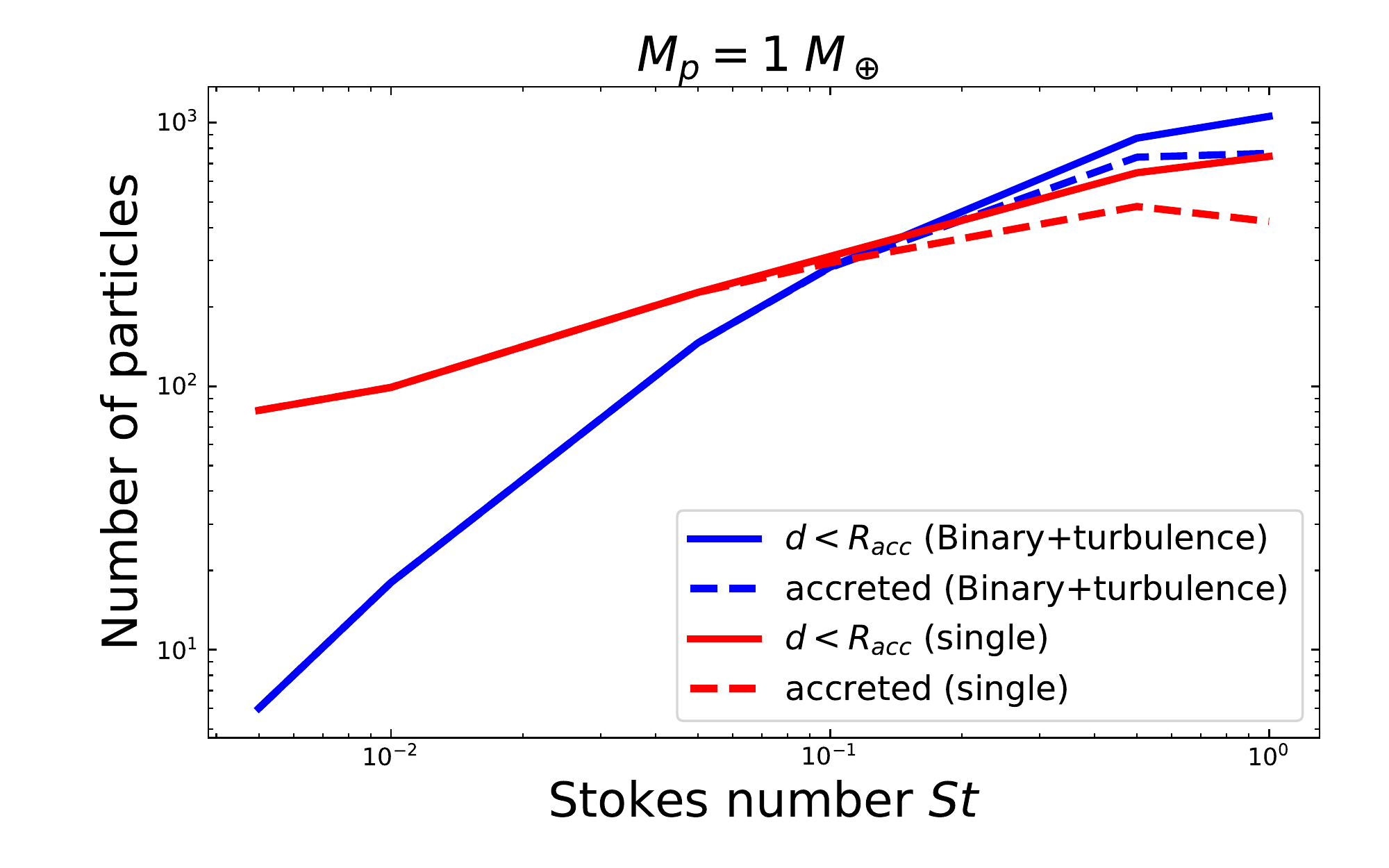}
\includegraphics[width=0.45\textwidth]{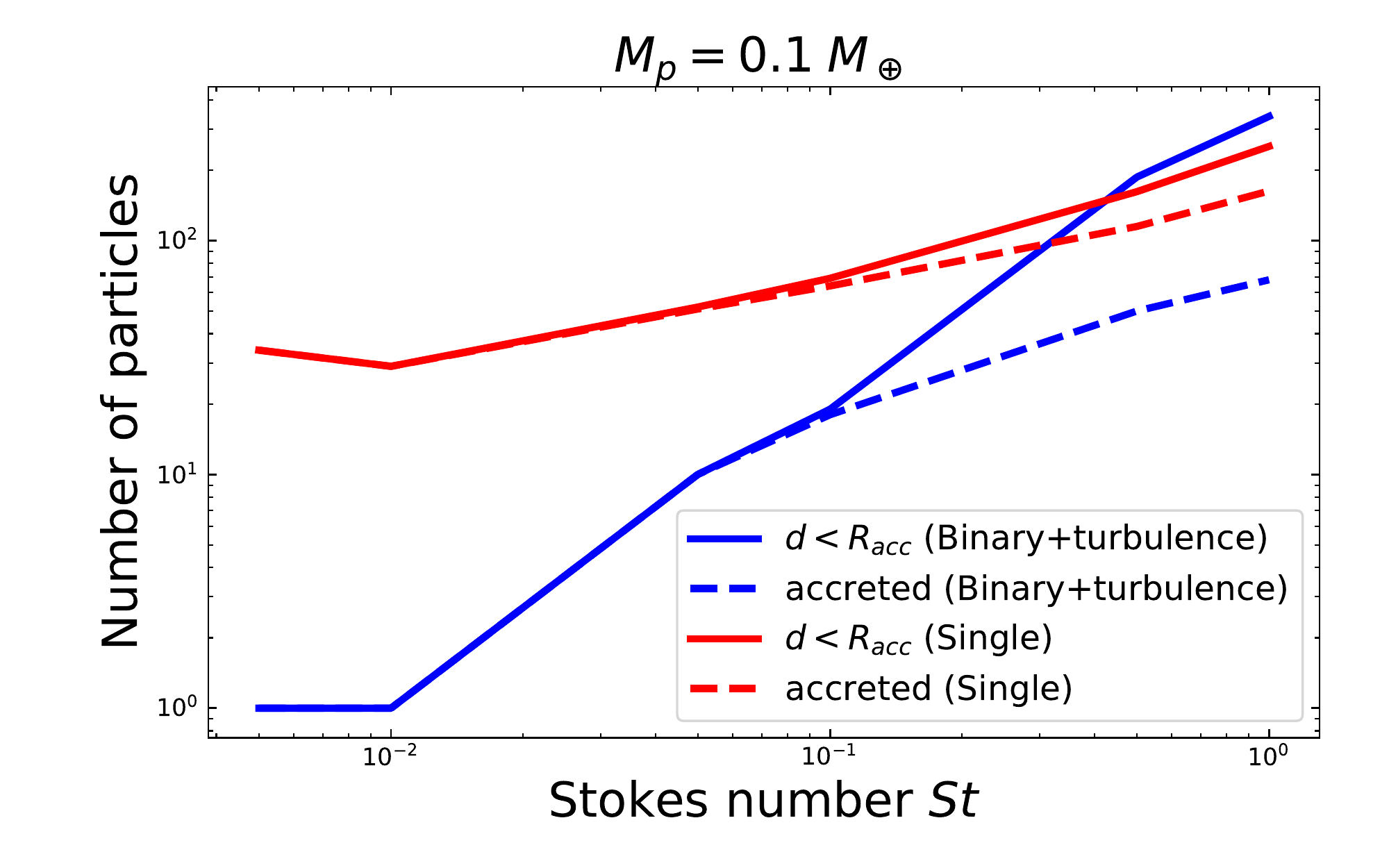}
\caption{{\it Top:} For $\msp=1\; M_\oplus$, number of particles that pass within $r_{\rm acc}$ and number of accreted particles as a function of Stokes number, in the turbulent and non-turbulent simulations.  {\it Bottom:} Same but for $\msp=0.1\; M_\oplus$.}
\label{fig:npart}
\end{figure}

The accretion rate, in a turbulent disc, $\dot M_{\rm 3D}$, is an inherently three dimensional process related to the two dimensional accretion rate in a laminar disc, $\dot M_{\rm 2D}$, by the following expression (Morbidelli et al. 2015):
\begin{equation}
\frac{\dot M_{\rm 3D}}{\dot M_{\rm 2D}}=\left(\frac{\pi r_{\rm acc}}{2\sqrt{2\pi} \Hd}\right).
\label{eq:mdot3d}
\end{equation}
In the right column of Fig.~\ref{fig:efficiency}, we compare the pebble accretion efficiency in the non-turbulent disc, but multiplied by a factor $\dot M_{\rm 3D}/\dot M_{\rm 2D}$, with the one obtained in the turbulent case. Decent agreement is found  when applying this procedure, which suggests that the smaller pebble accretion efficiency obtained in the presence of turbulence at least partly results from an increased pebble scale height. For core masses $\msp=0.1$ and $1 \; M_\oplus$, this is further demonstrated by Fig.~\ref{fig:npart} where we compare the number of particles that enter the accretion sphere with the number of accreted particles. For small pebbles, it is immediately evident that the number of particles that pass within a distance of $r_{\rm acc}$ from the protoplanet, $N_{d<r_{\rm acc}}$, is much smaller in the turbulent case. Comparing the $1$ $M_\oplus$ and $0.1$ $M_\oplus$ panels in the left column of Fig.~\ref{fig:efficiency} with Fig.~\ref{fig:npart} moreover reveals that such a difference in $N_{d<r_{\rm acc}}$ is of the order of the difference in the pebble accretion efficiency between the turbulent and non-turbulent runs. As mentioned above, this is of the order of $\sim r_{\rm acc}/\Hd$. Thus, the difference in $N_{d<r_{\rm acc}}$ between the turbulent and non-turbulent cases also scales as $\sim r_{\rm acc}/\Hd$, as expected.

\subsection{Impact of turbulent velocity kicks on pebble accretion}
As revealed by the lower panel in Fig.~\ref{fig:npart}, a result that emerges from the turbulent run with $\msp =0.1 \; M_\oplus$ is the significantly lower number of accreted particles with Stokes numbers $\st \gtrsim 0.1$ compared to the  number of pebbles that pass within the accretion sphere, which consequently gives rise to the lower pebble accretion efficiency at $\st \sim 1$ that is visible in the lower left panel of Fig.~\ref{fig:efficiency}. We interpret this as arising because the rms velocity arising from turbulent kicks, $\left< \delta \vp \right>$, is higher than the critical encounter velocity, $\vc$, below which pebble accretion is possible (Ormel \& Liu 2018). For pebble accretion to occur, the encounter time $t_{\rm enc}=r_{\rm acc}/v_{\rm enc}$, where $v_{\rm enc}$ is the encounter velocity, must be longer than $\sim 4 \ts$ (the factor of $4$ comes from our criterion ii) for pebble accretion described above), or equivalently the encounter velocity needs to be smaller than $\sim r_{\rm acc}/4\ts$. Using Eq.~\ref{eq:racc} for the accretion radius, this results in an estimate for the critical encounter velocity:
\begin{equation}
\vc \sim \left(\frac{\qp}{16\st}\right)^{1/3}\ap\Omega.
\end{equation}
Approximating the pebble rms velocity as $\left< \delta \vp\right> \simeq \Hd \Omega$, we estimate that for accreting cores with planet-to-binary mass ratios satisfying the condition
\begin{equation}
\qp\lesssim 16(\Hd/\ap)^3\st\sim 4\times 10^{-7}\st ^{0.4},
\end{equation}
stochastic kicks due to turbulence may prevent pebble accretion. A value of $\qp\sim 4\times 10^{-7}$ agrees with the significant decrease in the pebble accretion efficiency observed for $\st \sim 1$ in the run with $\msp=0.1 \, M_\oplus$ (for which $\qp=3\times 10^{-7}$). Moreover, the scaling $\qp\propto \st^{0.4}$ shows this effect is effective for pebbles that are only moderately coupled to the gas, which is also consistent with our results.

\section{Implication for the formation of circumbinary planets}
\begin{figure*}
\centering
\includegraphics[width=\textwidth]{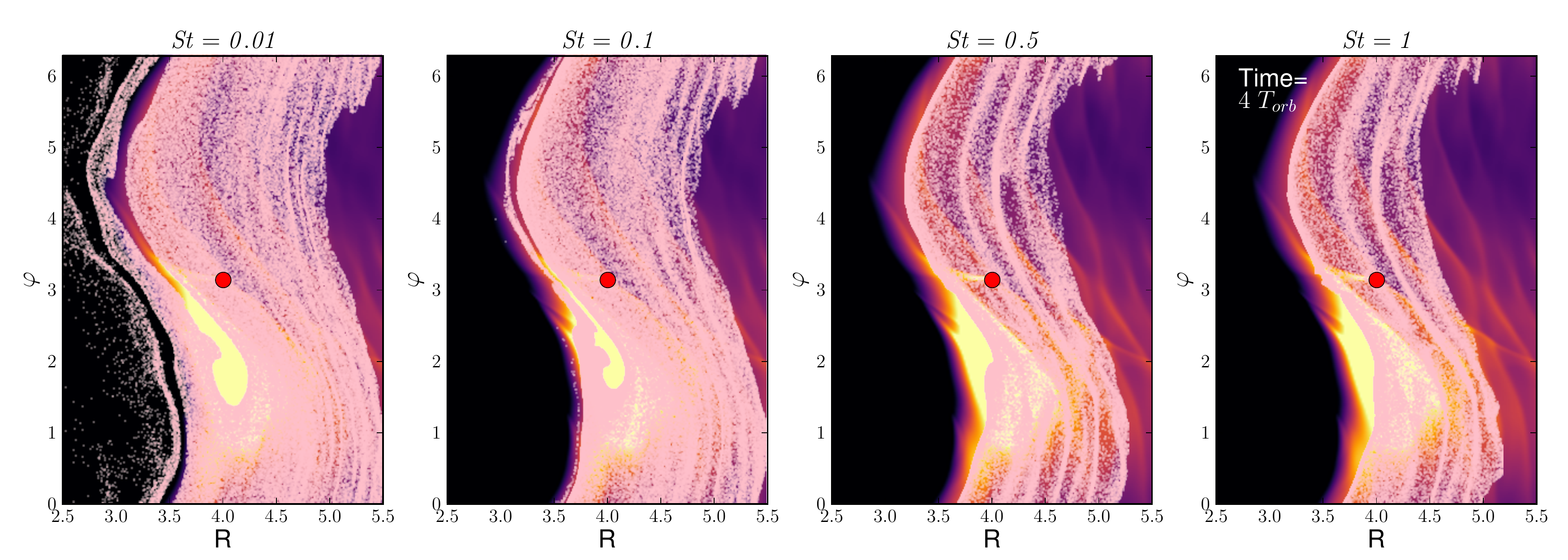}
\includegraphics[width=\textwidth]{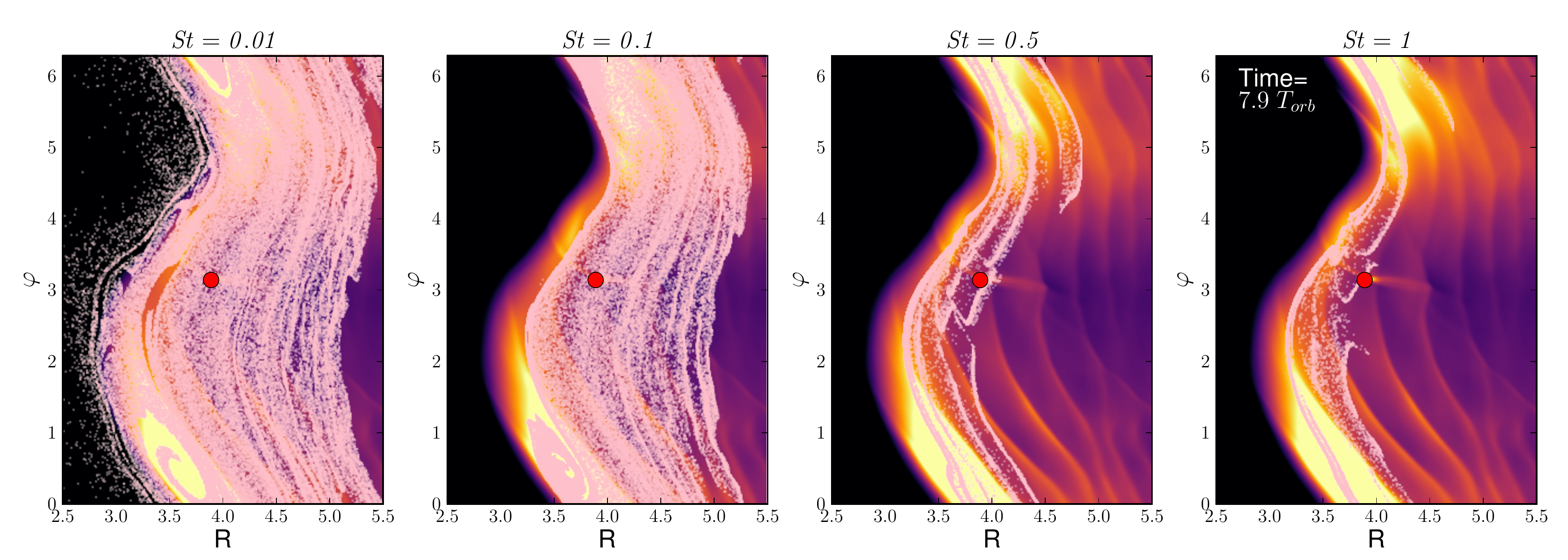}
\caption{Snapshots of the distribution of particles with different Stokes numbers at $t=4$ (top) and $t=7.9$~$T_{\rm orb}$ (bottom) for a pebble accretion simulation where an accreting 10 Earth mass planet is located just outside the cavity edge at $r=4$. Note the unit of time adopted in each panel is the orbital period at $r=6 \abin$.}
\label{fig:edge}
\end{figure*}

\begin{figure}
\centering
\includegraphics[width=\columnwidth]{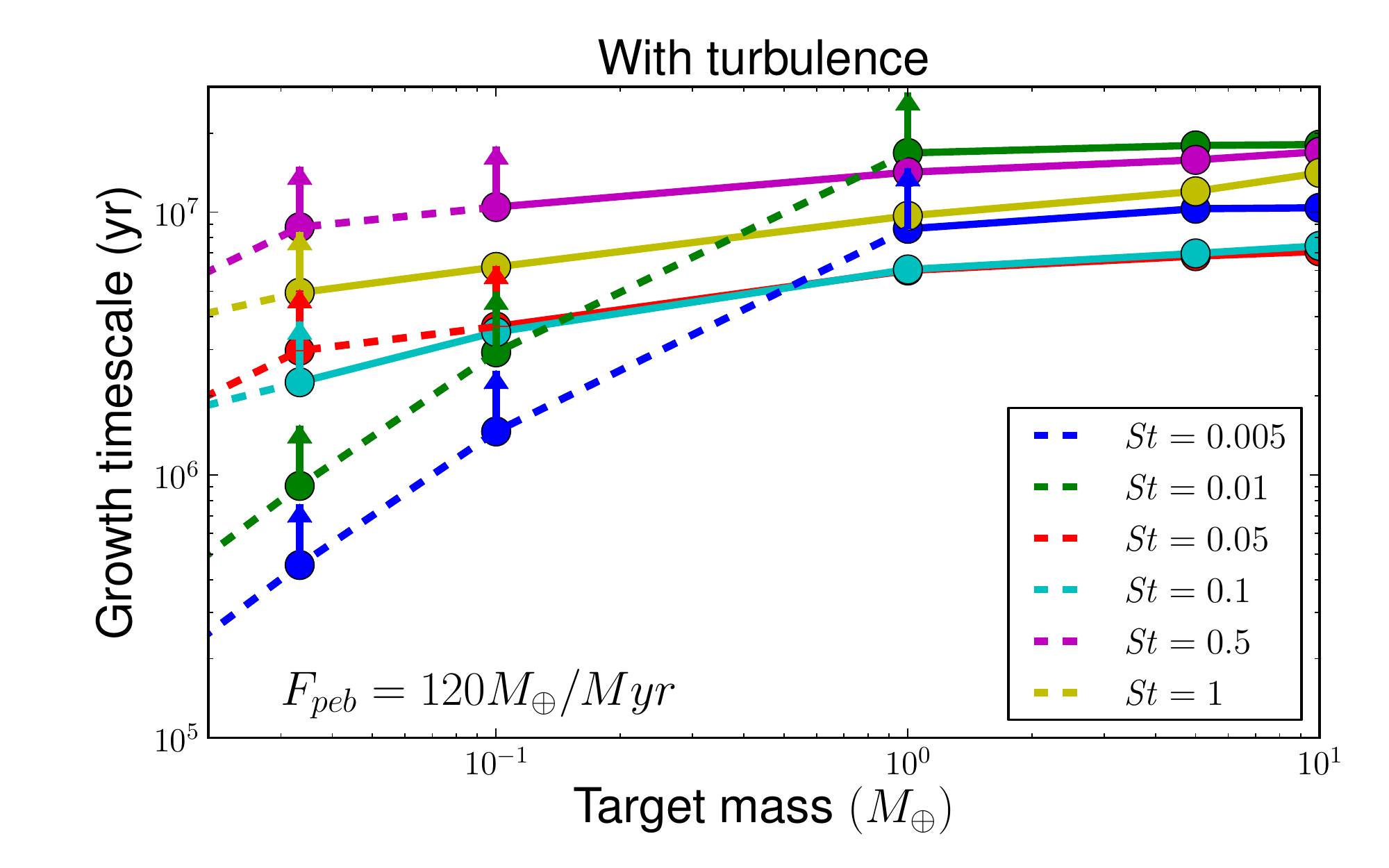}
\includegraphics[width=\columnwidth]{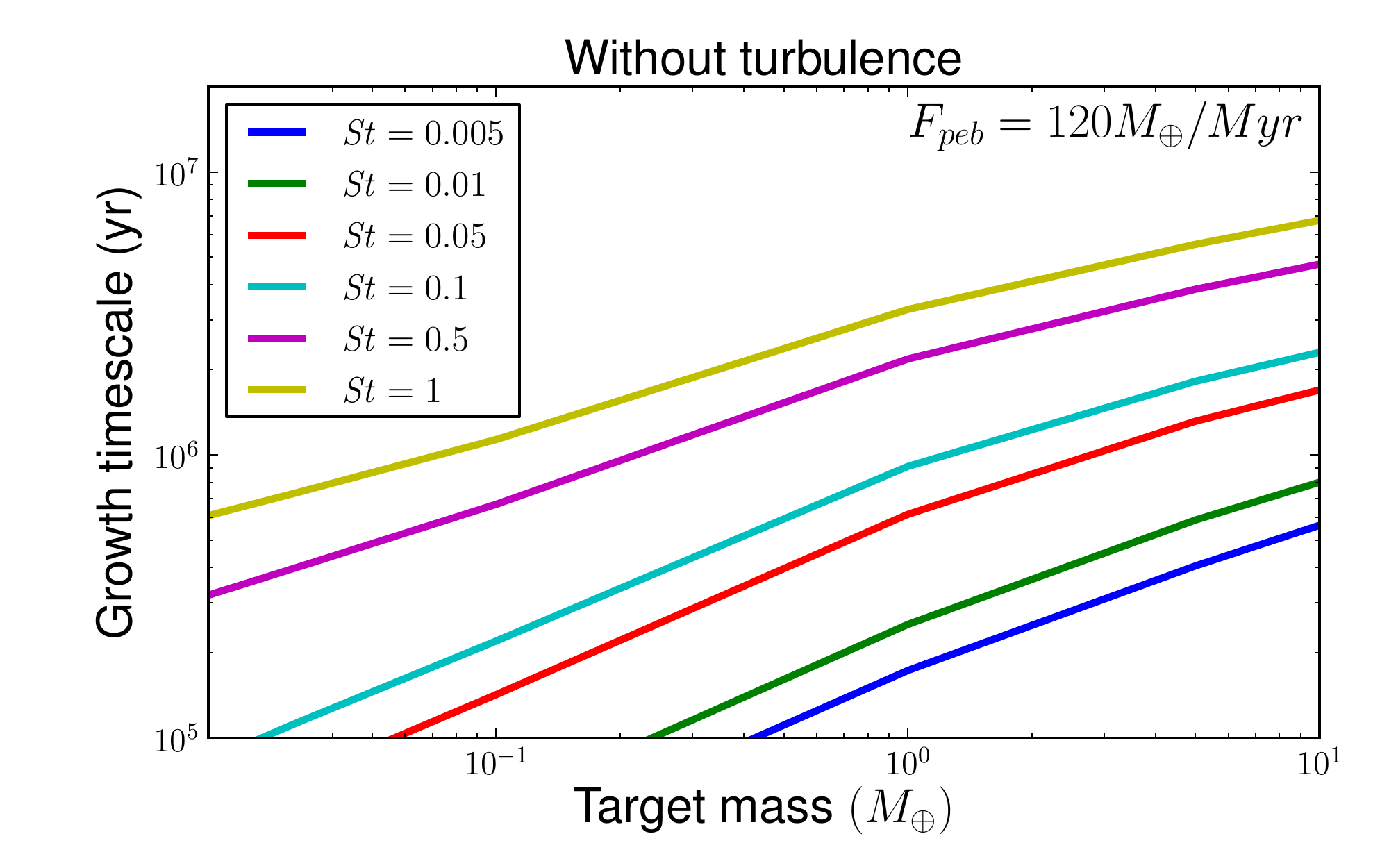}
\includegraphics[width=\columnwidth]{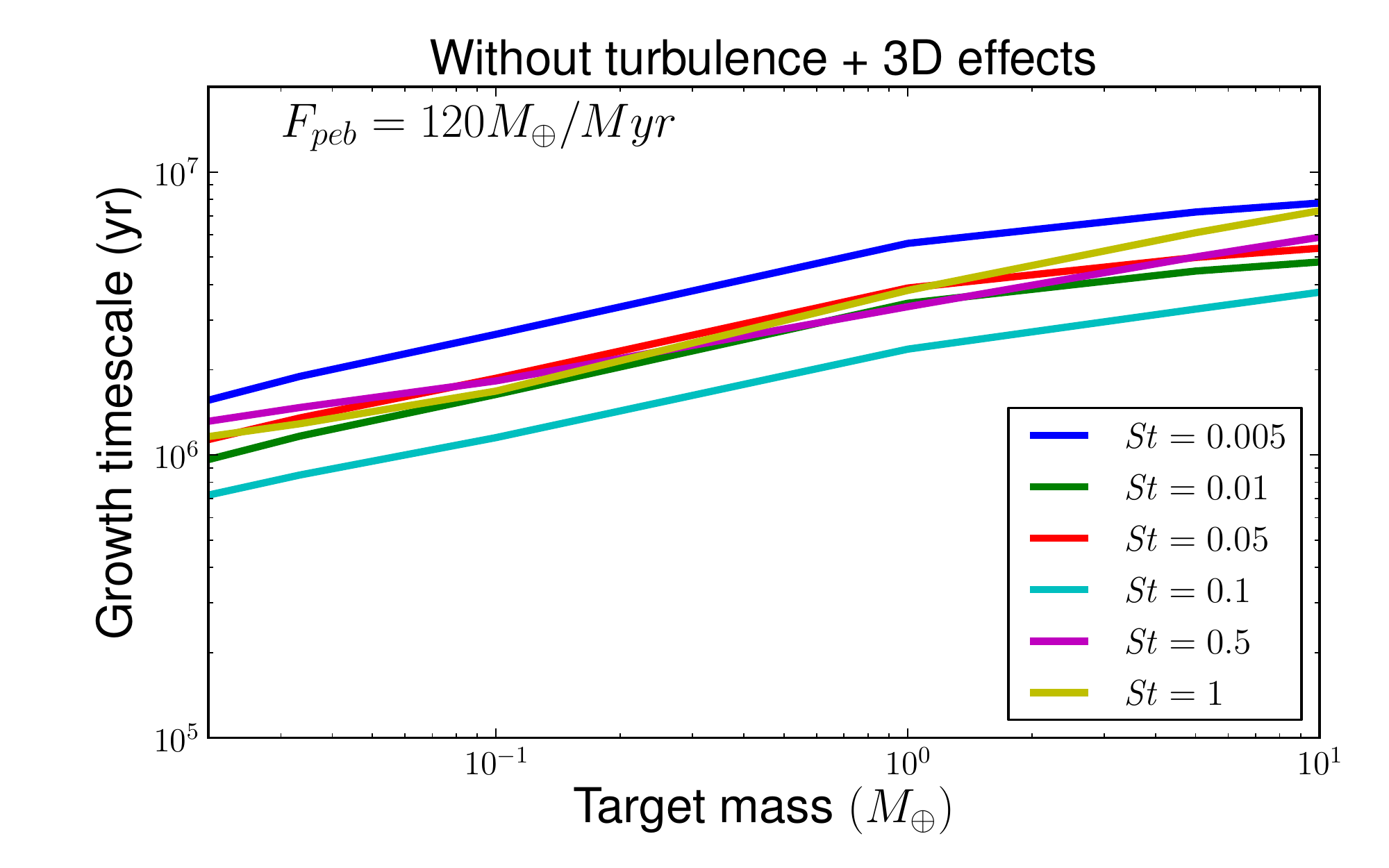}
\caption{{\it Top:} growth timescale as a function of target mass and Stokes number  in presence of turbulence. Accreting core masses that are considered in the turbulent runs are marked as filled dots. The mass located at the transition between the solid and dashed line corresponds to the one below no particle has been accreted during the course of the simulation. Below this mass critical mass, estimations of the growth timescale should be considered as lower bounds, and illustrated as the vertical arrows that are overplotted. {\it Middle:} same but  in a non-turbulent disc using the analytical expression of Liu \& Ormel (2018) for the pebble accretion efficiency. {\it Bottom:} same as middle panel but in the case where  the pebble accretion efficiency has been  scaled by a factor of $r_{\rm acc}/\Hd$. }
\label{fig:growthtime}
\end{figure}

As noted earlier, forming the observed circumbinary planets at their present locations through the accretion of km-size planetesimals is difficult because of the high collision velocities. The possibility of forming circumbinary planets in situ through a combination of streaming instability and pebble accretion has not yet been examined in detail, but the efficacy of these processes depends strongly on the level of turbulence in the disc. 

\subsection{Streaming instability}
Under favourable conditions, the settling and radial drift of dust grains and pebbles can lead to growth of the streaming instability and the formation of planetesimals (Youdin \& Goodman 2005, Johansen et al. 2009). In the following discussion, we denote the volume averaged dust-to-gas ratio in the circumbinary disc by $Z_0$, which we assume has the canonical value $Z_0=0.01$. We denote the vertically averaged dust-to-gas ratio at radius $R$ by $Z(R)$, which is defined by $Z=\Sigma_{\rm d}(R)/\Sigma(R)$, where $\Sigma_{\rm d}(R)$ and $\Sigma(R)$ are the dust and gas surface densities, respectively, and $Z=Z_0$ in the absence of the radial drift and concentration of dust. Fast growth of the streaming instability requires the \emph{local} dust to gas ratio to be close to unity or greater (i.e. $\epsilon=\rhod/\rho\gtrsim 1$). If $Z=Z_0$, then the condition for triggering the streaming instability translates to
\begin{equation}
Z \gtrsim \frac{\Hd}{H}.
\end{equation}
In laminar discs, non-linear simulations of the streaming instability indicate strong clumping of particles with $\st \in[10^{-2}, 1]$ occurs provided $Z\gtrsim 0.02$ (Carrera et al. 2015), whereas $Z\gtrsim 0.04$ is required for smaller particles with $\st=10^{-3}$ (Yang et al. 2017). 

Figure~\ref{fig:hd} shows the condition on $Z$ for non-linear clumping becomes quite severe in our turbulent disc models. $Z\gtrsim 0.1$ would be required for particles with $\st \in [10^{-3}, 0.1]$, while $Z\gtrsim 0.06$ would be required for solids with $\st=1$. Models and experiments of dust collisions and coagulation often invoke a bouncing barrier that limits growth to mm-cm sizes (Zsom et al. 2010), corresponding to $10^{-3} \le \st \le 10^{-1}$ in our models (see table~\ref{table2}). The turbulent stirring experienced by these particles means that $Z$ would need to be more than an order of magnitude larger than the canonical value $Z_0=0.01$ for planetesimals to form in situ via the streaming instability.  

The required value for $Z$ would be too high if it represented the volume average, however it could be achieved near the cavity edge where inward drifting solids might be trapped. This is illustrated in Fig. \ref{fig:edge}, which shows the distribution of particles near the cavity edge at two different times for a simulation containing a 10~$M_\oplus$ accreting core. There is a trend for the pebbles to become more radially concentrated at the cavity edge as the Stokes number is increased. This is consistent with the simulations of HD~142527 by Price et al. (2018) and of IRS~48 by Calcino et al. (2019), and arises in part because the finite run times of all these simulations are insufficient for the particles with smaller Stokes numbers to drift all the way to the cavity edge, although turbulent diffusion also plays a role in our simulations.

Although we are unable to run our 3D simulations for long enough to achieve a steady state, it is of interest to consider what this might look like. Let us assume gas accretes through the circumbinary disc at a steady rate and onto the central binary (Miranda et al. 2017), and consider the inwards drift of $\st \sim 0.01$ particles, corresponding to mm-cm sized pebbles (see table~\ref{table2}). As these pebbles concentrate towards the pressure maximum, the turbulence there will stir them, causing the particles to diffuse vertically and radially, and to collisionally evolve. Evidence for diffusion of $\st=0.01$ pebbles can be seen in Fig.~\ref{fig:edge}, where particles have clearly penetrated interior to the cavity edge, unlike those with larger Stokes numbers which do not diffuse as rapidly. Using equation~(10) from Ormel et al. (2008), and adopting $\alpha=3\times 10^{-3}$ for the turbulence generated by the parametric instability, an estimate of the mutual collision velocities between cm-sized particles gives $\sim 14$~m~s$^{-1}$, sufficient to erode the pebbles. As gas accretes onto the binary from the cavity edge, small tightly coupled particles resulting from pebble erosion will accrete with the gas, and solids will be lost from the region around the cavity (Zhu et al. 2012). The rate of collisions will increase with the local density of solids, providing a negative feedback on the  concentration level. The question of whether or not the streaming instability can occur locally then depends on whether or not a sufficient concentration of solids can build up despite the collisional grinding and loss of solids. A number of important factors will influence the outcome of this process, including how the gas accretes through the disc, and whether or not the back reaction of the pebbles on the gas modifies the turbulence as the concentration of solids increases. Exploring this scenario in a realistic manner goes beyond the scope of this paper, but needs to be examined in future work.

Figure~\ref{fig:edge} shows that two different modes of pebble concentration are found. Pebbles with $0.01\lesssim\st\lesssim 0.1$ are primarily concentrated in radius and azimuth at the location of the density maximum located at the disc apocenter. The density maximum there is not a vortex, but is a region where the fluid moves more slowly on its orbit. We expect particles aligned with the eccentric cavity to experience a traffic-jam effect, increasing their density at apocente, and the tendency to concentrate may be enhanced by drag forces operating in the higher density gas at apocenter. Particles with $\st \gtrsim 0.5$ collect in the spiral density waves launched by the binary, an effect that has been observed previously in simulations of gravitationally unstable discs (e.g. Dipierro et al. 2015). Particle over densities created through this process can collapse under self gravity to form protoplanets directly in the outer regions of protoplanetary discs (Gibbons et al.  2014). 

Concentrations of particles at the cavity edge or in the spiral arms could in principle collapse directly to form bound objects, provided the local particle density exceeds the Roche density given by
\begin{equation}
\rho_{\rm R} = \frac{9 M_\star}{4 \pi a_{\rm p}^3},
\label{eqn:rho_R}
\end{equation}
$a_{\rm p}$ is the semi-major axis of a particle clump. For $a_{\rm p} \sim 4.5 \abin$ (corresponding to $a_{\rm p} \sim 1$~au), $\rho_{\rm R} \simeq 4 \times 10^{-7}$~g~cm$^{-3}$. Our disc model has a midplane gas density of $\rho=2 \times 10^{-9}$~g~cm$^{-3}$ at 1~au, such that $\rho_{\rm R}/\rho=200$. This is a very high level of grain concentration, such that prior to this value being achieved we would expect the streaming instability to have already converted the dust to planetesimals during earlier evolution. Hence, it seems direct gravitational collapse of particle clumps caused by concentration in spiral waves, or at the cavity edge, is not a realistic means of forming circumbinary planets.


\subsection{Pebble accretion}
Planetesimals with sizes of a few hundred kilometres may be formed as a result of the streaming instability in a laminar disc (Johansen et al. 2015; Simon et al. 2016; Shaffer et al. 2017). The possibility of forming the circumbinary planet Kepler-16b, through the accretion of such large planetesimals, has been examined by Lines et al (2016). They found the eccentric circumbinary gas disc inhibits accretion inside $\sim 1.5$~au ($\sim 7\;\abin$) by forcing the planetesimal eccentricities, causing mutual collisions between planetesimals to be destructive. 

Pebble accretion onto planetesimals in this size range proceeds in the Bondi regime and is not expected to cause significant core growth. Turbulence arising because of the disc eccentricity may render the situation even worse. This is illustrated in Fig.~\ref{fig:growthtime} which displays, as a function of target mass $m_{\rm target}$ and Stokes number, the expected growth timescale, $t_{\rm growth}$, defined by:
\begin{equation}
t_{\rm growth}=\int_{m_{\rm Ceres}}^{m_{\rm target}}\frac{{\rm d} m}{\dot M},
\end{equation}
where the initial mass of the accreting body is assumed to be the mass of Ceres, $m_{\rm Ceres}=4\times 10^{-3}\;M_\oplus$.  From top to bottom, the panels show the growth timescales in the presence of turbulence, in a laminar disc where we employed the analytical expression of Liu \& Ormel (2018) for the pebble accretion efficiency, and in a laminar disc where the pebble accretion efficiency is scaled by a factor of $r_{\rm acc}/\Hd$.  We adopt a pebble flux of $F_{\rm peb}=120 M_\oplus/{\rm Myr}$ from Lambrechts \& Johansen (2014). 

To compute the growth timescale in the turbulent case (top panel), we performed some additional pebble accretion simulations with $\qp=10^{-7}$, $5\times 10^{-8}$ and $10^{-8}$.  In some simulations, these low mass planets fail to accrete  a single pebble over their run time ($\sim 10$ planetary orbits), particularly for the small pebble sizes. For a given value of $\st$, the critical mass below which no accretion event has been detected by the end of the simulations is shown in the upper panel of Fig. \ref{fig:growthtime} at the transition between the solid and dashed lines. Below this critical value, we assume that one particle has been accreted just after the end of the simulation, such that the growth timescale shown in the upper panel of Fig. \ref{fig:growthtime} is only a lower bound (represented as an overplotted vertical arrow).

The top panel of Fig.~\ref{fig:growthtime} shows between 6 and 20 Myr is required to form a $10$ $M_\oplus$ from the accretion of pebbles with Stokes numbers in the range $0.005 \le \st \le 1$ in a turbulent disc. These values should be contrasted with the 0.5 to 7 Myr growth timescales shown in the middle panel for a laminar disc. Hence, we see the turbulence generated by the binary significantly increases formation times, such that they exceed typical disc life times of 3 Myr (Haisch et al. 2001). The bottom panel of Fig.~\ref{fig:growthtime} shows the growth times obtained when the pebble accretion efficiency in the laminar disc is scaled by a factor of $r_{\rm acc}/\Hd$. The time required to form a $10 \, M_\oplus$ planet when accounting for 3D effects using this simple approach is between 4 and 8 Myr, again longer than typical disc life times.

In the context of the formation of the Solar System, it has been proposed that the dichotomy in mass between the terrestrial planets and the gas giants results from the fact that only mm-size chondrules/pebbles can cross the snowline and feed the inner disc (Morbidelli et al. 2015). In this region, the  bouncing barrier may prevent the growth of silicate chondrule-size solids to cm-sizes (Zsom et al. 2010). Assuming this can be transposed to the circumbinary case, such that the Stokes numbers of the pebbles entering the region close to the central cavity are $\st < 0.01$, we have demonstrated that turbulence may prevent efficient pebble accretion at the locations of known circumbinary planets. Hence, it would be difficult for an accreting body to reach the pebble isolation mass, which is $\sim 20$ $M_\oplus$ in a low viscosity disc with aspect ratio $h=0.05$ (Bitsch et al. 2018). Forming Kepler-16b in situ in a scenario that combines streaming instability plus pebble accretion appears to be very challenging unless very large pebble fluxes are invoked. 

In the outer regions of the disc, where the level of turbulence is much lower, the middle panel of Fig.~\ref{fig:growthtime} shows that forming a $\sim 10$ $M_\oplus$ planet through pebble accretion would be possible in less than 1 Myr, giving the resulting circumbinary planet time to migrate in to the cavity edge. Pierens \& Nelson (2013) have shown that a $20$ $M_\oplus$ planet that forms far from the cavity, and which migrates and accretes gas in a disc undergoing photoevaporation, can successfully reproduce the mass and semi-major axis of Kepler-16b. Hence, we conclude that this latter formation scenario is much more favourable for explaining the Kepler circumbinary planets than in situ formation.

\section{Conclusion}
In this paper, we have presented the results of three-dimensional global hydrodynamical simulations of circumbinary discs with binary parameters corresponding to the Kepler-16 system. We found that the significant disc eccentricity resulting from interaction with the central binary can trigger a parametric instability associated with the non-circular streamlines. The instability is essentially the same as that found by Papaloizou (2005a) and Barker \& Ogilvie (2014) in the case of Keplerian discs around single stars, and  involves the excitation of inertial-gravity waves that resonantly interact with the eccentric mode in the disc. Non-linear evolution of the instability generates turbulence, which transports angular momentum outwards with an effective viscous stress parameter $\alpha \sim 5\times 10^{-3}$, and with vertical velocity fluctuations that are a few tens of percent of the sound speed. Given that the auto-correlation timescale of the vertical velocity fluctuations is  $\taucor \sim 0.1 \Omega^{-1}$, this results in a Schmidt number of ${\it Sc}_z\sim 3$, where this is defined as the ratio between the turbulent viscosity coefficient that drives radial angular momentum transport and the vertical diffusion coefficient.

By following the evolution of Lagrangian particles that are characterized by their Stokes numbers $\st$, we examined the impact of turbulence on dust vertical settling. The particle vertical profile is found to reach a quasi-stationary state once turbulent diffusion counterbalances gravitational settling, showing a Gaussian distribution with scale height, $\Hd$, that is a few tenths of the gas pressure scale height, and which varies as $\Hd \propto \st^{-0.2}$ (see Fig. \ref{fig:hd}). This is consistent with previous findings for small particles that settle to the disc midplane in the presence of  MHD turbulence (Fromang \& Nelson 2006). The deviation from the classical prediction of $\Hd\propto \st^{-0.5}$ is likely a consequence of the presence of coherent vertical flows, rather than vertical velocity fluctuations that vary significantly with height as occur in MHD turbulence.

We also presented the results of pebble accretion simulations in which we measured the pebble accretion efficiency onto cores with masses $\msp\in[0.1,10]$~$M_{\oplus}$ located at orbital radii $R \le 6 \; \abin$, where $\abin$ is the binary semi-major axis. We find the impact of turbulence on pebble accretion is twofold:
\begin{enumerate}
\item Turbulence reduces the efficiency of pebble accretion by a factor of $\racc/\Hd$ compared to a laminar disc, which renders as highly inefficient the accretion of small pebbles that are subject to strong vertical stirring, particularly for low mass cores. This is a consequence of the geometrical aspect of pebble accretion, which is an inherently 3D process when the pebble layer has a finite thickness.
\item For core masses $\msp\lesssim 0.1$ $M_\oplus$, the velocity kicks induced by the turbulence can also lead to a decrease in the pebble accretion efficiency. For particles that enter the Hill sphere of the core, this occurs when the encounter time becomes of the order of the stopping timescale, such that particles which are moderately coupled to the gas are more sensitive to this effect. 
\end{enumerate}
 
Growing a Ceres-mass object to a $10$ $M_\oplus$ planet takes about one order of magnitude longer in a turbulent disc compared to a laminar one. For typical values of the pebble mass flux through the disc, this results in a growth timescale of $\gtrsim 8$~Myr, longer than typical disc life times. The main implication is that forming Kepler-16b close to the cavity edge through a scenario that invokes the streaming instability to build a seed object, followed by pebble accretion to grow that seed into a planetary core that accretes gas, seems to be very difficult.  In this context, a more appealing scenario would be to form the planet further from the binary where the streaming instability and pebble accretion are more efficient processes, due to a lower level of turbulence, followed by migration to the edge of the tidally truncated cavity formed by the binary. Previous studies have shown this latter scenario can satisfactorily explain the presence of circumbinary planets on orbits similar to the observed values. 

There are a number of caveats that must be addressed in future work. The excitation of inertial-gravity waves through the parametric instability is likely to be sensitive to the thermal evolution of the disc, and hence the turbulent flow may also change under a more sophisticated treatment of the disc thermodynamics.  This would in turn obviously affect both the dust settling and pebble accretion processes. We note that the solid accretion efficiency in radiative discs has been recently  examined by Zompas et al. (2020). These authors found that at distances $R\sim 5$ au from the central star, efficient radiative cooling leads to a small aspect ratio for the disc $h\sim 0.03$. As it scales as $\propto (h/0.05)
^3$ (Bitsch et al. 2018), the pebble isolation mass  may therefore be much smaller than $20$ $M_\oplus$ (Lambrechts et al. 2014). In the context of radiative circumbinary discs where the disc aspect ratio can be as small as $h\sim 0.02$ in the outer disc (Kley et al. 2019), the pebble isolation mass may even be smaller than $5$ $M_\oplus$ (Zormpas et al. 2020). Although further growth would be prevented at large distances from the binary, it cannot be excluded that pebble accretion is restarted as the planet reaches the cavity edge where the disc aspect ratio increases again. Growing circumbinary planets through this scenario seems to be plausible, as the decrease in pebble accretion efficiency due to turbulence is only modest for planets with mass $\lesssim 5$ $M_\oplus$ (see second row in Fig. \ref{fig:efficiency}).   The back reaction of pebbles on the gas may also be important, and has not been included in this study. Lin (2019) has shown dust settles more efficiently when the back reaction is included in a disc where turbulence originates because of the VSI, and a similar effect may occur in circumbinary discs, particularly if particles concentrate near the cavity edge. In this study, the pebble-accreting planets were kept on fixed circular orbits, whereas we would expect them to become eccentric through interaction with the central binary and the eccentric circumbinary disc. Inclusion of this effect is likely to change the pebble rates significantly.  These and other improvements to the models will be included in future studies.

\section*{Acknowledgments}
Computer time for this study was provided by the computing facilities MCIA (M\'esocentre de Calcul Intensif Aquitain) of the Universite de Bordeaux and by HPC resources of Cines under the allocation A0070406957 made by GENCI (Grand Equipement National de Calcul Intensif).  CPM and RPN acknowledge support from STFC through grants ST/P000592/1 and ST/T000341/1.

\appendix

\section{Particle settling in the absence of turbulence}

\begin{figure}
\centering
\includegraphics[width=\columnwidth]{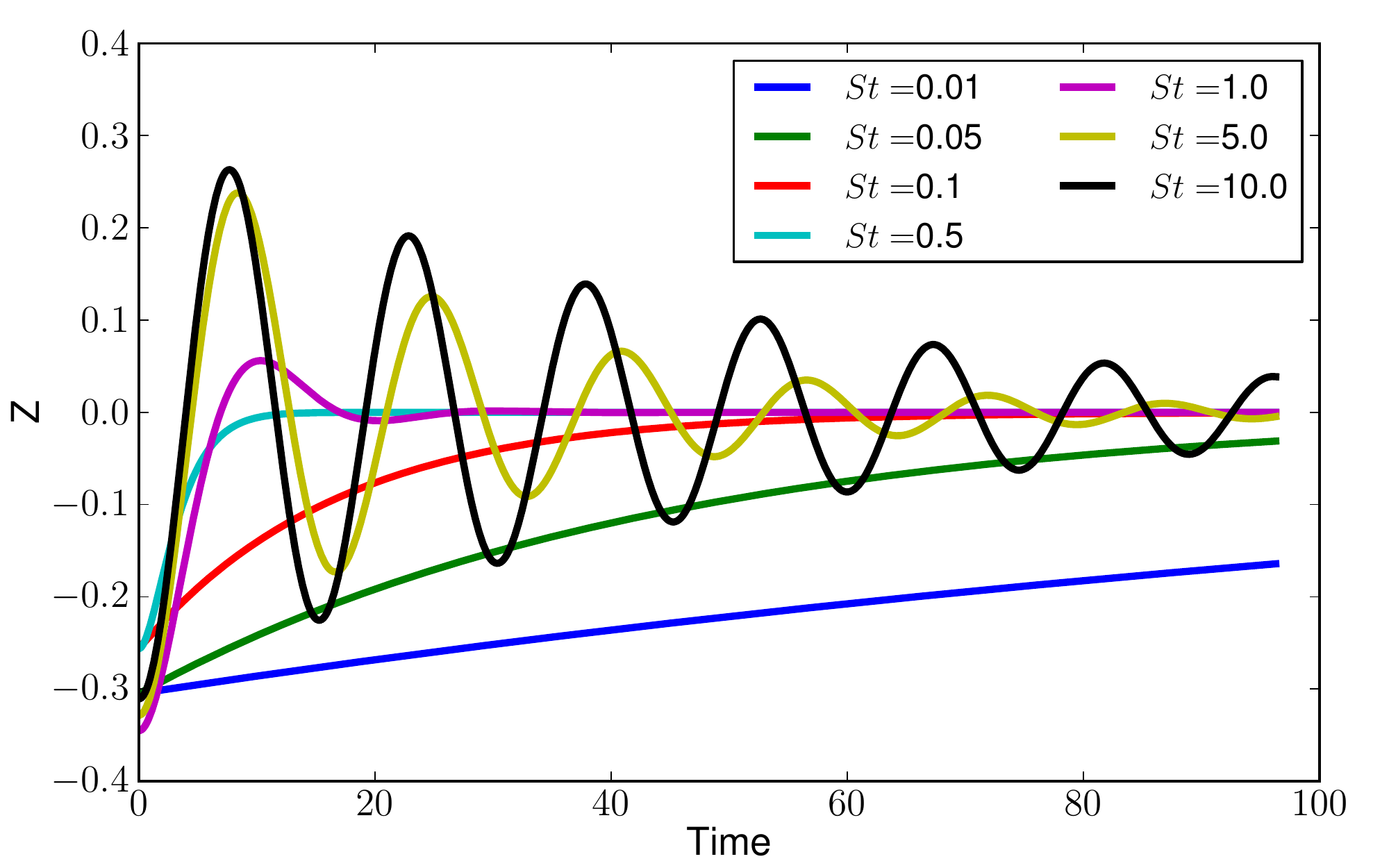}
\caption{{\it Upper panel:} Projected disc surface density along the line of sight characterized by the angles $(\theta,\phi)=(70^\circ, 80^\circ)$ for Model $7$ at $t=1000$ $T_{bin}$. }
\label{fig:test_settling}
\end{figure}

Here, we show that  the implementation of the particle solver is correct and accurately reproduces the behaviour of particles that settle in the disc 
midplane in the absence of turbulence. To this aim, we repeat the test presented in Stoll \& Kley (2016) and release particles with Stokes number in the range $[0.01,10]$ that are initially located one scale height below the midplane. For this particular test, the gas component does not evolve and the disc model that we used  corresponds to the initial conditions that were described in the main text.  In Fig. \ref{fig:test_settling}, we see that the analytical estimates are recovered, with particles whose Stokes number $St > 0.5$ undergoing damped oscillations around the disc midplane while those with $St<0.5$ suffering exponential decay of their vertical coordinate until they reach the midplane of the disc.

\end{document}